\documentstyle[praca,12pt,epsfig]{report}  

\pssilent

\title{Praca} 
\author{Andreas Freund}	
\mychapter{\large\bf \begin{center} \chapapp\ \thechapter \end{center} 
\begin{center} #1 \end{center} } {\large\bf \begin{center} #1 \end{center} }

\begin{document}	  
\pagestyle{plain}
%
%
\renewcommand{\thepage}{\roman{page}}
%
%
\thispagestyle{empty}
\dblesp

\noindent
\begin{center}
The Pennsylvania State University\\
The Graduate School\\

Eberly College of Science\\[1.5 cm]

{\large EXCLUSIVE, HARD DIFFRACTION IN QCD} \\[1.5 cm]

A Thesis in\\
Physics\\

by\\[1cm]
Andreas Freund\\[1.2 cm]
\snglsp
Submitted in Partial Fulfillment\\
of the Requirements\\
for the Degree of\\[1.2 cm]
Doctor of Philosophy\\[1.2 cm]

August 1998\\

\end{center}
%
%
%
%



%
%










%
%
\chapter*{Abstract}
\dblesp
\newcommand{\ABSTRACT}{
\indent

In the first chapter we give an introduction to hard diffractive 
scattering in QCD to introduce basic concepts and terminology, thus setting 
the stage for the following chapters.

In the second chapter we make predictions for nondiagonal parton distributions 
in a proton in the LLA.  We calculate the DGLAP-type evolution kernels in 
the LLA, solve the nondiagonal GLAP evolution equations with a modified 
version of the CTEQ-package and comment on the range of applicability 
of the LLA in the asymmetric regime. We show that the nondiagonal gluon 
distribution $g(x_{1},x_{2},t,\mu^2)$ can be well approximated at small $x$ 
by the conventional gluon density $xG(x,\mu^2)$.

In the third chapter, we discuss the algorithms used in the LO evolution 
program
for nondiagonal parton distributions in the DGLAP region and discuss the 
stability of the code.
Furthermore, we demonstrate that we can reproduce the case of the LO diagonal 
evolution within less than $0.5\%$ of the original code as developed by the 
CTEQ-collaboration.


In chapter 4, we show that factorization holds for the deeply virtual 
Compton scattering amplitude in QCD, up to power suppressed terms,
to all orders in perturbation theory. Furthermore, we show that the virtuality
of the produced photon does not influence the general theorem.


In chapter 5, we demonstrate that perturbative QCD allows one to calculate 
the absolute
cross section of diffractive exclusive production of photons
at large $Q^2$ at HERA, while the aligned jet model allows one to estimate 
the cross section for intermediate $Q^2 \sim 2 GeV^2$.  
 Furthermore, we find that the imaginary part of the amplitude for the 
production of real photons is larger than the imaginary part of the
corresponding  DIS amplitude, leading to predictions of a significant 
counting rate for the current generation of experiments at HERA. We also find
a large azimuthal angle asymmetry in $ep$ scattering for
HERA kinematics which allows one to directly measure the real part of the 
DVCS amplitude and hence the nondiagonal parton distributions.


In the last chapter, we propose a new methodology of gaining shape fits to 
nondiagonal parton distributions and, for the first time, to determine the 
ratio $\eta$ of the real to imaginary part of the DIS amplitude. We do this 
by using several recent fits 
to $F_2(x,Q^2)$ to compute the asymmetry $A$ for the combined 
DVCS and Bethe-Heitler cross section. The asymmetry $A$, isolates the 
interference term of DVCS and Bethe-Heitler in the total cross section, 
in other words, by isolating the real part of the DVCS amplitude through this 
asymmetry one has access to the nondiagonal parton 
distributions for the first time. Comparing the predictions for $A$ against 
experiment would allow one to make a prediction of the shape, 
though not absolute value, of nondiagonal parton distributions.

In the appendix, to illustrate an application of distributional methods as
discussed in chapter 4, we 
will show, with the aid of simple examples, how to make simple
estimates of the sizes of higher-order Feynman graphs. Our methods enable 
appropriate values of renormalization and factorization scales to be made.
They allow the diagnosis of the source of unusually large corrections that 
are in need of resummation.

}
\ABSTRACT

\dblesp

\tableofcontents 
\listoffigures	    

%
%

%
%

\pagebreak

\sloppy
\setcounter{page}{1}
\renewcommand{\thepage}{\arabic{page}}
%
%
\chapter{Introduction}
\indent

\section{What is Diffraction in QCD ?}
\label{ques}

The first question we will attempt to answer is: What is diffraction in 
QCD? The answer is, of course, not $42$ \cite{adams} but rather complicated 
and 
multi layered going to the heart of our understanding of QCD or lack thereof.

Before proceeding, we will introduce customary notation for variables used in 
describing inelastic and diffractive phenomena.

\begin{figure}
\centering
\mbox{\epsfig{file=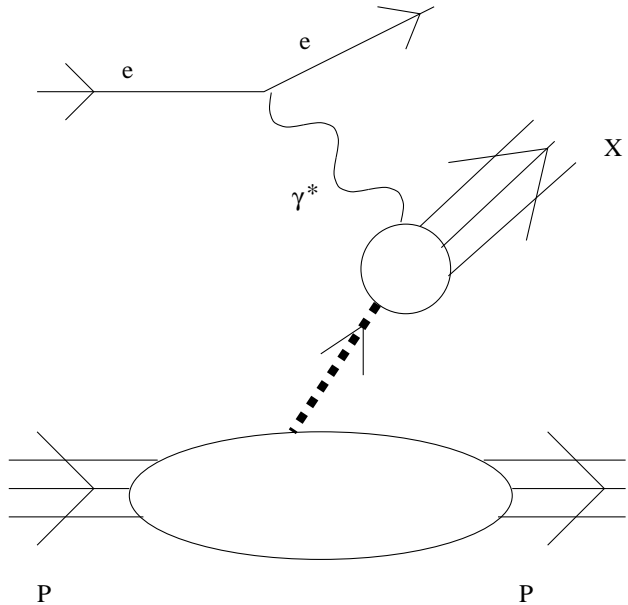,height=8cm}}
\vspace*{5mm}
\caption{Typical lepton-proton scattering with a final state $X$ clearly 
separated from the final state proton with the exchanged object between the 
virtual photon and the proton denoted by a dashed line.} 
\label{ep-exp}
\end{figure}

In the process
\bea
e(l) + p(p) &\rightarrow & e'(l') + X + P'(p')\nonumber\\
&\mbox{or} &\nonumber\\
p(l) + p(p) &\rightarrow & p(l') + p(p')\nonumber\\
&\mbox{or} &\nonumber\\
p(l) + p(p) &\rightarrow & X + p(p')
\label{diffracx}
\eea
with the momenta of the particles given in brackets. For defineteness and 
since this type of process will be of greatest relevance in the remainder of 
the thesis, we will introduce the relevant kinematics for the electron proton 
scattering. The center of mass energy squared is $s=(l+p)^2$ and the center of
mass squared of the hadronic system is
\beq
W^2 = (q+p)^2= - Q^2 + 2p\cdot q + m_{p}^2 = Q^2\left(-1+\frac{1}{x} \right )+ 
m_{p}^2,
\eeq
with $q=(-xp_+,Q^2/2xp_+,0_{\perp})$ \footnote{We define a vector in light 
cone coordinates by:
\begin{displaymath}
V^{\mu }=\left (V^{+},V^{-},V_{\perp } \right )=
\left ( \frac {V^{0}+V^{3}}{\sqrt {2}},
\frac {V^{0}-V^{3}}{\sqrt {2}},V^{1,2}\right ).
\end{displaymath}
} being the four momentum of the virtual photon exchanged between the lepton 
and the rest of the system (see Fig.\ \ref{ep-exp}) and $Q^2=-q^2$. 
The Bjorken $x$ is defined as:
\beq
x_{bj}=\frac{Q^2}{2p\cdot q}
\eeq
with $p$ being the four momentum of the incoming proton. 

$ep$ scattering, but also the rest of the reactions in (\ref{diffracx}), is 
said to be in the diffractive region if $X$ 
is sufficiently separated from the final state proton; the precise meaning of
this statement will be explained below.
$X$ is used in diffractive interactions to denote the part of the final state 
which might be in the case of $ep$ scattering, a real photon, a vector meson or
two jets. If the
final state proton dissociates as well, we will denote this state by $Y$ to 
indicate that it is still separated from the state $X$. The kinematics are the
following: The square of the momentum transfer at the proton vertex is 
\beq
t=(p-p')^2
\eeq
where $p'$ is the four momentum of the outgoing proton.
In the case of dissociation of the final state proton the proton momentum $p'$
 should be replaced by the four momentum of the outgoing system $Y$. A fourth 
Lorentz scalar is needed to define the kinematics of the hadronic part of the 
final state.

The fraction of the proton momentum carried by the pomeron\footnote{We will 
say more about the pomeron further down. For the moment we call the object 
exchanged between the proton and the incoming probe, the pomeron.} is
\beq
x_{P}=\frac{(p-p')\cdot q}{p\cdot q}=\frac{M_{X}^2 + Q^2 - t}{W^2 + Q^2 - m_{p}^2}
\eeq
where $M_{X}$ is the mass of the produced system. 
Note that the influence of $t$ and $m_{p}$ on 
$x_{P}$ is negligible for large $Q^2$ and $W^2$. 

The rapidity of a particle in the final state is given by
\beq
y=\frac{1}{2}\ln\left ( \frac{E+p_z}{E-p_z}\right ).
\eeq

Now consider the total hadronic cross section for processes like $pp,p\bar p,
\gamma p$, etc.\ as a function of the center of mass energy $\sqrt{s}$ 
\cite{rev}. Their total cross section can be described as
\beq
\sigma_{tot}^{h-h}=Ys^{-\eta}+Xs^{\epsilon}.
\label{sighad}
\eeq
which is the form predicted by Regge theory.
The first term describes the decrease of the cross section with $s$ at low 
energies and the second the slow increase at high energies. CDF \cite{cdf},
Cudell et al.\ \cite{cud} and originally Donnachie and Landshoff \cite{dl}
performed a fit using different hadronic data in each case to determine 
the values of $\eta$ and $\epsilon$ and thus obtain a universal description 
of the hadronic cross section. $\eta$ was found to be around $0.45$, with the 
results for $\epsilon$ varying on the data used for the fit with $0.08$ from
Donnachie and Landshoff to $0.112$ from CDF.

This behavior of the total cross section can be interpreted in terms of Regge
theory \cite{reg}. The hadronic reaction $A+B\rightarrow C+D$ can be described
by the $t$-channel exchange of a family of off-shell particles such that the 
relevant 
quantum numbers are conserved. There exists a theoretical relation between 
the spin $J$ and the mass squared ($t$) for these particles which has the 
the approximate form
\beq
J_i = \alpha_i (t) = \alpha_i (0) + \alpha_i 't.
\eeq
The particles lie on so-called ``Regge trajectories'', with intercept
$\alpha_i (0)$ and slope $\alpha_i '$ where i stands for the different 
particles.
The dependence of the elastic cross section with $t$ should behave as 
\beq
\frac{d\sigma_{el}}{dt}\sim \left ( \frac{s}{s_0} \right )^{2(\alpha_1(0)-1)}
e^{bt},~~\mbox{with}~~ b=b_0+2\alpha_1 ' \ln\left(\frac{s}{s_0}\right ).
\eeq

The term in Eq.\ (\ref{sighad}), dominant at low energies, was fitted in Ref.\
\cite{dl} and corresponds to the intercept $\alpha_{R}\simeq 0.5$ for reggeon 
exchange, i.e.\ , the degenerate $\rho,\omega,f$ and $a$ trajectories. The 
dominant term at high energies in Eq.\ (\ref{sighad}) corresponds to the 
so-called pomeron trajectory (see Ref.\ \cite{gribov}) with an intercept 
$\simeq 1.08$ according to 
the fit in \cite{dl}. The slope $\alpha_{P} '$ was fitted to be 
$\simeq 0.25~\mbox{GeV}^{-2}$, implying that the exponential $t$ distribution 
becomes steeper with increasing energy.

\begin{figure}
\centering
\mbox{\epsfig{file=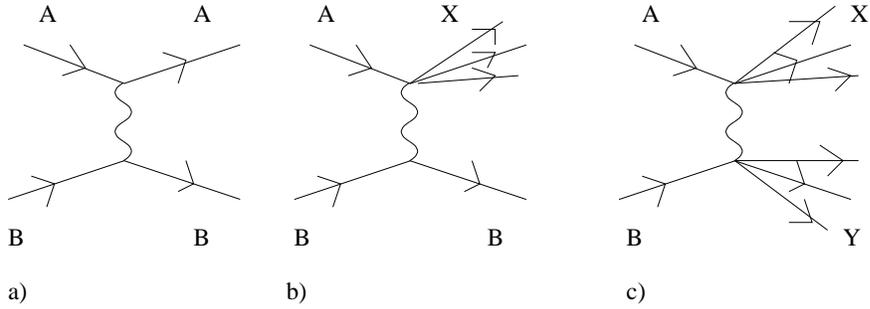,height=4cm}}
\vspace*{5mm}
\caption{a) elastic scattering. b) single diffractive scattering. c) double 
diffractive scattering.}
\label{class}
\end{figure}

The $d\sigma/dt$ distribution in p - p elastic scattering has a characteristic
 behavior with an exponential fall-off, a dip and a second exponential, which 
is reminiscent of the diffractive pattern of light by a circular aperture 
\cite{gul}, hence the name {\it Diffraction} was used to indicate a pomeron 
exchange. Processes mediated by pomeron exchange can be classified as elastic,
single diffractive and double diffractive as shown in Fig.\ \ref{class}. From
Fig.\ \ref{class}, due to the lack of additional final state particles created
through color interactions with the exchanged pomeron, it is also clear that 
the pomeron has to carry vacuum 
quantum numbers, i.e.\ , it has to be a color singlet and one expects to see 
a rapidity gap\footnote{A region in the detector with no particle tracks.} 
between the leading particle B and the system X.

The main open questions on the pomeron are: Does one have a universal ``soft''
pomeron or a continuum of different objects with different energy dependences 
etc.\ up to high energies ? Can it be treated as a particle with a partonic 
substructure \cite{ing} and if so what is the pomeron structure function? 
Does factorization hold or not?

In the following discussion on hard diffraction and the following chapters
some of these questions will be addressed.

\section{Hard Diffraction in QCD}
\label{harddiff}

Diffractive processes with one or several large scales, for example large 
$Q^2$ in deep-inelastic scattering (DIS), are called hard. Especially 
interesting are diffractive DIS processes at large $Q^2$ such as 
electro-production of longitudinally, polarized vector mesons and deeply
virtual compton scattering (DVCS), which we will treat in more depth later,
since they offer a unique opportunity to probe the interplay between soft and 
hard QCD physics.      
The fact that one observes a substantial number of diffractive like events at 
high $Q^2$ at the HERA, CDF and D0 experiments, is a signal for the interplay 
of hard and soft QCD phenomena. There are two effects which are expected to 
play an important role, color transparency for systems consisting of quarks 
and gluons contained within a small size configuration and color opacity for 
large
size configurations. Small size configurations lead to reactions dominated by 
hard processes with a cross section rising with energy, while reactions of 
large size configurations are dominated by soft processes. In order to test QCD
on a quantitative and qualitative level, the ability to separate these two 
regimes from one another becomes essential. In this context one can address 
the following issues:

\begin{itemize}

\item The dynamics of compact systems, color transparency and perturbative 
      color opacity in hard diffractive processes like electro-production of 
      longitudinally polarized vector mesons or DVCS.

\item QCD predictions for the high momentum behavior of wave functions of 
      hadrons consisting of light quarks and QCD physics of heavy quarkonia.

\item Violation of the DGLAP evolution equation. Due to unitarity 
      considerations the increase of parton densities at small $x$ will slow 
      down and it is believed that this effect 
      can be observed at HERA in hard diffractive processes, in the 
      measurements of the proton structure functions at moderate $Q^2$, as well
      as in the measurements of the structure functions of nuclei.

\item Semi-classical approximation to high energy interactions. In the limit
      of strong color fields which are typical in the HERA kinematical regime,
      new developments and tests of the semi-classical approach are possible
      \cite{heb}. 

\end{itemize}

\subsection{Theoretical Foundation}

In order to understand the issue of the interplay of soft and hard physics in
high energy reactions let us quickly classify them according to the number of 
scales involved \cite{amuell}.

\begin{itemize}

\item Soft QCD: Soft hadron collisions are usually considered to be processes 
      with a scale of $\sim 1$ fm. We have already discussed this topic
      in Sec.\ \ref{ques}.

\item Hard QCD evolution: This second class consists of hard processes which 
      are determined by two or more different scales in the interaction. DIS 
      as well as hard diffractive processes belong to this class. The hard 
      scale is provided by the virtuality of the photon or jets whereas the 
      soft scale is set by the size of the proton.

      In order to use pQCD one has to prove that the short distance part of the
      interaction factorizes from the soft one. This is achieved for the total
      cross section and special hard diffractive processes as for example
      diffractive vector meson production and DVCS \cite{C.F.S'96,ca} by
      proving QCD factorization theorems. These factorization theorems lead to
      BL and DGLAP type equations \cite{C.F.S'96,ca} describing the 
      evolution from large $Q^2$ to the scale of soft QCD processes, hence the
      soft physics ( in particular the soft pomeron ) enters as boundary 
      conditions to these equations.

      One has to mention that the characteristic attribute of two scale 
      processes is the violation of the pomeron pole factorization, i.e.\ , the
      description of the energy dependence of diffractive processes by a
      universal trajectory as mentioned before in Sec.\ \ref{ques}. Therefore
      one may observe different energy dependences for different external 
      particles and the energy dependence may change with $Q^2$.

\item BFKL Evolution: The original form of the BFKL equation was derived under
      the assumption of a small but fixed value of $\alpha_s$ with the scale
      of $\alpha_s$ set by the scale of the external particle. The BFKL 
      approximation therefore applies to processes with one large scale. HERA
      would offer a clean test by measuring high-$p_{\perp}$ forward jets with
      $p_{\perp}^2\simeq Q^2$ in low $x$ DIS as suggested by A.\ Mueller 
      \cite{muell91}. In diffractive DIS, semi-inclusive diffractive vector 
      production with large momentum transfer $t$ would be a place to look 
      for the BFKL pomeron.

\item Color Transparency and Perturbative Color Opacity: Color transparency
      is a phenomenon which describes within QCD the interaction of a small 
      size, color neutral parton configuration with a hadronic target. The
      essence of color transparency is expressed by the following formula
      \cite{blaet,fms,FRS} which follows from the factorization theorem for
      hard processes in QCD:
\beq
\sigma^{q\bar q}_{T} = 2 \alpha_s \pi^2 \frac{1}{N_{c}} Tr \left( \frac{L^2}{8}
\right ) xG(x,9/b^2),
\eeq
      where $L^2$ is the casimir operator of color $SU(3)$, b is the 
      transverse separation between the $q\bar q$ system and G stands for the 
      gluon distribution in the target.

      The name ``Color Transparency'' stems from the fact that high energy 
      processes are dominated by gluon exchange and that the cross section for
      a small size configuration is, through the gluon distribution, indirectly
      related to its size in the impact 
      parameter space. The decrease of the cross section with decreasing size 
      is partly compensated by the known increase of the gluon density. 
      Furthermore, at fixed b, the cross section increases with increasing 
      energy. Also, the interaction cross section at very high energies where 
      there is a large number of gluons in the target, becomes large, naturally
      leading to perturbative color opacity. Both phenomena are of particular
      relevance for photon induced high energy interactions as the photon
      fluctuates into a $q\bar q$ pair and most of the time these fluctuations 
      lead to a small size configuration with large $k_{\perp}$ between the 
      pair, effectively screening each other. 

\end{itemize}
          
A word of caution about this classification is in order. As always, the real 
world is more complicated and therefore there is not always a clear cut 
distinction between these four types of physics. We know, for example, that 
the spatial size of the known hadrons varies from the proton with a radius of 
$\simeq 0.8$ fm to the $\Upsilon$ with a radius of roughly $0.1$ fm. 
Therefore, $J/\Psi$ or $\Upsilon$ scattering of a proton belongs to the
class of hard QCD evolution, as mentioned above, over a wide kinematical range
\cite{lk}. However, with increasing energy 
the soft regime would dominate in most of the rapidity space of these reactions
as a result of diffusion from the large scale as given by the mass of the heavy
vector meson to the scale of soft QCD processes.

After the above classification and comments we are now ready to supplement the
bare statement from the previous section that diffractive events, whether hard
or soft, are characterized by a large rapidity gap. These large rapidity gaps 
can
be reasonably well described in terms of diffractive interactions as given by 
a phenomenological pomeron exchange. The basic idea behind this is the pomeron
having a parton content which can be probed in hard scattering processes 
\cite{ing}.

For small $q\bar q$ configurations, which usually occur at high energies, 
diffractive scattering is driven by a two gluon coupling, which is the 
simplest 
model of a pomeron, since two gluons form a color neutral object. It is 
important to realize that gluon radiation, which would fill the rapidity gap 
and coming from the pair of exchanged gluons, is suppressed. The issue here is
that of coherence in the radiation from a color neutral system when the color 
charges are almost at the same impact parameter. In this case, gluon radiation
 with small transverse momenta, required to generate long range color 
interactions capable of filling the rapidity gap, is known to be suppressed 
\cite{grib'67,zepp}, since such a gluon cannot resolve a colorless object. 
Radiation of gluons with large transverse momenta is suppressed as well, 
though in this instance due to the smallness of the coupling constant. This 
reasoning is also directly applicable to hard diffractive scattering 
initiated by a small size $q\bar q$ pair where the exchange of a colorless 
pair of a hard and a relatively soft gluon can be calculated in QCD. 
These processes are still of leading twist because the QCD factorization 
theorem is modified for processes with diffractive final states 
\cite{cfs'93,C97}. 

The above can be summarized in the following statement about the energy 
dependence of hard diffractive scattering: In the soft QCD regime which 
corresponds to large transverse separations of the $q\bar q$ pair, the parton 
model gives boundary conditions through e.g.\ the aligned jet model to the 
factorization theorem, hence it is substantially modified due 
to $Q^2$ evolution. In contrast to the parton model, one finds in QCD that the
 contribution of $q\bar q$ pairs with small b is only suppressed by 
$\alpha_s/k_{\perp}^2$ but the interaction cross section increases rapidly at 
small $x$ since it is proportional to the gluon density. Therefore, 
$\sigma_{\gamma^*N}$ may increase faster with $W^2$
compared to cross sections for hadron collisions since the probability of 
small size configurations in the wave function of a hadron is significantly 
smaller than in the photon wave function. 

\subsection{What kind of diffractive processes are calculable in QCD ?}

Recently, it has been understood and proved \cite{C.F.S'96,7,ca,jios,C97} that
perturbative QCD can be applied to photon induced, inclusive, hard 
diffractive scattering as observed at HERA \cite{C97} and photon induced,
exclusive hard diffraction as for example electro-production of longitudinally 
polarized vector mesons \cite{C.F.S'96}, DVCS \cite{7,ca,jios}, 
diffractive di-muon production \cite{ca} and high-$p_{\perp}$, di-jet 
production. In chapter 4, we will go into the details of factorization
of the DVCS and di-muon case and in chapters 5 and 6 we will study DVCS in 
more detail. 

\begin{figure}
\centering
\mbox{\epsfig{file=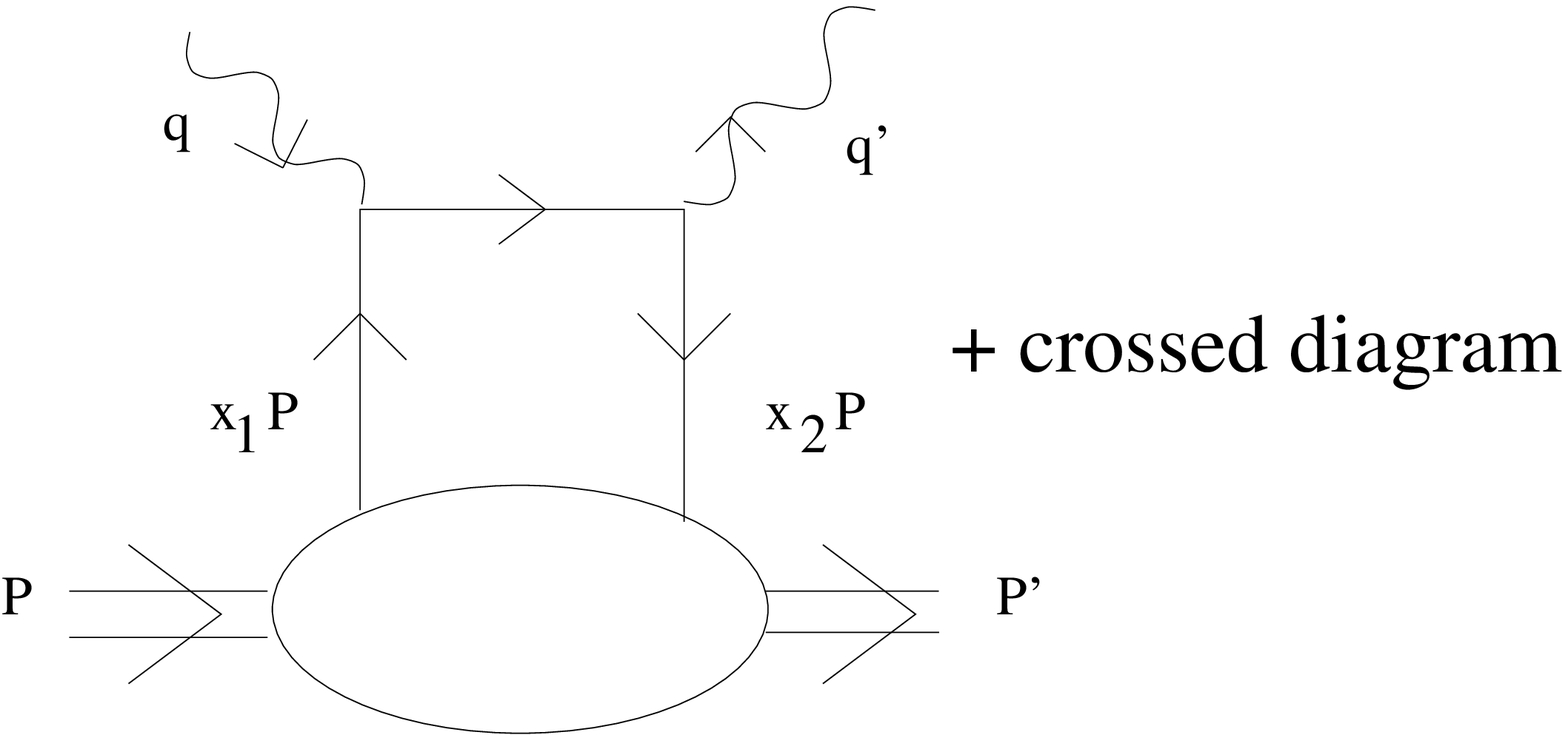,height=5.5cm}}
\caption{The lowest order handbag contribution to DVCS with $Q^2=-q^2$ and 
$q'^2=0$ as an example of a process where nondiagonal parton distributions 
are involved.}
\label{exampx}
\end{figure}

A feature particular to exclusive processes is that as a result of 
energy-momentum conservation, the fraction of proton momentum carried by the 
exchanged partons are not equal \cite{cfs'93,lm'89,Rad'96,Abram'95} leading
to nondiagonal parton distributions (see Fig.\ \ref{exampx}). In order to see 
this more clearly think 
of a parton being emitted from the proton with a fraction of the initial
proton momentum $x_1$ in the $+$ direction. In photon induced reactions the 
virtual photon adds the fraction $-x_{bj}$ to $x_1$ and in the case of DVCS 
for example, the produced final state which will be measured, i.e.\ , the real 
photon, does not carry any $+$ momentum away; thus the parton returning to
the proton has momentum $x_1-x$ which is not equal to the original momentum 
fraction $x_1$. 

The nondiagonal distributions, their leading order evolution and their 
applications particularly in the case of DVCS will be extensively studied in 
the remainder of this thesis. 

The investigation of these processes will provide novel information on the 
hadron structure and on the space-time development of QCD processes at high 
energies. They offer a unique possibility to measure generalized parton 
distributions in hadrons in a new way. They will also allow us to check QCD 
predictions for the asymptotic behavior of the light-cone wave functions of 
hadrons.

Chapter 2 is based on Ref.\ \cite{ours}, chapter 3 on Ref.\ \cite{AF97}, 
chapter 4 on Ref.\ \cite{ca}, chapter 5 on Ref.\ \cite{FFS}, chapter 6 on
Ref.\ \cite{new} and the appendix on Ref.\ \cite{apnd}.

\chapter{Nondiagonal Parton Distributions in the Leading Logarithmic Approximation}
\indent

\section{Introduction}

Due to the experimental possibility of probing nondiagonal distributions 
in hard diffractive electro-production processes, theoretical interest 
in this area in recent years 
\cite{Brod'94,FKS'95,Rad'96,C.F.S'96,Ji'96,7,Abram'95} has produced 
interesting results.
A pioneering analysis of the nondiagonal distributions for  
the diffractive photo production of $Z^0$-bosons in DIS where the 
applicability of PQCD is guaranteed was given by Bartels and Loewe in 1982 
\cite{Bartels} but went essentially unnoticed.

In this chapter, which is heavily based on Ref.\ \cite{ours}, we would like 
to complement these results by concrete predictions, albeit in the LLA, 
which can be tested by an experiment.  In Sec.\ \ref{sec:asym}
we shall demonstrate that in the limit of  small $x$ the amplitudes of
hard  diffractive processes can be calculated in terms of discontinuities 
of nondiagonal parton distributions. The real part of the amplitude will be 
calculated by applying a dispersion representation of the amplitude over
$x$. We will show that the term in the amplitude which cannot be calculated in 
terms of the discontinuities of nondiagonal parton distributions 
\cite{Rad'96,C.F.S'96} is suppressed by one power of $x$ in this
limit. This result will make it possible to calculate the evolution 
kernels in the LLA following the traditional methods \cite{Dok'80} and to 
compare them to results obtained in the QCD-string operator 
approach \cite{B.B'88}.  

In Sec.\ \ref{sec:kernel} we calculate the nondiagonal kernels and find them 
equivalent to those in \cite{Rad'96,Ji'96}. They are different from the 
evolution equations for nondiagonal parton densities which were presented
without derivation in \cite{Levin}. 
In Sec.\ \ref{sec:pred} we shall make predictions about the 
nondiagonal parton distributions by solving, numerically, the nondiagonal GLAP
evolution equations with the help of a modified version of the 
CTEQ-package. In Sec.\ \ref{sec:lim} we shall discuss the limitations of the 
approximation and the need for NLO-results. Future directions will be
discussed in the conclusions.

\section{Nondiagonal parton distributions and hard diffractive processes.}
\label{sec:asym}

It has been recently understood that the major difference in QCD
between leading twist effects in DIS and higher-twist effects in 
hard diffractive processes is to be attributed to the fact that the 
latter, initiated by highly virtual, longitudinally polarized
photons, can be calculated in terms of nondiagonal, rather than diagonal, 
parton distributions \cite{Abram'95}.

In order to calculate unambiguously hard two-body processes, 
it is necessary to calculate nondiagonal parton distributions in a 
nucleon. This implies knowledge of the non-perturbative nondiagonal parton 
distributions in the nucleon which have not been measured so far. Therefore,
the
aim of this section is to express the nondiagonal parton distributions in the 
nucleon through quantities being maximally close to the diagonal
parton distributions. Our second aim is to elucidate the kinematics of the 
nondiagonal parton distributions in the nucleon needed to describe hard 
diffractive processes. We shall also discuss the expected limiting 
behavior of the nondiagonal parton distributions. 

For the leading twist effects QCD evolution equations have traditionally 
been discussed in terms of parton distributions entering the imaginary part of
the amplitude. This is because the bulk of experimental data available is on the 
total cross section of inclusive processes.
This form of the evolution equation can be generalized to the case of hard 
diffractive processes \cite{FKS'95} and 
hard two-body processes \cite{C.F.S'96}. The analysis of the QCD evolution 
equation for the nondiagonal parton densities shows that the evolution equation
contains two terms. The first one is described by a GLAP-type evolution 
equation  \cite{FKS'95,Rad'96,C.F.S'96,Abram'95}, whereas the second term,
found in Ref.\ \cite{Rad'96} for vector meson production at 
small $x$, cannot be interpreted in terms of parton distributions. The QCD 
evolution of this term is governed by the Brodsky-Lepage evolution equation
\cite{Rad'96,7}.

\subsection{GLAP evolution equation for hard diffractive processes}
 
The aim of this section is to prove that for hard diffractive processes
in general, the $Q^2$-evolution
at any $x$ in the DGLAP-region as discussed below, is described by a 
nondiagonal GLAP-type evolution equation with asymmetric DGLAP-type kernels 
and that these processes can be 
calculated through the discontinuity of hard amplitudes. This property is 
important for the quantitative
calculations since the dispersive contribution has a relatively simple
physical interpretation and a deep relation with the conventional
parton densities. First, we shall deduce a relationship 
between amplitudes of hard two-body processes and parton densities,
and we will find an additional term which has no probabilistic interpretation.
We will restrict ourselves to the $Q^2$-region where the parton distributions 
are still rising and the additional term is of no importance as discussed 
below.

\begin{figure}
\centering
\mbox{\epsfig{file=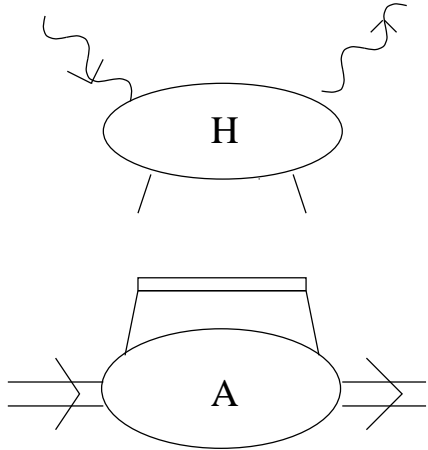,height=6cm}}
\vspace*{5mm}
\caption{Diagrammatic form of a factorized process in QCD.}
\label{factorx}
\end{figure}

The QCD factorization theorem for hard processes (see Fig.\ \ref{factorx})
means that the hard
part of the process can be factorized from the collinear, non-perturbative part
up to terms suppressed by powers of 
$Q$ \cite{Muell'89}. The topologically dominant Feynman diagrams for small $x$ 
processes correspond to attachments of only two gluons to the hard part.
Although our analysis is rather general, for certainty we shall
restrict ourselves to the case of diffractive processes where diagrams 
with two-gluon exchange dominate. \footnote{ 
Hard collisions due to the exchange of 2 quarks are numerically small in the 
LLA at small $x$.}
It is convenient to decompose the 
momentum of the exchanged gluon $k$ in Sudakov-type variables: 
\beq
k=x_{1} \tilde p  +\beta \tilde q  +k_{t},
\eeq
where 
\beq
\tilde p^2=\tilde q^2=0\ \ \mbox{and}\ \ (k_{t} q)=(k_{t} p)=0.
\eeq 
To express the amplitude in terms of nondiagonal parton distributions,
the contour of integration over $\beta$ should be closed over the 
singularities of the amplitude in gluon-nucleon scattering at fixed $x_1$ 
and $x_1-x$. The singularities over $\beta$ are located in the complex plane
of  discontinuities over the virtualities of the vertical propagators: 
${i\epsilon\over x_1}$ and ${i\epsilon\over x_1-x}$, and from the
$s$- and $u$- channel discontinuities: ${-i\epsilon\over(1-x_1)}$ and
${i\epsilon\over(1+x_1-x)}$. 
The amplitude differs from 0 depending on the different contributions of the 
poles given a certain contour of integration. This causality condition 
restricts the region of integration to: 
\beq 
-1+x\leq x_{1} \leq 1.
\eeq
Our main interest is in the amplitude within the physical region where 
$-t\geq 0$ but small as compared to other relevant scales of the process
under consideration. In this region the amplitude can be represented 
as the sum of terms having $s$- or $u$-channel singularities only.
For the $s$-channel contribution to the amplitude of hard 
diffractive processes,
relevant for the region $1\geq x_{1} \geq x$, the integral over $\beta$  can 
only be closed over the discontinuity in the amplitude of the gluon-nucleon 
scattering in the variable $s$. 

Therefore, this contribution to the amplitude is expressed through the
imaginary part  of the amplitude for gluon-nucleon scattering.
The QCD evolution of this term is described by a GLAP-type 
evolution equation where the kernel accounts for the off-diagonal kinematics. 
One also has to add a similar term corresponding to $u$-channel singularities.

The contribution of the region $x \geq x_1 \geq 0$ has no direct 
relationship to the conventional parton densities. This is 
because the integral over $\beta$  cannot be closed for $s$- or $u$-channel 
discontinuities but it may be closed for the  discontinuities over the 
gluon "mass". In Ref.\ \cite{Rad'96} the analogy of this term with the wave
function of a vector meson has been suggested. The presence of this
piece which cannot be evaluated in terms of parton densities
introduces theoretical uncertainties into the treatment of hard 
two-body processes at large and moderate $x$.

However, for the imaginary part of the amplitude, more severe 
restrictions on the region of integration apply: 
\beq
1 \geq x_{1} \geq x.
\eeq  
We only have to consider the discontinuity of the hard amplitude in the $s$-
channel. This restriction on $x_1$ for the $s$-channel contribution follows 
from the positivity requirement for the energy of the intermediate state 
in the $s$-channel cut. 
This result helps to prove that the piece which cannot be evaluated in terms
of parton densities is
inessential for hard diffractive processes. Let us now
apply a dispersion representation over the variable $s$ which reconstructs the
real part of the amplitude according to the following formula for small $x$
\cite{reim1,reim2}
\beq
Re A = Im A\frac{\pi}{2}\frac{d\,\ln(x Im A)}{d\, \ln{1\over x}}.
\label{reala}
\eeq   
The only term
which cannot be reconstructed in terms of a dispersion relation,
i.e.\ , in terms of discontinuities of parton densities,
is the subtraction constant\footnote{This constant is independent of $s$.} 
in the real part. The
contribution of the subtraction term to the amplitude with a positive 
signature, i.e.\ , symmetric under the transposition of  $s\rightarrow u$, 
is suppressed by an additional power of $s$ or, equivalently, by an additional
power of $x$. For the processes with negative charge parity in the 
crossed channel, i.e.  electro-production of a neutral pion, the amplitude is 
antisymmetric under the transposition\footnote{This 
corresponds to a negative signature.} $s\rightarrow u$.
This amplitude has no subtraction terms at all, since, in QCD, it increases 
with energy slower than\footnote{An odderon-type contribution in PQCD
is suppressed by an additional power of $Q^2$.} s. Therefore, in this case, 
a dispersion
representation gives the full description. To summarize let us point out once
more that the small $x$ behavior of hard diffractive processes is described 
through the discontinuities of hard amplitudes.

\subsection{Small $x_{i}$ behavior of the nondiagonal gluon distribution}

We want to stress that the slope of the $x$ dependence of the amplitudes for 
diffractive processes, however not their residue, should be independent of
the asymmetry between fractions $x_1$ and $x_2$ of the nucleon momentum carried
 by the initial and final gluons. This is due to the fact that
the  $x_i$ of the partons in the ladder are essential, but not the external 
$x$, and increase with the length of the parton ladder. Hence, the 
asymmetry between the 
gluons may be important in one or two rungs of the ladder but not 
in the whole ladder. Therefore, at sufficiently small $x$, it is legitimate to 
neglect $x$ in most of the rungs of the ladder as compared to the 
$x_{i}$. This means that the asymmetry between the gluons influences the 
residue but not the slope of the $x$ dependence.

Let us now discuss the small $x_{i}$ behavior of 
$g(x_{1},x_{2})$ -- the nondiagonal gluon density in a nucleon.
The factorization theorem -- Eq. (3) of  \cite{C.F.S'96}-- is the basis for 
the formal definition of the nondiagonal gluon density as the matrix
element of gauge-invariant bilocal operators (cf. Eq. (6) of \cite{C.F.S'96}):
\begin{eqnarray}
   g_{g/p}(x_{1},x_{2},t,\mu )  &=&
   - \int _{-\infty }^{\infty } \frac {dy^{-}}{4\pi }
   \, \frac {e^{-ix_{2}p^{+}y^{-}}}{p_{+}^2}
   \;
   \langle p'|T G_{\nu }{}^{+}(0,y^{-},{\bf 0}_{T}) \,
     {\cal P} \, G^{\nu +}(0)|p\rangle.
\label{eq.gluondis} 
\end{eqnarray}

Here $\cal P$ is a path ordered exponential of a gluon field along 
the light-like
line joining the two gluon operators, $t$ is the square of the 
invariant momentum transferred to the target, and $\mu$ describes the scale 
dependence. The sum over transverse gluon polarizations is implied. 
Actually Eq.\ (\ref{eq.gluondis}) coincides with the definition given in 
\cite{Rad'96,C.F.S'96} for the same quantity. 

For $x_{1}=x_{2}$, $g_{g/p}(x_{1},x_{2})$ is related to the diagonal gluon 
distribution as $xG_{{\rm diag}}(x)=g_{{\rm nondiag}}(x,x)$. Within the 
leading $\alpha_s \ln x$ approximation
where the difference between  $\ln x_{i}$  and $\ln x$ 
can be neglected, the nondiagonal distribution coincides with the diagonal one 
\cite{Brod'94}:  
\beq 
G_{{\rm leading\, \alpha_s \ln x}}(x_{1},x_{2},t=t_{\mbox{min}},\mu^2)= 
xG(x,\mu^2).
\eeq 

We want to stress here that at fixed Bjorken variable  $x$, the cross 
sections of 
hard diffractive processes are expressed through $g(x_{1},x_{2},t,\mu^2)$  
where $x_1-x_2=x$. This can be proved by calculating the
high energy limit of hard diffractive processes and then applying Ward
identities similar to Ref.\ \cite{Brod'94}. This means that the region of 
integration near $x_2=0$  ($x_2\ll x_1$) gives only a
small contribution to the amplitudes of hard diffractive processes.

\begin{figure}
\centering
\mbox{\epsfig{file=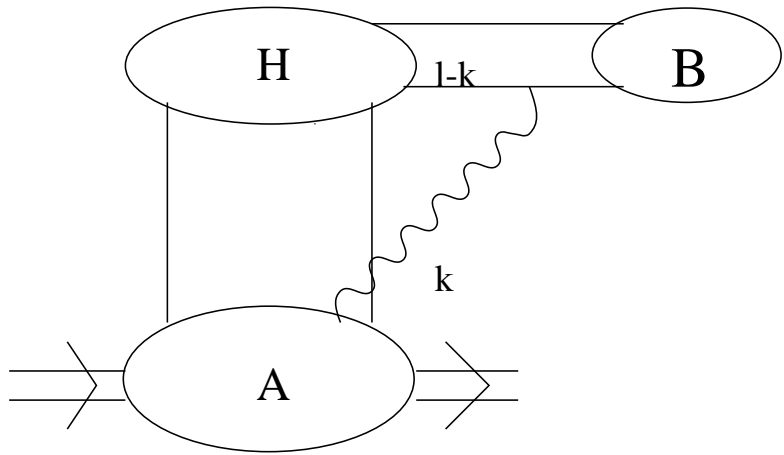,height=6cm}}
\vspace*{5mm}
\caption{Possible configuration in a parton distribution which could yield an 
enhancement for $x_2\rightarrow 0$ for a soft momentum $k$.}
\label{parton}
\end{figure}

Alternatively, one can examine the leading regions of the integrals 
in the calculation of the distribution of a parton in a parton (see 
Fig.\ \ref{parton}) which is imperative 
in finding the correct hard scattering coefficients for the desired process.
This calculation is necessary since one has not only ultraviolet divergences 
in the partonic cross sections from which one wants to extract the 
Wilson coefficients, but also infrared divergences stemming from initial-state 
collinearities of the participating partons (see Ref.\ \cite{Muell'89}
for further details) which are cancelled by the perturbatively calculated 
expansion of the parton distribution. The claim is that the region of 
$x_{2}=0$ does not give a leading contribution. This can be seen by 
using a simple argument that proper Feynman diagrams have no singularity at 
$x_{2}=0$, and the region of integration over the exchanged gluon momenta
$x_{2}=0$ forms an insignificant part of the permitted phase volume.

In the first step one has to show that a gluon with $x_2=0$ corresponds 
to a soft gluon and then one can use the argument by Collins and Sterman 
\cite{ColSter} first introduced for proving factorization in inclusive 
$e^+e^-$-reactions:
\begin{itemize}

\item For clarification, the quark-loop to which the gluons attach connects
      both the hard part and the part whose momenta are parallel to 
      the vector meson and of the order ${\nu\over m_N} = {2q\cdot p_N\over 
      m_N}$.

\item The minus component $l_-$ \footnote{We define a vector in light cone 
      coordinates by:
      \begin{displaymath}
      V^{\mu }=\left (V^{+},V^{-},V_{\perp } \right )=
      \left ( \frac {V^{0}+V^{3}}{\sqrt {2}},
      \frac {V^{0}-V^{3}}{\sqrt {2}},V^{1,2}\right ).
      \end{displaymath} and use the Breit frame as our frame of reference.}
      of the quark-loop  is of order $Q$. The minus component of the gluon 
      momentum $k$ is:
      \beq k_- \simeq \frac{m_N^2}{Q}\ll l_-.
      \eeq 
      Thus we can neglect $k_-$ with respect to $l_-$ in the
      calculation of the leading term of the amplitude corresponding to 
      the leading power in the energy of the process.

\item The transverse momentum in the quark loop ($l_t$) is cut off by the 
      vector meson wave function and thus $l_t^2\ll Q^2$ in stark contrast 
      to DIS, whereas the $k_t^2$ of the gluon is only restricted by the 
      virtualities in the photon wave function which can be as high as $Q^2$
      \footnote{ The similarity to DIS will be restored at extremely large 
      $Q^2$ as a consequence of both a Sudakov-type form factor in the photon 
      vertex and a slow decrease, with increasing $k_t$, of the vector meson
      wave function.}.          
      However one has to satisfy the Ward-identity $k_{\mu}T^{\mu\nu} = 0$
      where $T$ stands for the amplitude. Using Sudakov-variables this becomes:
      \beq
      x_2p_{\mu}T^{\mu\nu} + k_{t\mu}T^{\mu\nu} = 0.
     \eeq
      For $x_2\rightarrow 0$ the transverse momentum of the gluon is very
      small and can be safely neglected as compared to $l_t$.

\item One concludes from the above said that the $x_2=0$ region corresponds 
      to a soft gluon ($k^2\sim 0$) and we can now use the argument by 
      Collins and Sterman, as will be explained in the statements below.   

\end{itemize}

Keeping the above said in mind, the $k$-integral involved in the determination 
of the leading regions is of the form (up to overall factors):
\begin{eqnarray}
& &\int_{soft\, k}d^4 k\frac{1}{((l-k)^2+i\epsilon)(k^2+i\epsilon)}f(l-k,p)
\simeq \nonumber\\
& &\int_{soft\, k}dk_+ \frac{1}{((l_+ - k_+)l_{-}-l^2_t + i\epsilon)
(2k_+k_{-}-k^2_t + i\epsilon)}f(l-k,p),
\end{eqnarray}
where $f(l-k,p)$ is the amplitude of gluon-nucleon scattering and the 
integrals over $k_{-}$ and $k_{t}^2$ are suppressed for convenience.
If one now integrates over the remaining $k_+$ momentum one will have the 
following situations:
\begin{itemize}
\item $k_+k_-\geq k^2_t$: There are no obstructions in the 
      deformation of the integration contour since $l_t\gg k_t$.
      In other words this region does not give a leading contribution.
\item $k_+k_-\ll k^2_t$:  There are no obstructions to the contour 
      deformation since $l_t\gg k_t$, hence one only has one pole in $k_+$.
      Therefore, this region does not give a leading contribution either.
\end{itemize}

In conclusion, one has proved that if one of the gluons attaching the soft to 
the hard 
part has $x_{2}=0$, it will be soft and thus, according to the above 
reasoning, the $x_2=0$ region of integration does not give a leading 
contribution to the parton distribution.

\section{Kernels in the LLA}
\label{sec:kernel}

There are several possible ways of calculating the evolution kernels to 
leading order in QCD. We first used the traditional approach of calculating 
the evolution kernels in the LLA via the method of decay cells of e.g.
a quark decaying into a quark \cite{Dok'92}\footnote{
Changes appropriate to the nondiagonal case were made.
} and using cut-diagram techniques to calculate the appropriate Feynman graphs.

As a cross-check we calculated the first order corrections to the bi-local 
quark and gluon operators on the light-cone, which not only yielded the 
nondiagonal kernels for the DGLAP equation but also the nondiagonal 
Brodsky-Lepage kernels, since we were calculating the whole amplitude, not only
its imaginary part. However, since we are not interested in those kernels at 
the moment we will not comment on this fact further, let it be said though 
that our results on the Brodsky-Lepage kernels agree with those of
 Ref.\ \cite{Ji'96,7}.

We performed the calculation of the cross-check in a planar gauge 
i.e $q' \cdot A = 0$ with $q'^2\neq 0$\footnote
{ The advantage of such a physical gauge being that no gluons couple to the 
operators to first order, simplifying the calculations considerably. }
and used once more Sudakov-variables:
\beq
k = \beta p' + \alpha q'  +k_{t},
\eeq
where 
\beq
p'^2 = q'^2=0\ \ \mbox{and}\ \ (k_{t} q)=(k_{t} p)=0.
\eeq 
Since one is neglecting the proton mass one can set $p' = P$ where $P$
is the proton momentum.

The insertion of the appropriate bi-local operators for quarks and gluons 
on the light cone into the Feynman graphs for first order corrections to those
operators, short circuits the $+$-momentum in the graph, which means that the 
loop variable $k$ has not a $+$-momentum of $\beta P$ but rather $x_1 P$ 
(see \cite{Muell'89} for more details on calculating one loop corrections to 
parton distributions.). This fact rids us from the duty of taking the 
integral over $\beta$. In the calculation of the kernels, it remains to take 
the integral over $\alpha$ which can be done by taking the residues and then 
isolating the leading term multiplying $dk^2_{t}/k^2_{t}$ and the tree level 
amplitude. This will then yield the kernels in 
the leading logarithmic approximation.\footnote{ Note that the quark to quark 
and gluon to gluon kernels also need the self-energy diagrams to regulate 
the two possible collinear singularities, of course, after proper 
renormalization.} 

In the integral over $\alpha$, one finds three different residues. Two 
residues stemming from the vertical quark or gluon propagators which yield 
$\alpha = \frac{k^2_{t}}{x_1 s}$ and $\alpha = \frac{k^2_{t}}{x_2 s}$ giving
a contribution in the Brodsky-Lepage region, i.e.\ , the Brodsky-Lepage 
kernels, and one stemming from the horizontal propagator yielding 
$\alpha = \frac{k^2_{t}}{(y_1-x_1)s}$ which contributes to the DGLAP region,
i.e the DGLAP kernels. This is analogous to the statements made in 
Sec.\ \ref{sec:asym}.     

After having taken proper care of the definitions of our quark and gluon 
distributions in the amplitudes, we find the following expressions for the
nondiagonal evolution kernels, where $\Delta$ is given by $x_{1} - x_{2}$  
and corresponds to $x_{B_{j}}$ of e.g., vector-meson production.  
For the quark $\rightarrow$ quark transition we find:
\beq
P_{qq}(\frac{x_1}{y_1},\frac{\Delta}{y_1})=\frac{\alpha_s}{\pi} C_f \left [ 
\frac{\frac{x_1}{y_1} + \frac{x_1^{3}}{y_1^3} - 
\frac{\Delta}{y_1} (\frac{x_1}{y_1} + \frac{x_1^{2}}{y_1^2})}
{(1 - \frac{\Delta}{y_1}) (1 - \frac{x_1}{y_1})}_+ +\delta 
(1-\frac{x_1}{y_1})\frac{3}{2} \right ],
\eeq
The other kernels are computed the same way from the appropriate diagrams:
\begin{eqnarray}
P_{qg} (\frac{x_1}{y_1}, \frac{\Delta}{y_1}) &=& \frac{\alpha_s}{\pi} N_F 
\frac{[\frac{x_1^{3}}{y_1^3} + \frac{x_1}{y_1}(1 - \frac{x_1}{y_1})^{2} 
- \frac{x_1^{2} \Delta}{y_1^3}]}{(1 - \frac{\Delta}{y_1})^2},\\
P_{gq}(\frac{x_1}{y_1}, \frac{\Delta}{y_1}) &=& \frac{\alpha_s}{\pi} C_F 
\frac{[ 1 + (1 - \frac{x_1}{y_1})^{2} - \frac{\Delta}{y_1}]}{1 - \frac{\Delta}
{y_1}},\\
P_{gg}(\frac{x_1}{y_1}, \frac{\Delta}{y_1}) &=& \frac{\alpha_s}{\pi}N_{c} 
[2\frac{(1 - \frac{x_1}{y_1})^2 + (\frac{1}{2} - \frac{x_1^{2}}{y_1^2})
\frac{(x_1 - \Delta)}{y_1}}{(1 - \frac{\Delta}{y_1})^2} - 1 - \frac{x_1}{y_1}
+ \frac{1}{1 - \frac{x_1}{y_1}}_+\nonumber\\ 
& & + \frac{\frac{x_1 - \Delta}{y_1}}{(1 - \frac{x_1}{y_1})(1 - \frac{\Delta}
{y_1})}_+  + \delta (1-\frac{x_1}{y_1})\frac{\beta_0}{2N_C}],
\label{eq.kern}
\end{eqnarray}
with $\beta_0 = 11 - \frac{2n_f}{3}$.
A word concerning our regularization prescription is in order. In 
convoluting the 
above kernels, after appropriate scaling of $x_1$ and $\Delta$ with $y_1$,
with a nondiagonal parton density, one has to replace $z_1$ and $z_2$ in the 
regularization integrals with $z_1 \rightarrow (y_1 - x_1)/y_1$ and $z_2 
\rightarrow (y_1-x_1)/(y_1-\Delta)$. This leads to the following 
regularization prescription as employed in the modified version in the CTEQ 
package in the next section and in agreement with Ref.\ \cite{7}:
\begin{eqnarray}
\int^{1}_{x_1}\frac{dy_1}{y_1}\frac{f(y_1)}{1-x_1/y_1}_{+} &=& 
\int^{1}_{x_1}\frac{dy_1}{y_1}\frac{y_1f(y_1)-x_1f(x_1)}{y_1-x_1} 
+ f(x)\ln (1-x_1)\\
\int^{1}_{x_1}dy_1\frac{(x_1-\Delta)f(y_1)}{(y_1-x_1)(y_1-\Delta)}_{+} &=& 
\int^{1}_{x_1}\frac{dy_1}{y_1}\frac{y_1f(y_1)-x_1f(x_1)}{y_1-x_1}
-\int^{1}_{x_1}\frac{dy_1}{y_1}\frac{y_1f(y_1)-\Delta f(x_1)}
{y_1-\Delta}\nonumber\\ 
& & + f(x_1)\ln \left ( \frac{1-x_1}{1 - \Delta} \right ).
\end{eqnarray}  

For $\Delta = 0$ one obtains, necessarily, the diagonal kernels, however for
the distributions $q = x_1 Q(x_1,Q^2)$ and $g = x_1 G(x_1,Q^2)$, since we
chose the definitions of our nondiagonal distributions to go into $q=xQ(x,Q^2)$
and $g=xG(x,Q^2)$ 
rather then $Q$ and $g$ as in \cite{7}. We have cross-checked these results 
with those of 
Ref.\ \cite{B.B'88} via the conversion formulas given by Radyushkin in a 
recent paper \cite{Rad'96}. The formulas given in a recent paper by 
Ji \cite{Ji'96} do not seem to agree with ours but this is only due to a 
different choice of independent variables used by Ji. After appropriate 
transformations, the formulas of \cite{Ji'96} agree with our results 
\cite{7}. It should be noted however that the kernels from Ref.\ 
\cite{Rad'96,Ji'96,7} are given for $Q$ and $g$ and not for $q$ and $g$ 
as we do. One just has to multiply the kernels given by those authors for the 
quark evolution equations with $x_1/y_1$ after appropriate changes for 
independent variables of course. Conversion formulas between the different 
notations can be found in Ref.\ \cite{7}.  

The evolution equation for the quantities $g(x_1,\Delta)$
and $q(x_1,\Delta)$ take the following form in our notation:
\begin{eqnarray}
& &\frac{dg(x_{1},\Delta,Q^2)}{d\ln{Q^2}}=\int^{1}_{x_{1}} 
\frac{dy_{1}}{y_{1}}\left [ P_{gg}g(y_{1},\Delta,Q_0^2)+P_{gq}q(y_{1},
\Delta,Q_0^2)\right ]
\nonumber\\
& &\frac{dq(x_{1},\Delta,Q^2)}{d\ln{Q^2}}=\int^{1}_{x_{1}} 
\frac{dy_{1}}{y_{1}}\left [ P_{qq}q(y_{1},\Delta,Q_0^2)+P_{qg}g(y_{1},
\Delta,Q_0^2)\right ].
\label{eq:evol}
\end{eqnarray} 

We are interested in the calculation of the asymptotic
distribution in terms of the symmetric distribution in the limit 
of small $x$ and large $Q^2$. The reason why this is possible
is due to the fact that in this limit the main contribution originates 
from the nondiagonal distributions at 
$\tilde x_{1},\tilde x_{2}=\tilde x_{1}-\Delta$ with 
$\tilde x_1 \gg x_1$. In the case 
$\tilde x_1, \tilde x_2 \gg \Delta$ deviations from the diagonal 
distribution are small and can be neglected. 

In the following section we will present the results of our numerical
study and show that for the case of $x_{1}\gg x_{2}\simeq 0$
in the kinematic region of practical interest the diagonal and nondiagonal 
distribution will coincide for large $Q^2$ up to about a factor of $2$.

\section{Predictions for nondiagonal parton distributions}
\label{sec:pred}
Utilizing a modified version of the CTEQ-package, we calculate the 
evolution of the nondiagonal 
parton distributions, starting from a low $Q_{0}=1.6 GeV$ and 
with rather flat initial distributions for the diagonal and 
nondiagonal case by using the most recent global CTEQ-fit CTEQ4M \cite{cteq4}.
We chose $g(x_1,x_2) = x_1 G(x_1)$ in the normalization point, in accordance 
with our earlier argument that the possible difference in the distributions at 
small $x$ and large $Q^2$ is only given by the $Q^2$-evolution of 
$g(x_1,x_2,t,\mu^2)$.

We have only considered light quarks, since we are interested in a
proton as the initial state hadron and the $s$-quarks 
are only considered to give a small correction. 
The following figure (see Fig.\ \ref{Fig.1}) shows the ratio of the 
nondiagonal distribution $g(x_1,x_2)$ to the diagonal distribution $xG(x)$
from $Q =7~\mbox{GeV}$ to $Q =110~\mbox{GeV}$ and 
$x_2$ from $\frac{x_1}{100}$  
to $x_1$ with $x_1$ = $1.1\,10^{-4}$, $1.1\,10^{-3}$, $1.1\,10^{-2}$. 
 
The nondiagonal and diagonal distributions agree for 
$x_2 \rightarrow x_1$, i.e. for vanishing asymmetry, as expected, and within
a deviation of a factor between $0.2$ and $2$, they agree for 
$x_2\ll x_1$. The expectation that there is no $\ln x_2$ contribution in the 
parton distribution, which would give a 
singularity for $x_2 \rightarrow 0$, is also supported by our numerical 
calculations. 
   
Note that at large $Q^2$ and fixed $\Delta\ll 1$, $g(x_{1},x_{2})$ is 
determined by the initial parton distributions 
at $x_1,x_2 \gg \Delta$ where the validity of the diagonal approximation for 
$g(x_{1},x_{2})$ does not depend on our argument in Sec. \ref{sec:asym}.
The numerical calculation finds that the ratio of nondiagonal to diagonal 
distribution is larger than $1$ as 
anticipated by Radyushkin \cite{radpriv1} based on general arguments about 
the nature of the double distribution which he discusses in \cite{7}.

To see whether our numbers, i.e.\ , our numerical methods, could be trusted, 
we used a MATHEMATICA program to calculate the first iteration and the first 
derivative of the evolution to see how good or bad our numbers were. 
As it turns out our integration routines produce a very good agreement with 
the numbers from MATHEMATICA with a relative difference of $5\%$. This leads 
us to believe that our numbers can be trusted to high accuracy for $x_2$ of 
$O(x_1)$ and within $5\%$ at $x_2$ down by two orders of magnitude as compared
to $x_1$.

A few words about the nature of the modifications to the CTEQ-package are in 
order at this point. The basic idea we employed, was the following:
In the CTEQ package the parton distributions are given on a dynamical $x$- and 
$Q$-grid of variable size where the convolution of the kernels with the initial
distribution is performed on the $x$-grid. Due to the possibility of singular 
behavior of the integrands, we perform the convolution integrals by first 
splitting up the region of integration according to the number of grid 
points, analytically integrating between two grid points $x_i$ and $x_{i+1}$ 
and then adding up the contributions from the small intervals. We can do the 
integration analytically between two neighboring grid points by approximating 
the distribution function through a second order polynomial $ay^2 + by +c$, 
using the fact that we know the function on the grid points $x_{i-1},x_i$ and 
$x_{i+1}$ and can thus compute the coefficients a,b,c of the polynomial. 
This approximation is warranted if the function is well behaved and the 
neighboring grid points are close together. We treat the last integration 
between the points $x_1$ and $x_2$ (which are not to be confused with the 
$x_1$ and $x_2$ of the parton ladder) by taking the average of $x_1$ and 
$x_2$ and the values of the function at $x_1$ and $x_2$ and using those 
averages together with $x_1$, $x_2$ and the value of the function at $x_1$ 
and $x_2$ to compute the coefficients of the polynomial\footnote{See the next
chapter for an updated prescription. The results of the code are, however unchanged.}. 
The coefficients are
computed in the new subroutine NEWARRAY and the integration of the different 
terms in the kernels is performed in the new subroutine NINTEGR. 
Appropriate changes in the subroutines NSRHSM, NSRHSP and SNRHS were made
to accommodate the fact that the kernels and also the integration routines 
changed from the original CTEQ package. A detailed description of the code
will be in the next chapter; also see Ref.\ \cite{AF97}.

\section{Limitations of the LLA in the nondiagonal case}
\label{sec:lim}

The LLA approach of the previous sections accounts for the
contribution of a certain rather limited range of integration in the 
parton distributions. Regions outside these limits might 
contribute to the leading power. Looking at some other physical
quantities such as $F_{2}$, where one finds substantial 
modifications due to the NLO-terms, we are forced to assume that this may
be also true in our case. This results in the urgent need to carry out a NLO 
calculation and numerical 
study of the evolution equation, which will be the next step of our program.

\section{Conclusions and Outlook}

In summary, we have calculated the evolution kernels for nondiagonal parton
distributions in the LLA using traditional methods and found agreement with
the results of \cite{Rad'96,Ji'96,B.B'88} deduced by other methods. 
It was important to show that the 
traditional approach can still be applied and thus traditional methods can
be used to calculate systematically hard diffractive processes within 
the NLO approximations. We have also proved the similarity between the
diagonal and nondiagonal parton distributions. The latter ones determine
the cross sections of hard diffractive processes in the small $x$ region.
We have made predictions about the nondiagonal parton distributions
within the LLA with the help of a modified version of the 
CTEQ-package. Numerical calculations found the diagonal and 
nondiagonal gluon distributions, which dominate hard diffractive processes, 
to be very similar at small $x$ as expected from the previous discussion.

\begin{figure}
\centering
\mbox{\epsfig{file=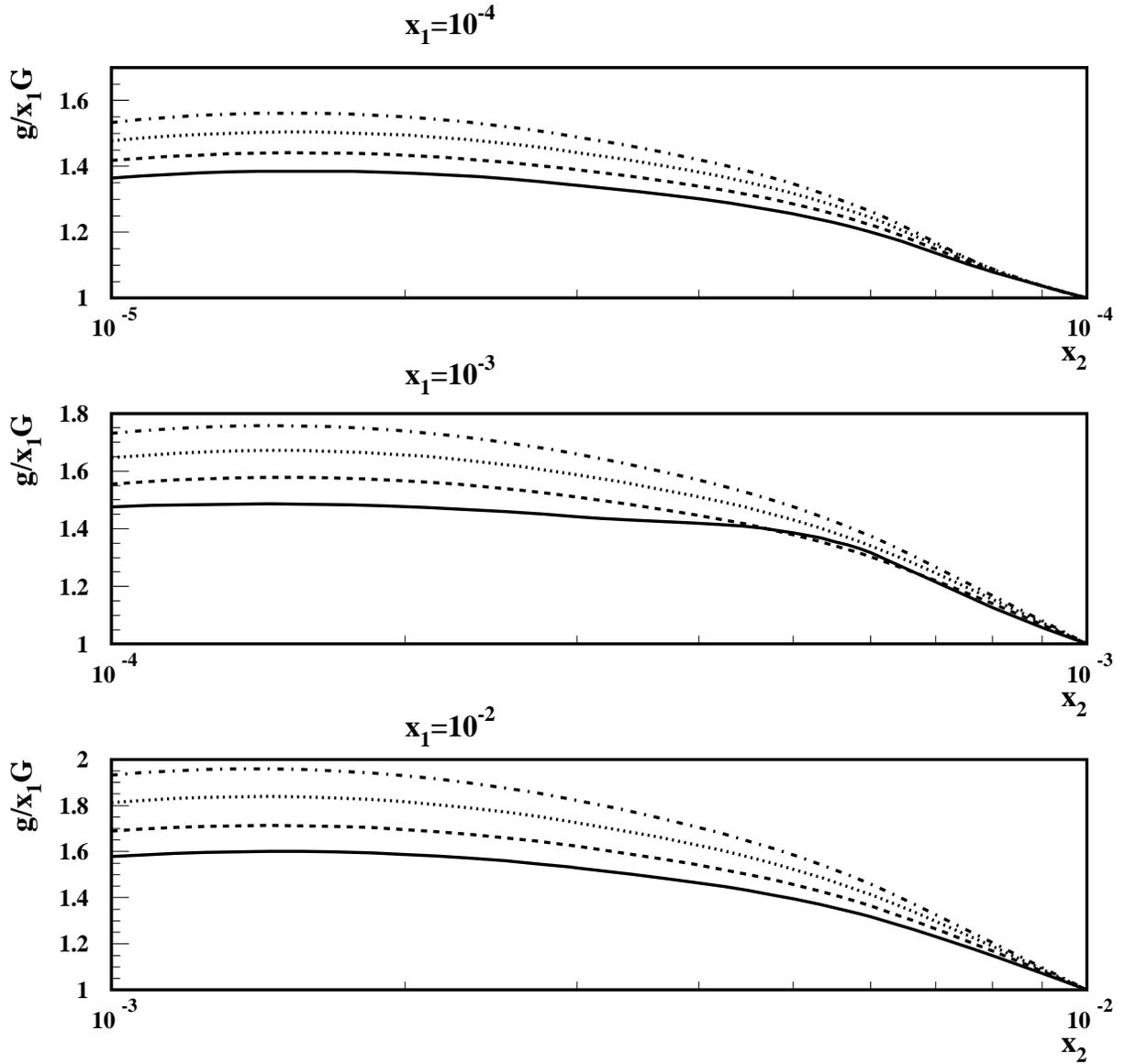,height=19.5cm,width=14cm}}
\caption{The fraction $g(x_1,x_2)/x_1G(x_1)$ as a function 
of $x_2$ for fixed $x_1$ and $Q=9.3~\mbox{(solid line)}$, $~17.1~
\mbox{(dotted line)}$, $~40.3~\mbox{(dashed line)}$ and $110~
\mbox{(dashed-dotted line)}~\mbox{GeV}^2$.}
\label{Fig.1}
\end{figure}

\chapter{Methods in the LO Evolution of Nondiagonal Parton 
Distributions: The DGLAP Case} 
\indent

\section{Introduction}
\label{intro}

In this chapter, which is based on Ref.\ \cite{AF97}, we give an exposition 
of the algorithms used to numerically 
solve the generalized GLAP-evolution equations of the last chapter. 
The main part of the evolution program was taken over from the CTEQ package 
for the diagonal parton distributions
from inclusive reactions. At this point in time the evolution kernels for
generalized parton distributions are known only to leading order in 
$\alpha_s$, as pointed out previously and thus our analysis will be a leading 
order one.

This chapter is organized in the following way. In Sec.\ \ref{def} we will 
quickly review the formal expressions for the parton distributions and the 
evolution equations together with the explicit expressions for the kernels 
and a first comment on the resulting numerical problems. In Sec.\ \ref{num1} 
we 
will explain the difference between our algorithms and the ones used in the 
original CTEQ package and then give a detailed account of how we implemented 
our algorithms. In 
Sec.\ \ref{num2} we demonstrate the stability of our code and show that we 
reproduce the case of the usual or diagonal parton distributions within $0.5\%$
for a vanishing asymmetry factor. Sec.\ \ref{concl} contains concluding
remarks.

\section{Review of Nondiagonal Parton Distributions, Evolution Equations and
Kernels}
\label{def}

\subsection{Nondiagonal Parton Distributions}
\label{ndpd}

\begin{figure}
\centering
\mbox{\epsfig{file=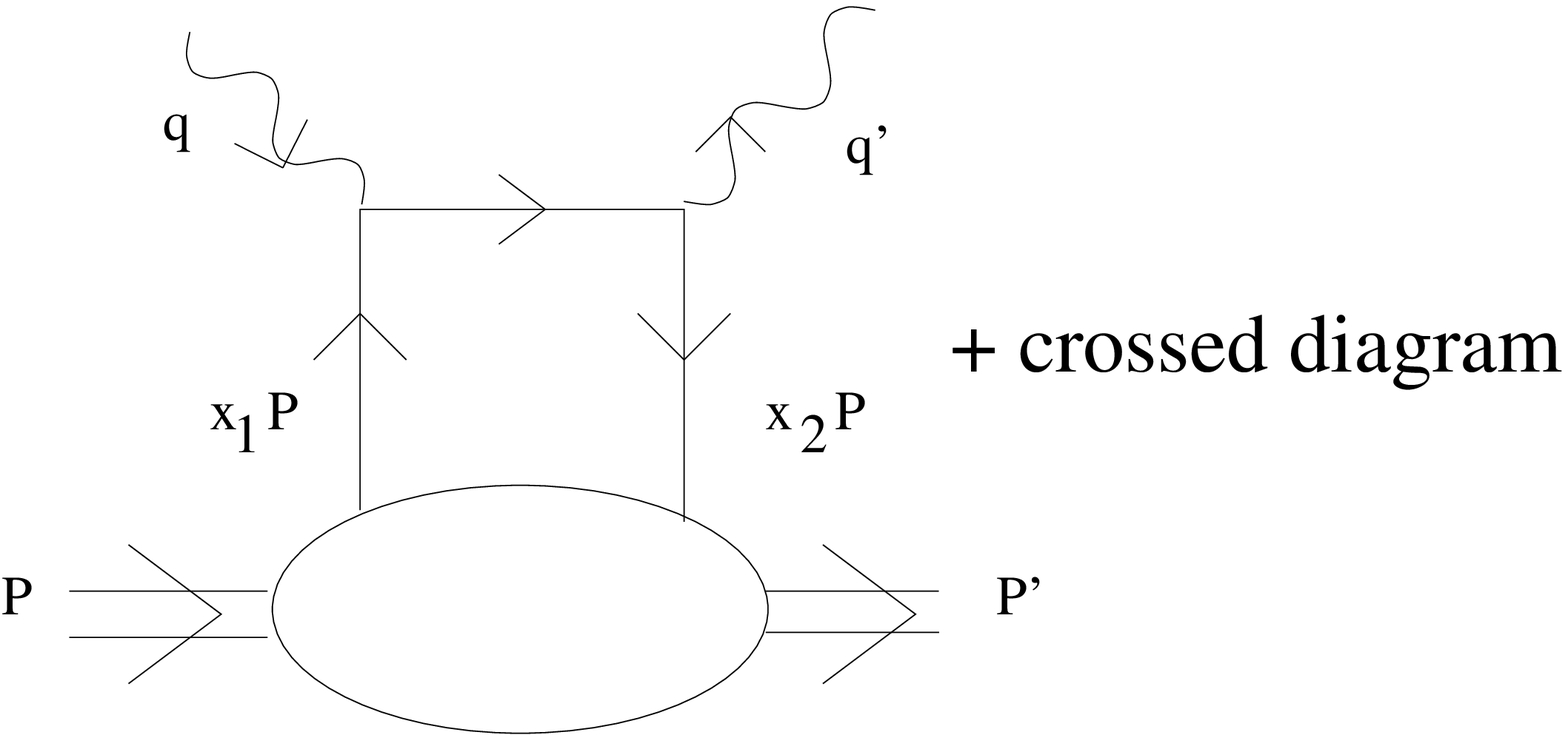,height=5.5cm}}
\caption{The lowest order handbag contribution to DVCS with $Q^2=-q^2$ and 
$q'^2=0$.}
\label{hand}
\end{figure}

As explained in the previous chapter, generalized or nondiagonal parton 
distributions occur for example in exclusive, hard diffractive $J/\psi$ or 
$\rho$ meson production or 
alternatively in deeply virtual Compton scattering (DVCS), where a real photon
is produced\footnote{We will say more about DVCS in the following chapters. 
See also Ref.\ \cite{Ji'96,7,muell,Ji1,BM,DGPR,chen,FFS,man,ca,jios}}. 
As mentioned in Sec.\ \ref{intro}, since one imposes the condition
of exclusiveness on top of the diffraction condition, one has a kinematic 
situation in which there is a non-zero momentum transfer onto the target 
proton as evidenced for example by the lowest order ``handbag'' diagram of 
DVCS in Fig.\ \ref{hand}.

The nondiagonal quark and gluon distributions have the following formal
definition as matrix elements of bilocal, path-ordered and renormalized quark 
and gluon operators sandwiched between different momentum states of the proton
as in the factorization theorems for exclusive vector meson 
production \cite{C.F.S'96} and DVCS \cite{7,ca,jios}:
\bea
& &f_{q/p}= \int^{\infty}_{-\infty}\frac{dy^-}{4\pi}e^{-ix_2p^+y^-}\langle p|
T\bar \psi(0,y^-,{\bf 0_{\perp}})\gamma^+{\it P}\psi(0)|p'\rangle,\nonumber\\
& &f_{g/p}= -\int^{\infty}_{-\infty}\frac{dy^-}{2\pi}\frac{1}{x_1x_2p^+}
e^{-ix_2p^+y^-}\langle p|
T G_{\nu}^+(0,y^-,{\bf 0_{\perp}}){\it P}G^{\nu+}(0)|p'\rangle.
\eea
with $x_2 = x_1 - \Delta$ where the asymmetry or nondiagonality parameter 
$\Delta$ is usually $x_{Bj}$ in, for example, DVCS or exclusive vector meson
production however not in diffractive di-muon production where 
$\Delta = x_{Bj} + \xi_{1}$ and $\xi_1$ is the longitudinal momentum fraction
of the produced time-like photon decaying into a $\mu^{+}\mu^{-}$-pair.
   
\subsection{The GLAP-Evolution Equations for Nondiagonal Parton Distributions}
\label{ee}

Reprising on the discussion of chapter 2, the GLAP-evolution equations follow 
from the usual renormalization group 
transformation of the parton distributions and lead to the following evolution
equations for the singlet(S) and non-singlet(NS) case \cite{ours,Ji'96,7}:
\bea
\frac{dq_{NS}(x_{1},\Delta,Q^2)}{d\ln{Q^2}}&=&\int^{1}_{x_{1}} 
\frac{dy_{1}}{y_{1}}P^{NS}_{qq}q_{NS}(y_{1},\Delta,Q_0^2),\nonumber\\
\frac{dg_S(x_{1},\Delta,Q^2)}{d\ln{Q^2}}&=&\int^{1}_{x_{1}} 
\frac{dy_{1}}{y_{1}}\left [ P^{S}_{gg}g_S(y_{1},\Delta,Q_0^2)+P^{S}_{gq}
q_S(y_{1},\Delta,Q_0^2)\right ],\nonumber\\
\frac{dq_S(x_{1},\Delta,Q^2)}{d\ln{Q^2}}&=&\int^{1}_{x_{1}} 
\frac{dy_{1}}{y_{1}}\left [ P^{S}_{qq}q_S(y_{1},\Delta,Q_0^2)+P^{S}_{qg}
g_S(y_{1},\Delta,Q_0^2)\right ].
\label{evol}
\eea
Note that $q_{S,NS} = x_1 f_{q/p}$, $g_S = x_1x_2f_{g/p}$ and the kernels 
to leading order\footnote{For more details on the derivation of the kernels 
to leading order see for example \cite{ours,Ji'96,7}.} were given in the 
previous chapter together with a discussion on the the generalized $+$ 
regularization prescription.

\section{Differences between the CTEQ and our Algorithms}
\label{num1}

Let us point out in the beginning that our code is to $99\%$ the original
CTEQ-code (for more on this code see Ref.\ \cite{cteq}). 
We only modified the subroutines NSRHSM, NSRHSP and SNRHS within
the subroutine EVOLVE and added the subroutines NEWARRAY and NINTEGR. These
routines are only dealing with the convolution integrals but not with, for 
example, the $Q^2$-integration or any other part of the CTEQ-code which remains
unchanged. This is due to the fact that the main difference between the 
diagonal and nondiagonal evolution stems from the different kernels which only
influence the convolution integration and nothing else. 

In order to make the simple changes in the existing routines more obvious we 
will first deal with the new subroutines.

\subsection{NEWARRAY and NINTEGR}
\label{nsubs}

Due to the increased complexity of the convolution integrals as compared to
the diagonal case as pointed out in Sec.\ \ref{ee}, we were forced to 
slightly change the very elegant and fast integration routines employed
in the original CTEQ-code. The basic idea, very close to the one in the 
CTEQ-code, is the following:
Within the CTEQ package, the parton distributions are given on a dynamical 
$x$- and 
$Q$-grid of variable size where the convolution of the kernels with the initial
distribution is performed on the $x$-grid. Due to the possibility of singular 
behavior of the integrands, we perform the convolution integrals by first 
splitting up the region of integration according to the number of grid 
points in $x$, analytically integrating between two grid points $x_i$ and 
$x_{i+1}$ where $i$ runs from $1$ to the specified number of points in $x$ 
and then adding up the contributions from the small intervals as exemplified 
in the following equation:
\beq
\int^1_{x_1}\frac{dy_1}{y_1}Kf(x_1/y_1,\Delta/y_1,y_1) = \Sigma_{i=0}^{N}
\int^{x_{i+1}}_{x_i}\frac{dy_1}{y_1} Kf(x_1/y_1,\Delta/y_1,y_1),
\label{split}
\eeq
where $Kf(x_1/y_1,\Delta/y_1,y_1)$ is the product of the initial distribution 
for each evolution
step and an evolution kernel with $x_0=x_1$, $x_N=1$. We can do the 
integration analytically between two neighboring grid points by approximating 
the distribution function $f(y_1)$ through a second order polynomial 
$ay_1^2 + by_1 +c$, using the fact that we know the function on the grid points 
$x_{i-1},x_i$ and $x_{i+1}$ and can thus compute the coefficients a,b,c of 
the polynomial in the following way, given the function is well behaved and 
the neighboring grid points are close together \footnote{The parton 
distributions functions are smooth and well behaved thus one just has to use 
enough points in $x$.}:
\bea
f(x_{1+1}) &=& ax^2_{i+1}+bx_{i+1}+c\nonumber\\
f(x_i) &=& ax^2_i+bx_i +c\nonumber\\
f(x_{i-1}) &=& ax^2_{i-1}+bx_{i-1}+c
\eea
which yields a $3\times 3$ matrix relating the coefficients of the polynomial 
to the values of the distribution functions at $x_{i-1},x_i$ and $x_{i+1}$.
Inverting this matrix in the usual way one obtains a matrix relating the $x$
values of the distribution function to the coefficients making it possible to 
compute them just from the knowledge of the different $x$ values and the 
value of the distribution function at those $x$ values. This calculation is 
implemented in NEWARRAY where the initial distribution is handed to the 
subroutine and the coefficient array is then returned. The coefficient array
in which the values of the coefficients for the integration
are stored, has $3$ times the size of the user-specified
number of points in $x$ since we have $3$ coefficients for each bin in $x$.  
We treat the last integration between the points $x_0$ and $x_1$ by approximating the 
distribution in this last bin through a second order
polynomial. However, for this last bin, the coefficients are computed using
the last three values in $x$ and of the distribution at those points, since the
point $x_{-1}$ which would be required according to the above prescription for
calculating the coefficients, does not exist.

After having regrouped the terms appearing in the convolution integral in such
a way that all the necessary cancelations of large terms occur within the 
analytic expression for the integral and not between different parts of the 
convolution integral, the integration of the different terms is performed in 
the new subroutine NINTEGR with the aid of the coefficient array from NEWARRAY.

As mentioned above the convolution integral from $x_1$ to $1$ is split up 
into several intervals in which the integration is 
carried out analytically. To give an example of this procedure we consider 
the convolution integral of $P_{qg}(x_1/y_1,\Delta/y_1)$ with the 
parton distribution $g_S(y_1)$:
\beq
\int^1_{x_1}\frac{dy_1}{y_1}P_{qg}g_S = \int^1_{x_1}\frac{dy_1}{y_1}
\frac{x^2_1(x_1-\Delta)}{y_1(y_1-\Delta)^2}g_S(y_1)+\int^1_{x_1}
\frac{dy_1}{y_1}\frac{x_1(y_1-x_1)^2}{y_1(y_1-\Delta)^2}g_S(y_1) 
\label{examp}
\eeq
suppressing presently irrelevant factors in front of the integral. 
The two parts
in Eq.\ (\ref{examp}) are calculated in different parts of NINTEGR
and then put together in either NSRHSM, NSRHSP or SNRHS.

In NINTEGR the integrals are split up according to Eq.\ (\ref{split}) and then 
analytically evaluated in the different $x$-bins \footnote{The general 
analytic expressions for the convolution integrals in an arbitrary $x$-bin 
were obtained with the help of MATHEMATICA.}. 
If the dependence of
the integrand on $\Delta$ is only of a multiplicative nature it is enough
to compute the integral for each bin once. To get the value of the 
convolution integral for a term with such a $\Delta$ \footnote{The value of 
$\Delta$ is specified in NINTEGR.} dependence, it is enough 
to store the result of the integration in the bin from $x_{N-1}$ to $x_N$
in the output array for this term at the position $N-1$ \footnote{The value of
the output at position $N$ is always $0$ since in this case the upper and the 
lower bound of the integral coincide.}, 
add to this result the value of the integral in the bin from $x_{N-2}$ to 
$x_{N-1}$ and store it at the 
position $N-2$ and so forth. In this manner one only has to calculate $N-1$
integrals, however if the integrand has a more complicated dependence on
$\Delta$ like $x_1-\Delta$ one needs to compute $N(N-1)/2$ integrals. 
For example in order to find the integration value for the $x_{N-1}$ bin
with $x_1=x_{N-1}$ one needs only one integral but at $x_{N-2}$ we have to 
redo our integral for the $x_{N-1}$ bin since $x_1=x_{N-2}$ plus we need to 
add the contribution from the $x_{N-2}$ bin to get the correct answer for the
output array at position $N-2$ and so forth. This need for additional
evaluations of integrals slows the program down but in the end it turns
out to be only about a factor of $4-5$ slower than the original CTEQ-code 
which is speed optimized. The integral with the regular + - prescription
is evaluated using the routine HINTEG from the original CTEQ-code whereas the 
generalized + - prescription is evaluated according to the methods described
above due to its nontrivial dependence on $x_1$ and $\Delta$. 

In the case of $\Delta<<x_1$ 
the analytic expressions obtained for the above general case are expanded to 
first order in $\Delta$ and then the same methods as above for evaluating the 
integrals are applied. The last case also allows us to go to the diagonal case
by setting $\Delta=0$ without using the integration routines from the 
original CTEQ-code giving us a valuable tool to compare our code to the 
original one.

\subsection{Modifications in NSRHSM, NSRHSP and SNRHS}
\label{mod}

The modification in the already existing routines NSRHSM, NSRHSP and SNRHS 
of the original CTEQ package are 
rather trivial. The most notable difference is that the subroutine NEWARRAY 
is called every time either of the three subroutines is called since the 
distribution function handed down on an array changes with every call of
NSRHSM, NSRHSP and SNRHS. In NSRHSM and NSRHSP, NEWARRAY is only called once
since one is only dealing with the non-singlet part containing no gluons, 
whereas in SNRHS the subroutine for the singlet case, one needs a coefficient
array for both the quark and the gluon. Besides this change, the calls for 
INTEGR are replaced by NINTEGR according to how the convolution integral has 
been regrouped as explained in Sec.\ \ref{nsubs}. The different regrouped 
expressions are then added, after integration for different $x$-values, to obtain the final 
answer in an 
output array which is handed back to the subroutine EVOLVE. 
The method is the same as in the 
original CTEQ-code but the terms themselves have changed of course.

\section{Code Analysis}  
\label{num2}

As a first step we tested the stability and speed of convergence of the code 
and found that by doubling the number of points in the $x$-grid, which is only
relevant for the convolution integral, from $50$ to $300$ the result of our 
calculation changed by less than $0.5\%$, hence we can assume that our code 
converges rather rapidly. We also found the code to be stable down to an 
$x_2=10^{-10}$ beyond which we did not test. Furthermore we can reproduce
the result of the original CTEQ-code, i.e.\ the diagonal case in LO within 
$0.4\%$ giving us confidence that our code works well since the analytic
expressions for the diagonal case are the expansions of the general case 
of non-vanishing asymmetry up to, but not including, $O(\Delta^2)$.  

In the following figures (Fig.\ \ref{nddratio}-\ref{nddratio5}) 
we compare, for illustrative purposes, the diagonal and nondiagonal case by 
plotting the ratio
\begin{eqnarray}
R_g(x_1,x_2,Q^2) = \frac{g(x_1,x_2,Q^2)}{x_1G(x_1,Q^2)}\nonumber\\
R_q(x_1,x_2,Q^2) = \frac{q(x_1,x_2,Q^2)}{x_1Q(x_1,Q^2)},
\end{eqnarray}
for various values of $x_1$, $Q^2$ and $\Delta=x_{Bj}$ \footnote{We also plot 
the same ratio for $\Delta = 0$ to demonstrate the deviation from our code in 
the diagonal limit from the CTEQ-code.},i.e.\ , varying $x_2$, using the CTEQ4M and CTEQ4LQ 
\footnote{CTEQ4LQ gives 
the best fit at low $Q^2$ whereas CTEQ4M gives the best $\chi^2$-fit for a 
large range of $Q$ and $x$.} parameterizations \cite{cteq4}. We assume  the 
same initial conditions for the diagonal and nondiagonal case 
(see Ref.\ \cite{ours} for a detailed physical motivation of this ansatz).

The reader might wonder why only CTEQ4M and CTEQ4LQ and not GRV or MRS were 
used. The answer is not 
a prejudice of we against GRV or MRS but rather the fact that a 
comparison of CTEQ4M and CTEQ4LQ shows the same characteristic as comparing,
for example, CTEQ4M and GRV at LO. The observation is the following:
CTEQ4LQ is given at a different, rather low, $Q$, as compared to CTEQ4M and
hence one has significant corrections from NLO terms in the evolution at 
these scales. This leads to a large difference between CTEQ4LQ and CTEQ4M 
(see Fig.\ \ref{nddratio6}), if one evolves the CTEQ4LQ set 
from its very low $Q$ scale to the scale at which the CTEQ4M distribution is
given, making a sensitivity study of nondiagonal parton distributions for
different initial distributions impossible at LO. Of course, the inclusion
of the NLO terms corrects this difference in the diagonal case but since there
is no NLO calculation of the nondiagonal case available yet, a study of the
sensitivity of nondiagonal evolution to different initial distributions has to
wait.

The figures themselves suggest the following. First, the lower the starting 
scale, the stronger the effect of the difference of the nondiagonal evolution
as compared to the diagonal one and also that most of the difference between
nondiagonal and diagonal evolution stems from the first few steps in the 
evolution at lower scales.Secondly, under the assumption that the NLO 
evolution in the nondiagonal case will yield the same results for the parton
distributions at some scale $Q$, irrespective of the starting scale $Q_0$, in 
analogy to the diagonal case. One can say that the NLO corrections to the 
nondiagonal evolution will be in the same direction and same order of 
magnitude as the diagonal NLO evolution. If, in the nondiagonal case, the NLO 
corrections were in the opposite direction, which would lead to a 
marked deviation from the LO results, compared to the diagonal
case, the overall sign of the NLO nondiagonal kernels would have to change for
some $\Delta \neq 0$ since in the limit $\Delta \rightarrow 0$ we have to 
recover the diagonal case. This occurance is not likely for the following 
reason: First, the Feynaman diagrams involved in the calculation of the NLO 
nondiagonal kernels are the same as in the diagonal case, except for the 
different kinematics, therefore, we have a very good idea about the type 
of terms appearing in the kernels, namely polynomials, logs and terms in need 
of regularization such as $\ln(z) \frac{\ln (1-z)}{(1-z)}$. Moreover, the 
kernels, as 
stated before, have to reduce to the diagonal case in the limit of vanishing 
$\Delta$ which fixes the sign of most terms in the kernel, thus the only type
of terms which are allowed and could change the overall sign of the kernel are
of the form 
\beq
\frac{\Delta}{y_1} f(x_1/y_1,\Delta/y_1) 
\label{term}
\eeq
which will be numerically 
small unless $y_1 \simeq \Delta$ in the convolution integral of the 
evolution equations. Moreover, we know that in this limit the contribution of 
the regularized terms in the kernel give the largest contributions in the 
convolution integral and therefore sign changing contributions in the 
nondiagonal case would have to originate from regularized terms. This in turn
disallows a term like Eq.\ (\ref{term}) due to the fact that regularized terms
are not allowed to vanish in the diagonal limit, since the regularized 
terms arise from the same Feynman diagrams in the both  
diagonal and nondiagonal case. Therefore, the overall sign of the contribution
of the NLO nondiagonal kernels will be the same as in the diagonal case.  
 
A word should be said about how the results of Ref.\ \cite{MR97} compare to 
ours. For the case of the same $\Delta =10^{-3}$ similar starting scales and 
almost identical values of $Q$ we find good agreement with their 
numbers for $R_g$ at $x_1\simeq \Delta$ \footnote{This was also the case in 
Ref.\ \cite{ours} where we initially put the energies as $Q^2$ 
where in fact they are given as $Q$, which led to some confusion in the 
comparisons of this first study to Ref.\ \cite{MR97}.}
and are slightly higher at larger $x_1$. The observed differences are due to 
the fact that the 
quark distributions are included in our evolution as compared to \cite{MR97}
and their initial distributions are slightly different.
We also find very similar ratios to \cite{MR97} if one changes the starting
scale to a lower one. The slight difference of a few percent in the ratios 
between us and \cite{MR97} can again be attributed to the fact that they used 
the GRV distribution as compared to our use of the CTEQ4 distributions, 
hence a slight difference in the starting scales and their lack of 
incorporating quarks into the evolution.

\section{Conclusions}
\label{concl}

We modified the original CTEQ-code in such a way that we can now compute
the evolution of nondiagonal parton distributions to LO. We gave a detailed
account of the modifications and the methods employed in the new or modified
subroutines. As the reader can see, the modifications and methods themselves 
are not something magical but rather a straightforward application of well
known numerical methods. We further demonstrated the rapid convergence and
stability of our code. In the limit of vanishing asymmetry we reproduce
the diagonal case in LO as obtained from the original CTEQ-code within 
$0.4\%$.  We also have good agreement with the results in Ref.\ \cite{MR97}.
In the future, after the NLO kernels for the nondiagonal case have
been calculated, we will extend the code to the NLO level to be on par with
the diagonal case.

\begin{figure}
\centering
\vskip-3cm
\mbox{\epsfig{file=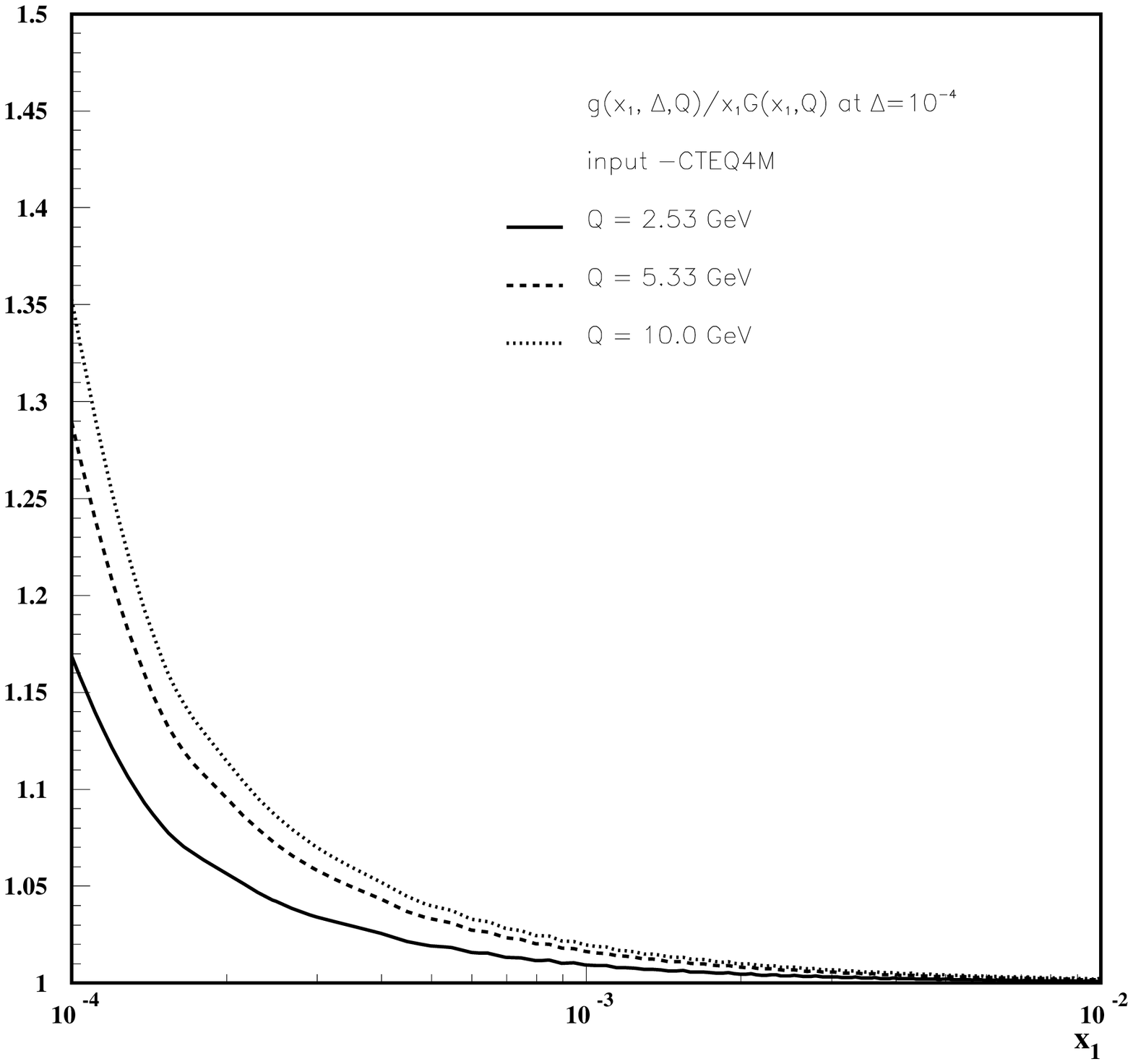,width=14cm,height=14cm}}
\vskip-3cm
\mbox{\epsfig{file=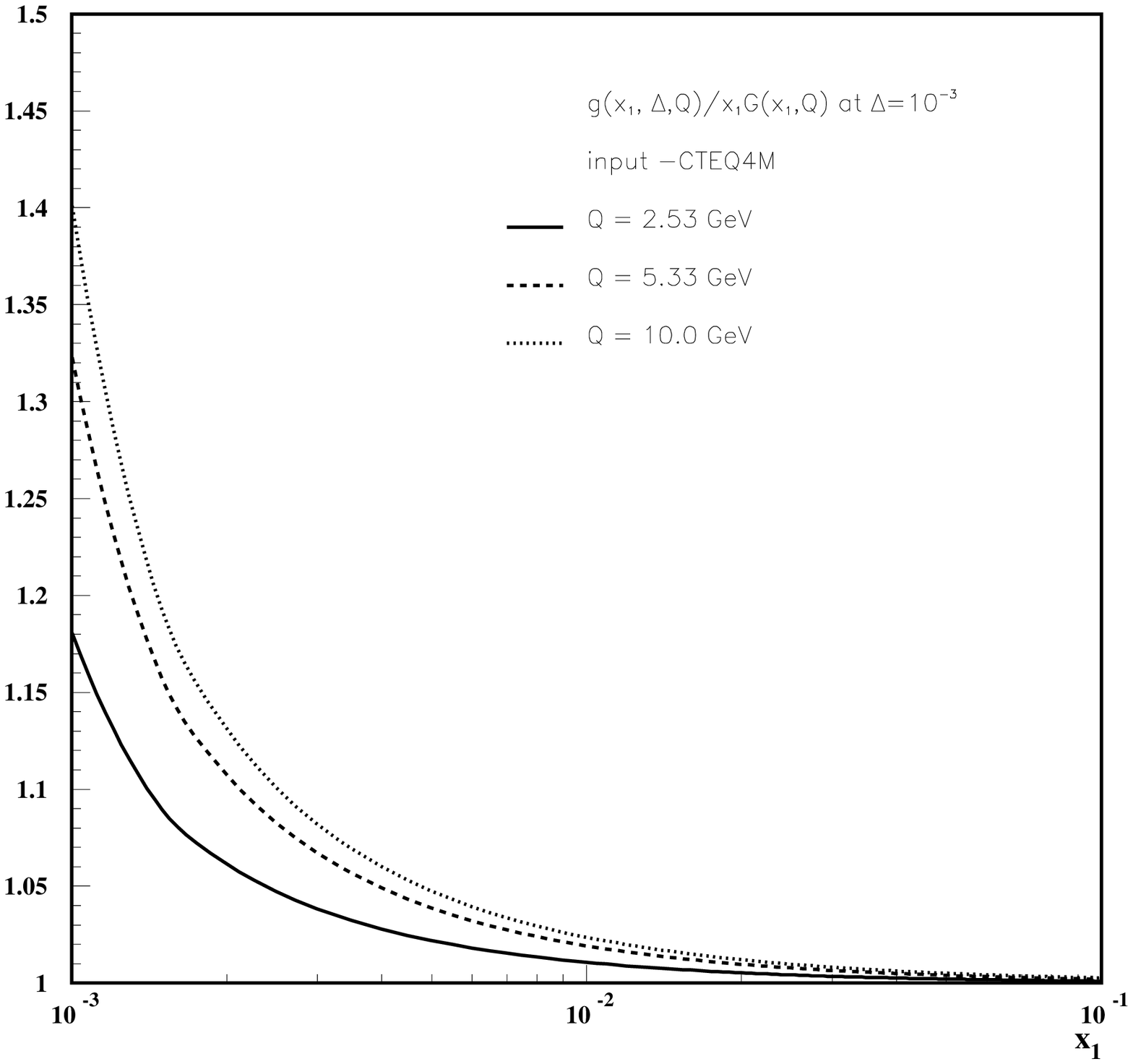,width=14cm,height=14cm}}
\vskip-2cm
\caption{$R_g$ is plotted versus $x_1$ for fixed $\Delta$ using the CTEQ4M 
parameterization with $Q_0=$\mbox{1.6 GeV} and $\Lambda$ =\mbox{ 202 MeV}.}
\label{nddratio}
\end{figure}
\newpage
\begin{figure}
\centering
\vskip-3cm
\mbox{\epsfig{file=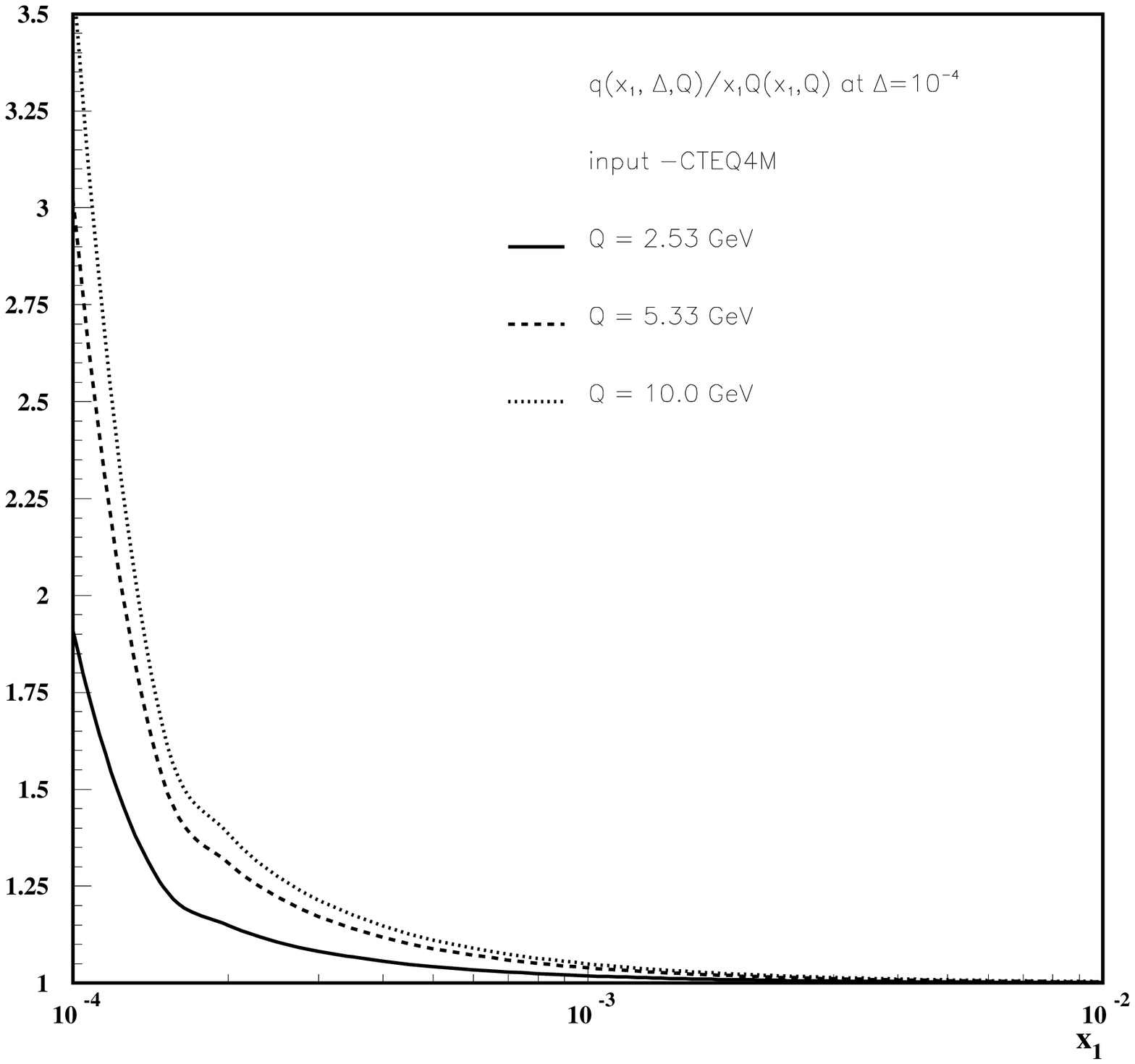,width=14cm,height=14cm}}
\vskip-3cm
\mbox{\epsfig{file=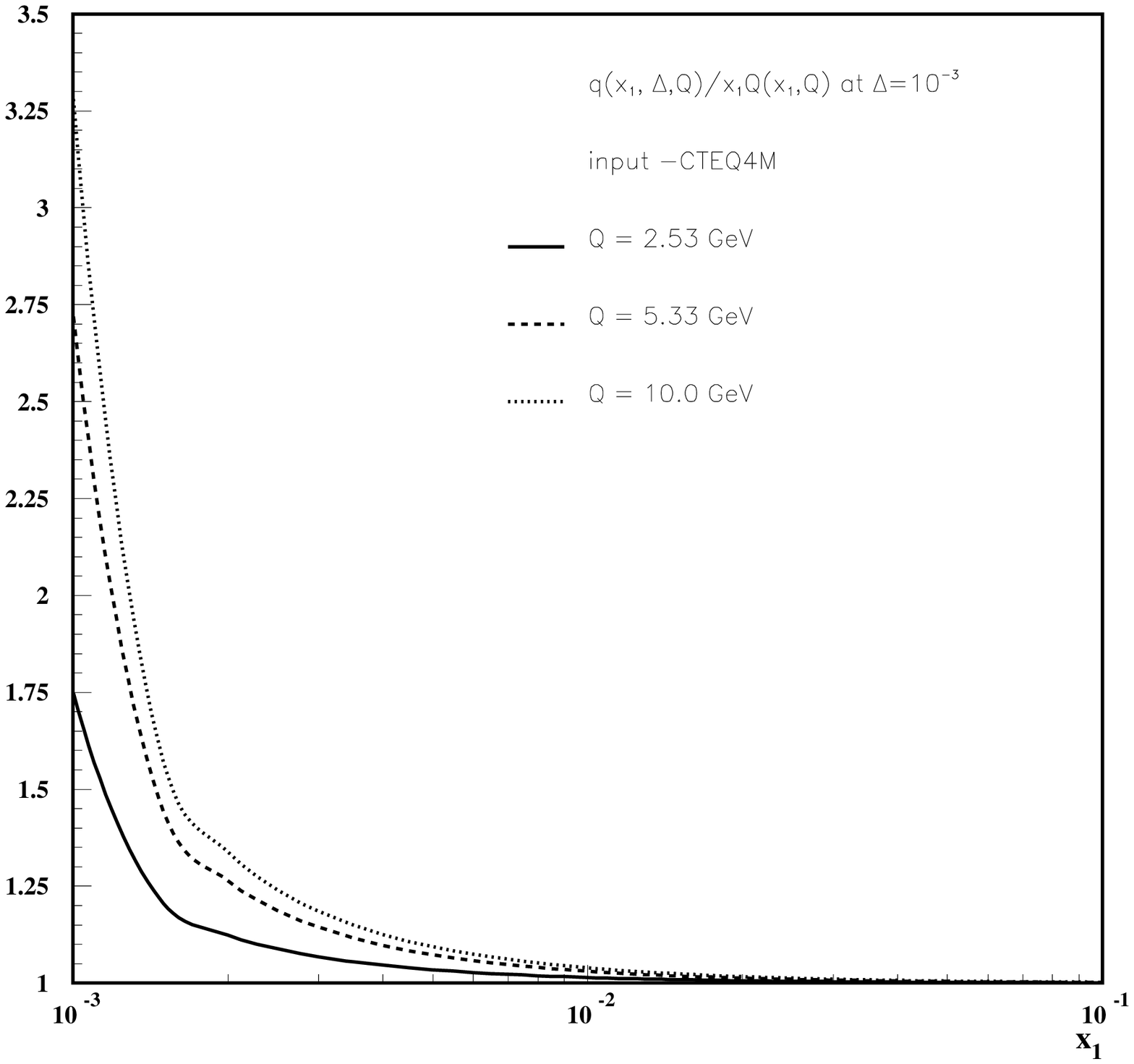,width=14cm,height=14cm}}
\vskip-2cm
\caption{$R_q$ is plotted versus $x_1$ for fixed $\Delta$ using the CTEQ4M 
parameterization with $Q_0=$\mbox{1.6 GeV} and $\Lambda$ =\mbox{202 MeV}.}
\label{nddratio1}
\end{figure}
\newpage
\begin{figure}
\centering
\vskip-3cm
\mbox{\epsfig{file=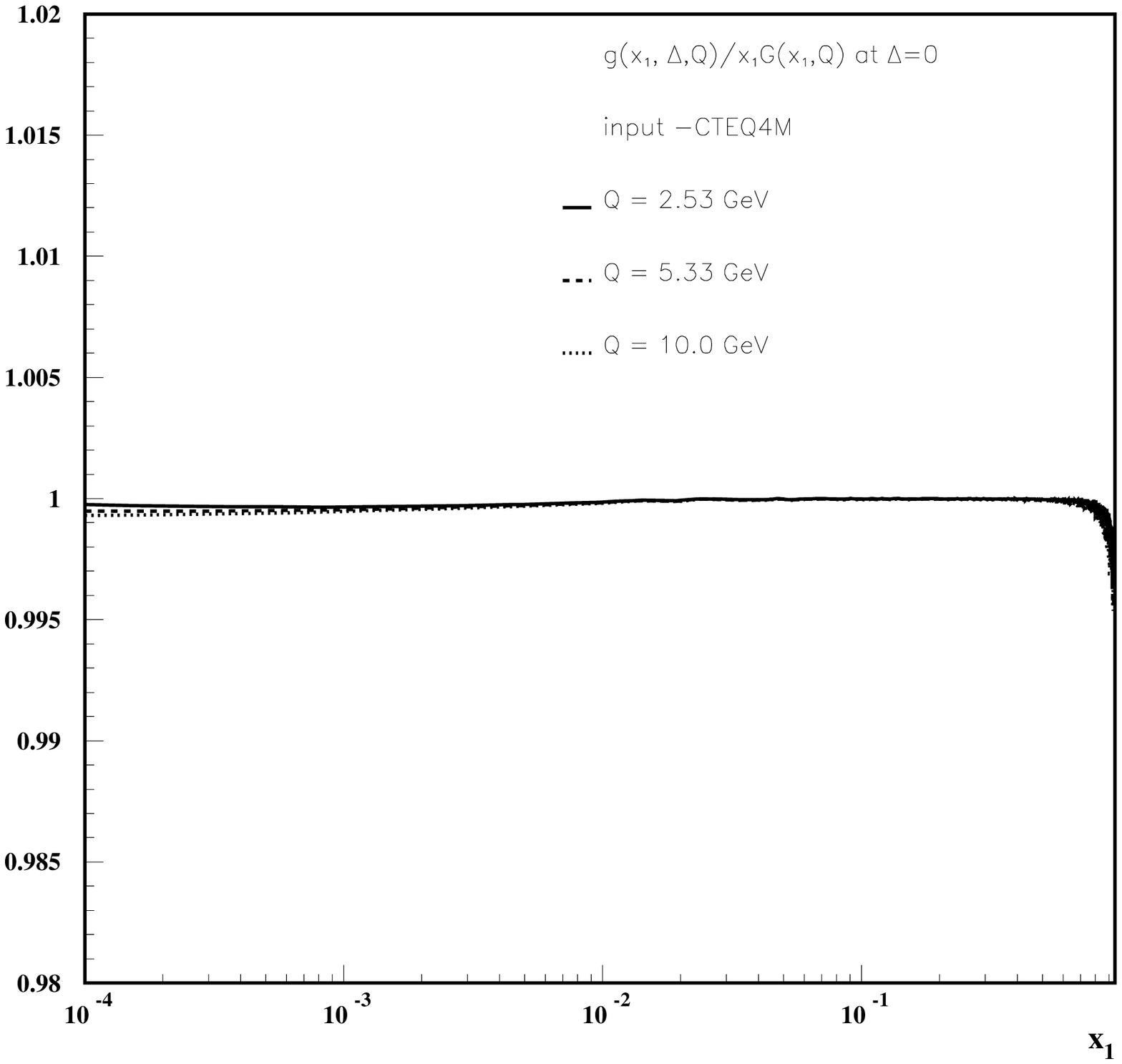,width=14cm,height=14cm}}
\vskip-3cm
\mbox{\epsfig{file=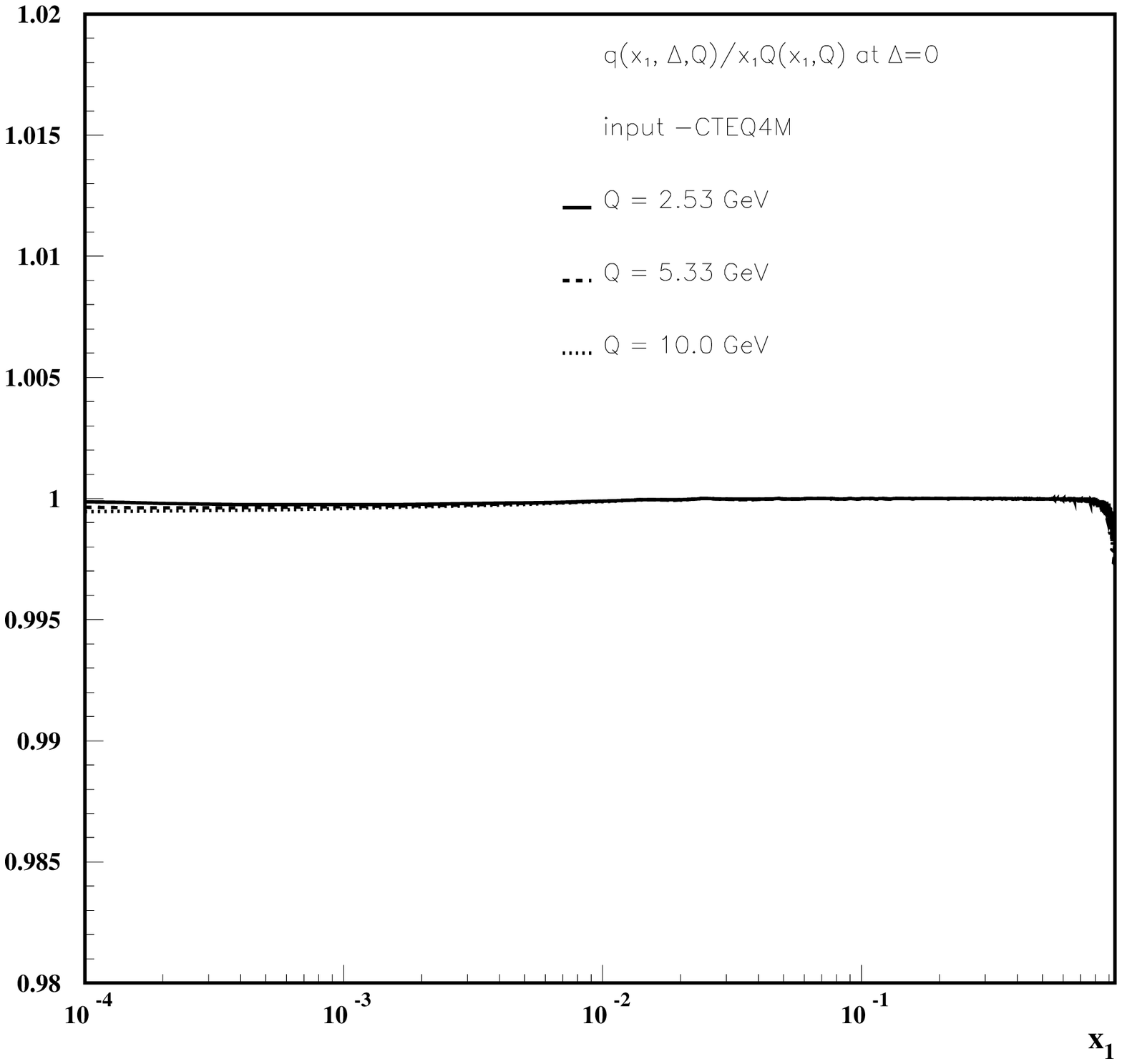,width=14cm,height=14cm}}
\vskip-2cm
\caption{$R_g$ and $R_q$ are plotted versus $x_1$ for $\Delta = 0$ using the 
CTEQ4M parameterization with $Q_0=$\mbox{1.6 GeV} and $\Lambda$ =\mbox{ 202 MeV}.}
\label{nddratio2}
\end{figure}
\newpage
\begin{figure}
\centering
\vskip-3cm
\mbox{\epsfig{file=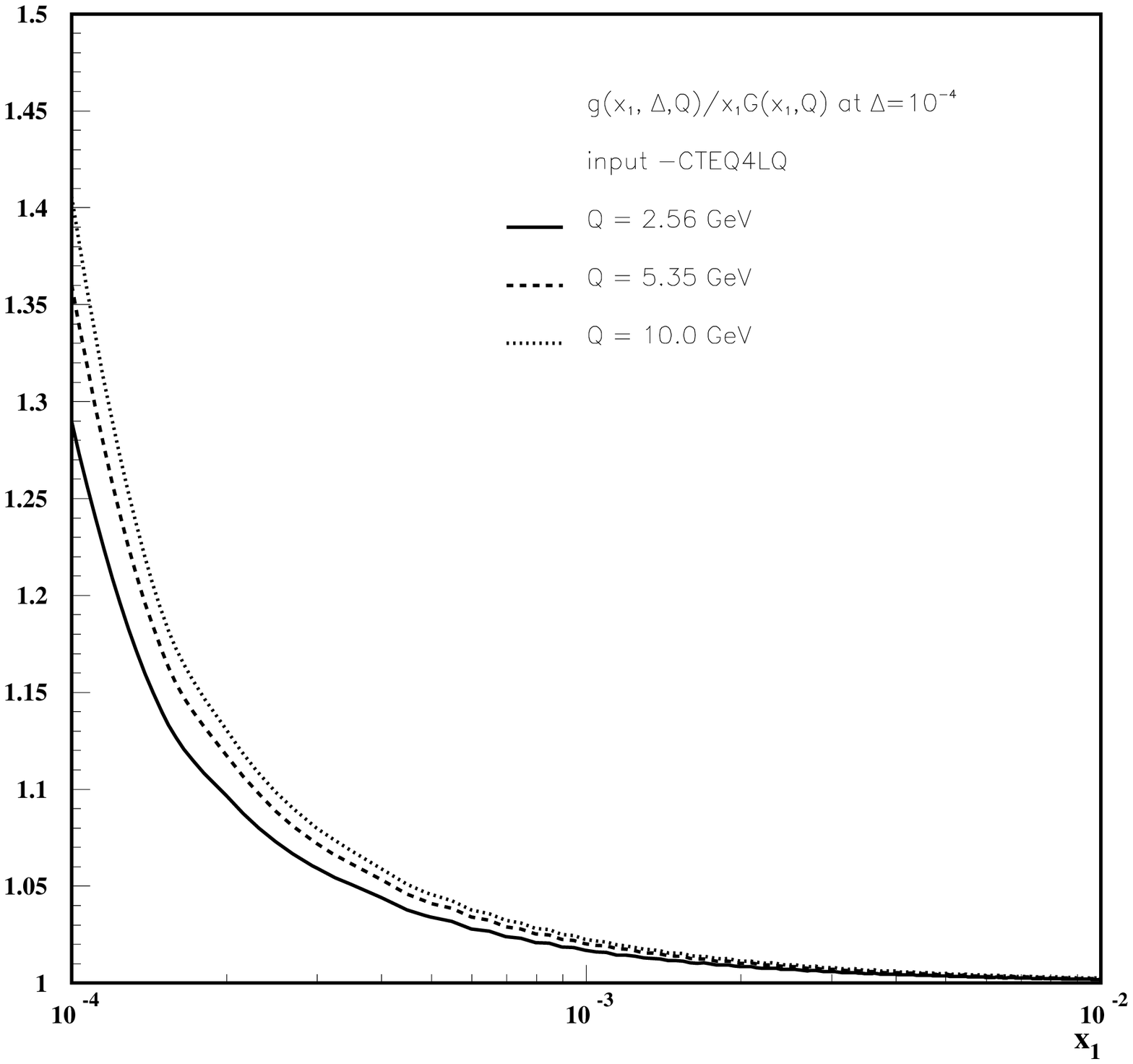,width=14cm,height=14cm}}
\vskip-3cm
\mbox{\epsfig{file=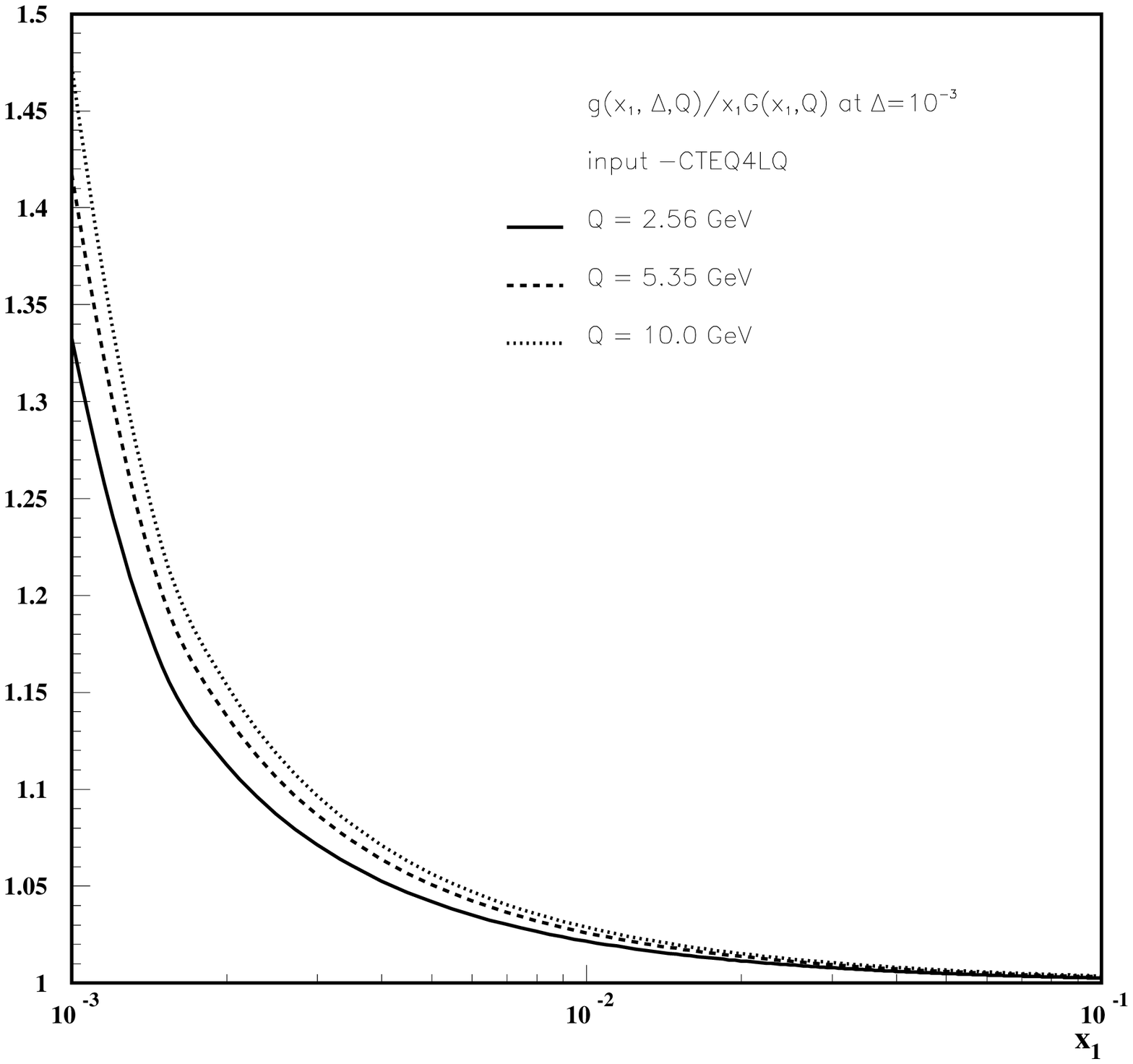,width=14cm,height=14cm}}
\vskip-2cm
\caption{$R_g$ is plotted versus $x_1$ for fixed $\Delta$ using the 
CTEQ4LQ parameterization with $Q_0=$\mbox{0.7 GeV} and 
$\Lambda$ =\mbox{174 MeV}.}
\label{nddratio3}
\end{figure}
\newpage
\begin{figure}
\centering
\vskip-3cm
\mbox{\epsfig{file=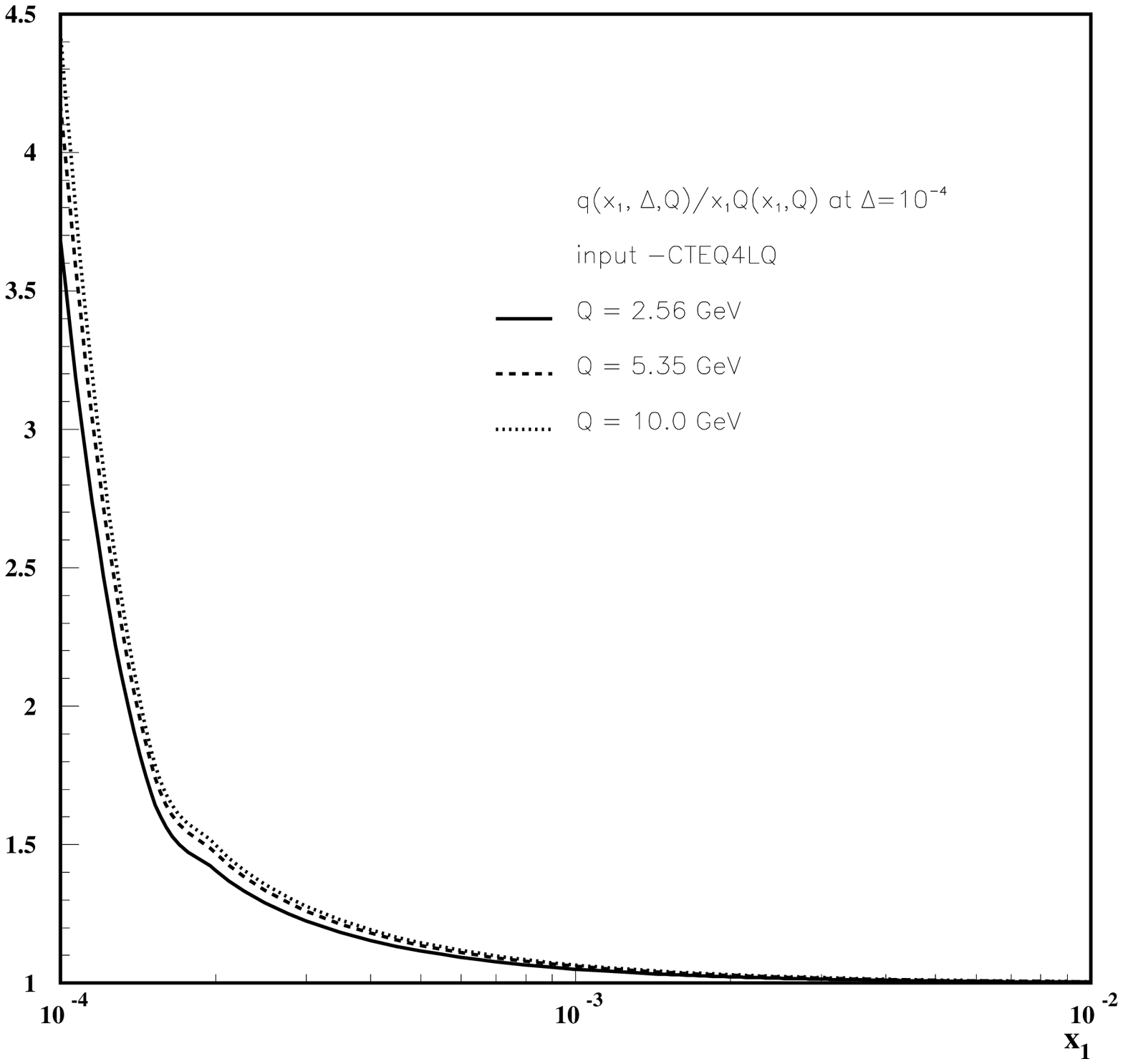,width=14cm,height=14cm}}
\vskip-3cm
\mbox{\epsfig{file=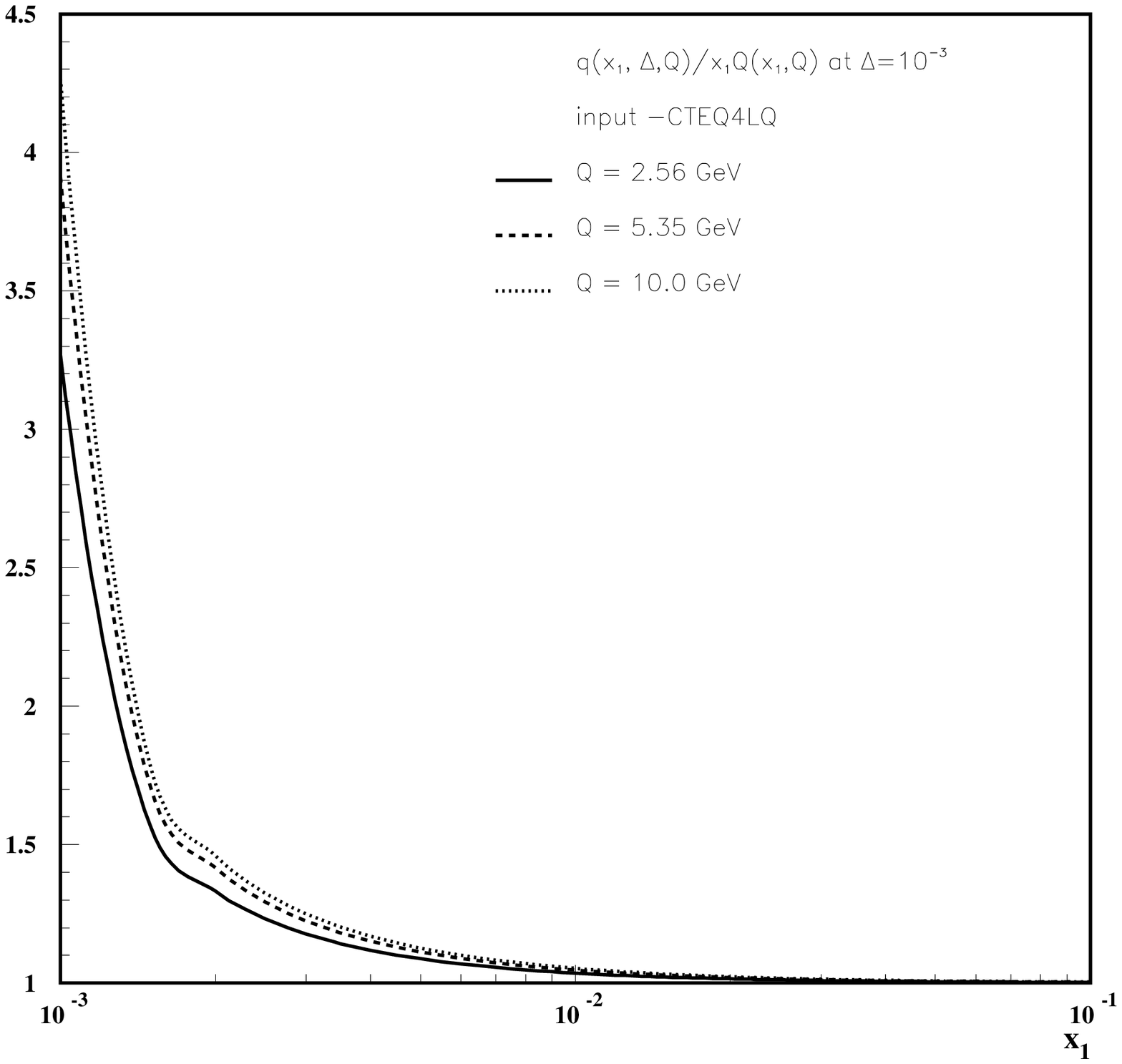,width=14cm,height=14cm}}
\vskip-2cm
\caption{$R_q$ is plotted versus $x_1$ for fixed $\Delta$ using the 
CTEQ4LQ parameterization with $Q_0=$\mbox{0.7 GeV} and 
$\Lambda$ =\mbox{174 MeV}.}
\label{nddratio4}
\end{figure}
\newpage
\begin{figure}
\centering
\vskip-3cm
\mbox{\epsfig{file=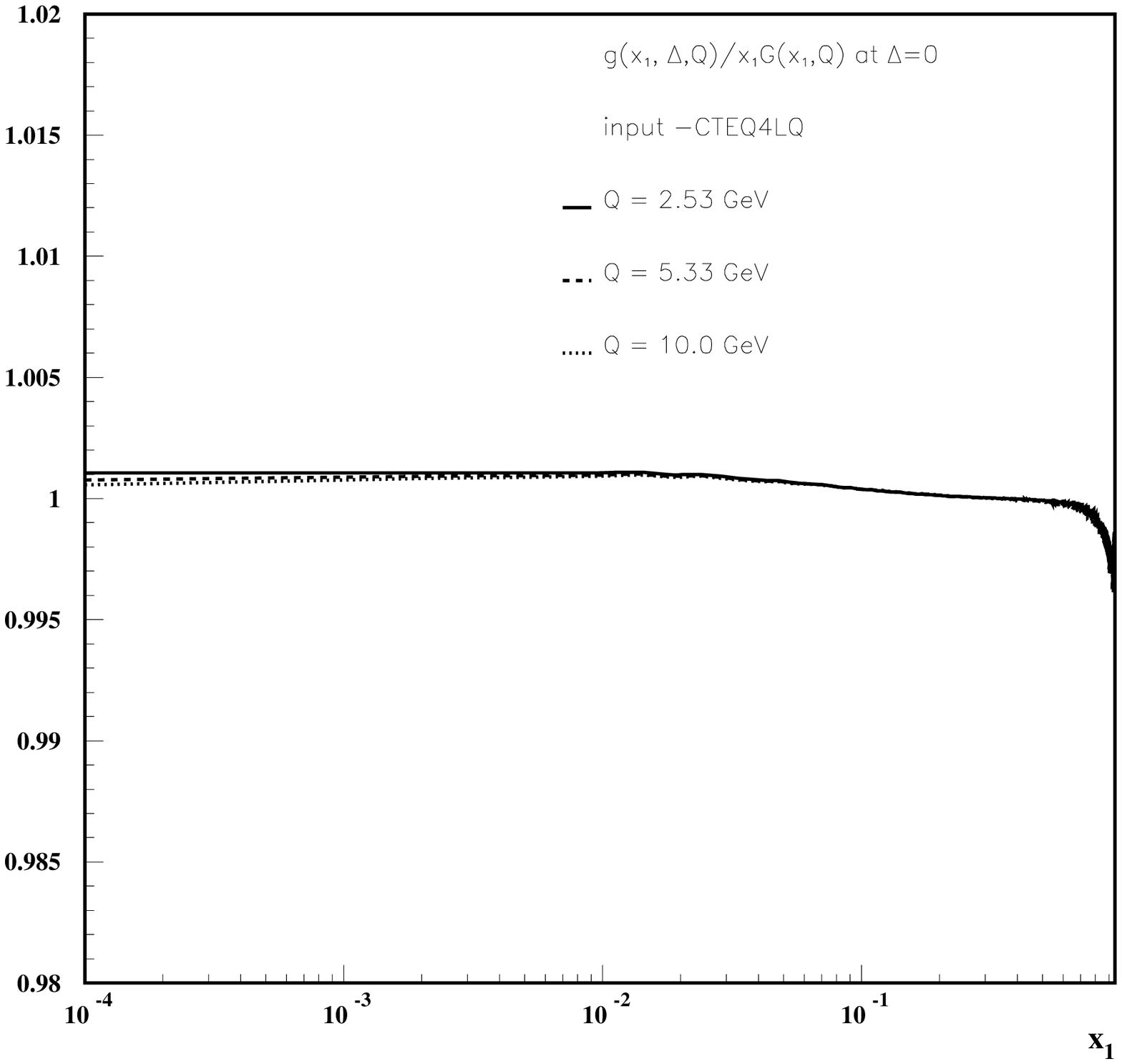,width=14cm,height=14cm}}
\vskip-3cm
\mbox{\epsfig{file=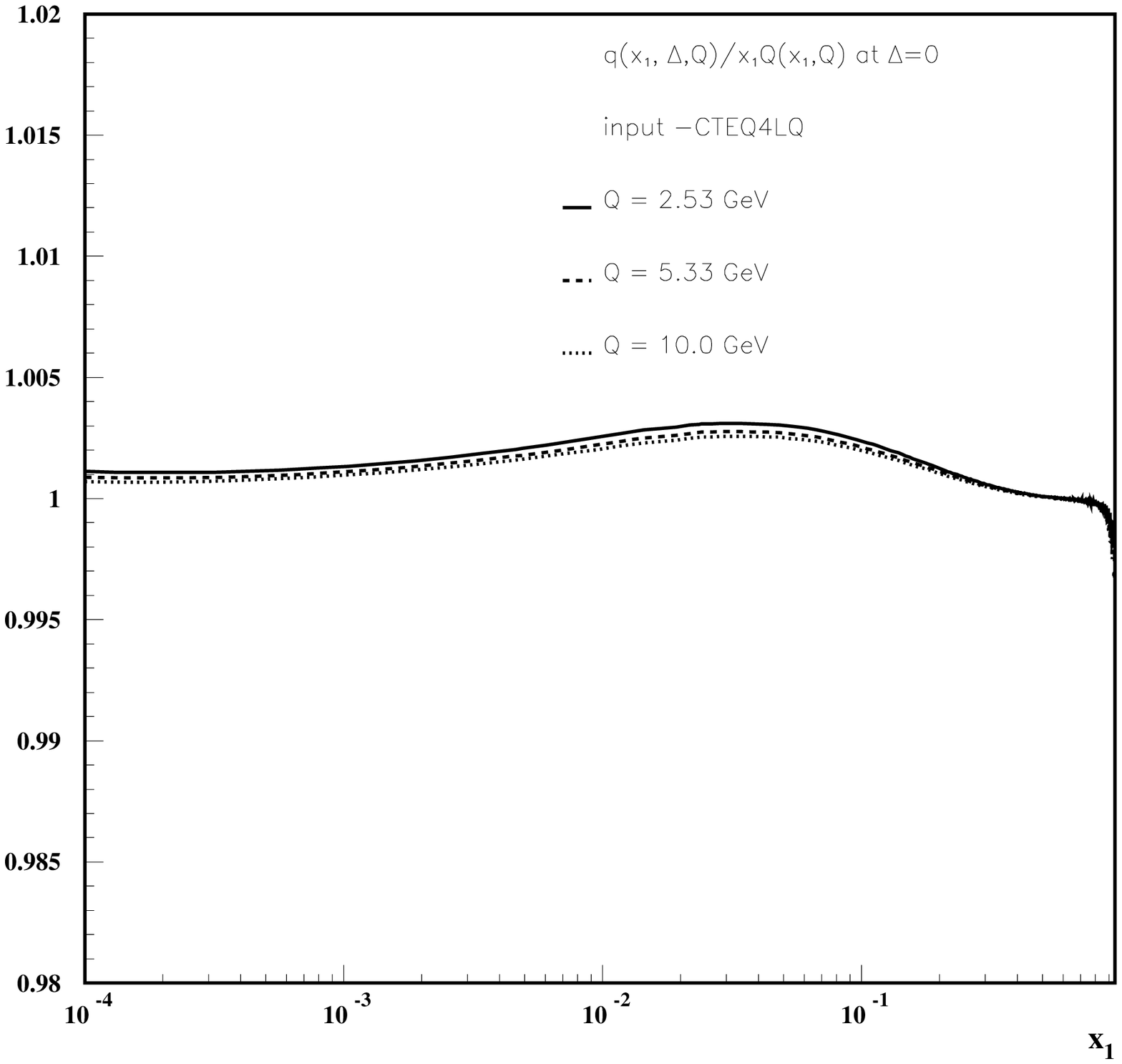,width=14cm,height=14cm}}
\vskip-2cm
\caption{$R_g$ and $R_q$ are plotted versus $x_1$ for $\Delta = 0$ using the 
CTEQ4LQ parameterization with $Q_0=$\mbox{0.7 GeV} and 
$\Lambda$ =\mbox{174 MeV}.}
\label{nddratio5}
\end{figure}
\newpage
\begin{figure}
\centering
\mbox{\epsfig{file=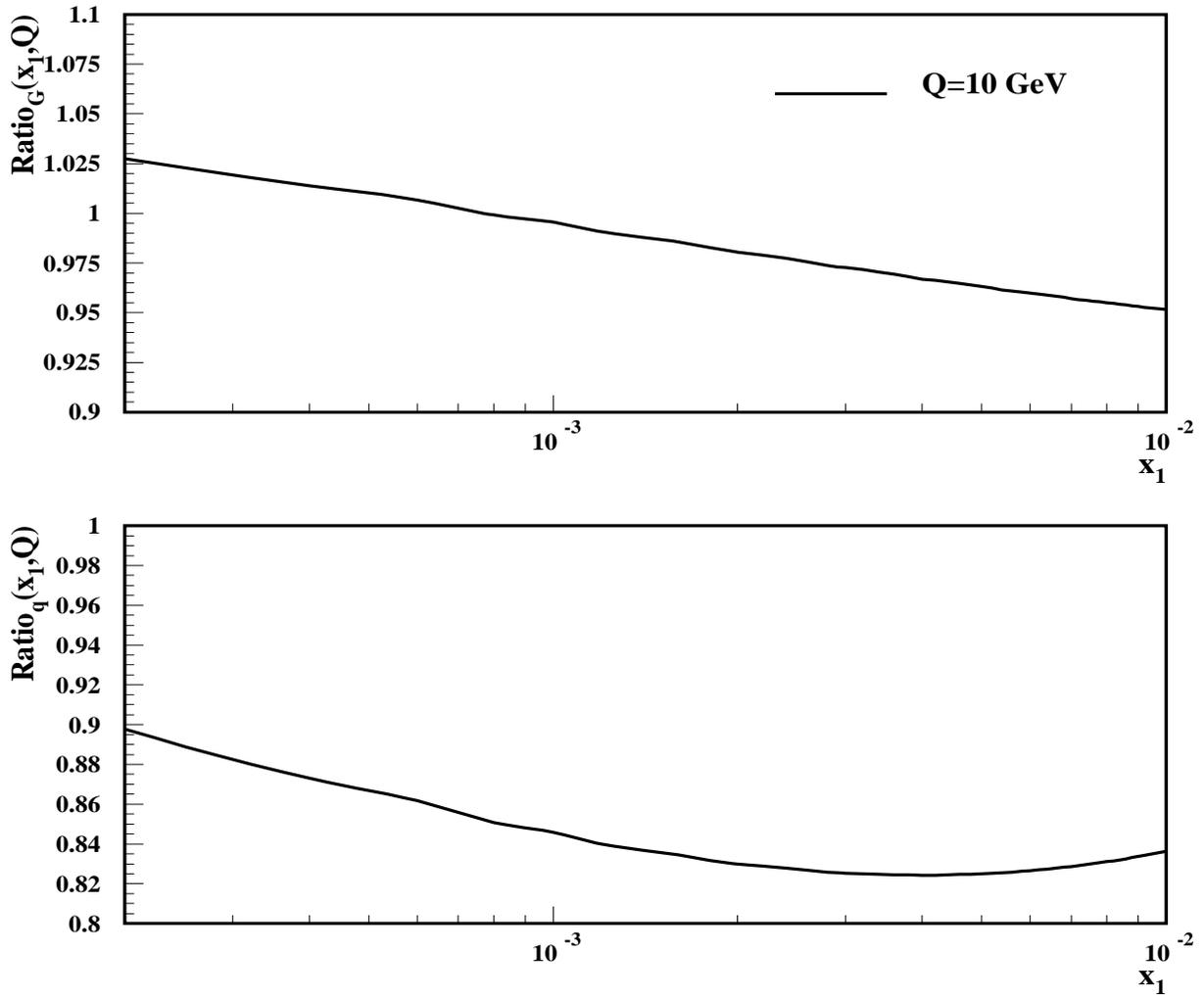,width=14cm,height=18cm}}
\vspace*{5mm}
\caption{The ratios for CTEQ4M to CTEQ4LQ for gluons and quarks 
in the diagonal case is plotted to demonstrate the difference between the 
LO evolution for these 
parameterizations.}
\label{nddratio6}
\end{figure}

\chapter{Proof of Factorization for Deeply Virtual Compton Scattering in QCD}
\indent

\section{Introduction}

In this chapter, based on Ref.\ \cite{ca}, we prove factorization for the 
deeply virtual Compton scattering (DVCS) amplitude in QCD up to power 
suppressed terms, to all orders in perturbation theory. This proof is
important because of the recent great interest in DVCS
\cite{Ji'96,Rad,muell,Ji1,BM,DGPR,chen,FFS,man}. One important use
of DVCS is as a probe of off-forward (or nondiagonal)
distributions \cite{Ji'96,Rad,OFPD,C.F.S'96,ours}. These differ from
the usual parton distributions probed in inclusive
reactions by having a non-zero momentum transfer between the
proton in the initial and final state.

A related process which is also used to probe off-diagonal parton
densities is exclusive meson production in deep-inelastic
scattering \cite{Brod'94,Ryskin}, for which a proof of factorization was
given in \cite{C.F.S'96}.  Compared with this process, DVCS is simpler
because the composite meson in the final state is replaced by an
elementary particle, the photon, and thus there is no meson wave
function in the factorization formula.

Ji \cite{Ji'96} and Radyushkin \cite{Rad} have provided the key
insights that indicate that a factorization theorem is valid for
DVCS and Radyushkin also gave an all-orders proof \cite{Rad}. 
In this chapter, we provide an alternative proof, and give a new treatment of
some problems that were touched upon in Ref.\ \cite{Rad} but that were not 
fully solved. The proof follows the general lines of
proofs of factorization for other processes \cite{Muell'89,ColSter},
and the most noteworthy feature is that the proof is
simpler than for any other process.  Even for ordinary
deep-inelastic scattering one needs to discuss the cancelation
of soft gluon exchanges and of final-state interactions, whereas
these complications are not present in the leading power for
DVCS.

The chapter is organized in the following way: First we will state
the theorem to be proved in Sec.\ \ref{sec:factorx} and then explain the 
steps necessary to prove it, as adapted from Ref.\ \cite{C.F.S'96}, in 
Sec.\ \ref{sec:proof}. The
complications mentioned above concern the situation when one of the
two lines connecting the parton density to the hard scattering carries
zero longitudinal momentum, and these are given a detailed treatment in 
Sec.\ \ref{subt}. In the last section we will make concluding remarks.

\section{Factorization Theorem}
\label{sec:factorx}

The process under consideration is DVCS,
which is the elastic scattering of virtual photons:
\begin{equation}
\gamma ^{*}(q) + P(p) \rightarrow \gamma ^{(*)}(q') + P'(p-\Delta )
\label{dvcs}
\end{equation}
where the diffracted proton $P'$ may also be replaced by
a low-mass excited state and the
final-state photon can be either real or time-like. This process is the
hadronic part of $ep \rightarrow e \gamma p'$ for a real photon or of
$ep \rightarrow e\mu ^{+}\mu ^{-}p'$ for a time-like photon.

It is convenient to use light-cone coordinates with respect to the collision
axis\footnote
{We define a vector in light cone coordinates by:
\begin{displaymath}
V^{\mu }=\left (V^{+},V^{-},V_{\perp } \right )=
\left ( \frac {V^{0}+V^{3}}{\sqrt {2}},
\frac {V^{0}-V^{3}}{\sqrt {2}},V^{1,2}\right ).
\end{displaymath}
}.
The momenta in the process then take the form:
\begin{eqnarray}
   p^{\mu } &=& \left ( p^{+},\frac {m^{2}}{2p^{+}},{\bf 0}_{\perp } \right ),
\nonumber\\
   q^{\mu } &\simeq& \left ( -xp^{+},\frac {Q^{2}}{2xp^{+}},{\bf 0}_{\perp }
\right ),
\nonumber\\
   q'^{\mu } &\simeq& \left ( xp^{+}\frac {\Delta ^{2}_{\perp }+\alpha
Q^{2}}{Q^{2}},
      \frac {Q^{2}}{2xp^{+}},{\bf \Delta }_{\perp }
   \right ),
\nonumber\\
   \Delta ^{\mu } &\simeq& \left ( x(1+\alpha )p^{+},
      -\frac {\Delta ^{2}_{\perp } + m^{2} (1+\alpha ) x}{2(1-x-\alpha
x)p^{+}},
      {\bf \Delta }_{\perp } \right ).
\label{momenta}
\end{eqnarray}
Here, $x$ is the Bjorken scaling variable,
$Q^{2}$ is the virtuality of the initial photon,
$m^{2}$ is the proton mass, $t=\Delta ^{2}$ is
the momentum transfer squared, and
$\alpha $ is a parameter that specifies the virtuality of the
outgoing photon: ${q'}^{2} = \alpha Q^{2}$.  Thus, $\alpha =0$ for a
real photon and $\alpha >0$ for a time-like photon. Finally,
$\simeq$ means ``equality up to power suppressed terms''.

The theorem to be proved is that the amplitude for the process (\ref{dvcs})
is:
\begin{equation}
T = \Sigma_{i} \int_{-1+x}^{1} dx_{1} f_{i/p}\left ( x_{1},x_{2},t,\mu \right )
H_{i}\left (x_{1}/x,x_{2}/x,\mu \right )\, + \, 
\mbox{power-suppressed corrections},
\label{theorem}
\end{equation}
where the $f_{i/p}$ is a nondiagonal parton distribution and
$H_{i}$ is the hard-scattering coefficient for scattering off a
parton of type $i$.
We let $x_{1}$ be the momentum fraction of parton $i$ coming from
the proton, so that $x_{2}=x_1-(1+\alpha)x$ is the momentum fraction which is 
returned
to the proton by the other parton line joining the parton
distribution and the hard scattering.  There is implicit
polarization dependence in the amplitude.
$\mu $ is the usual
renormalization/factorization scale which should be of order $Q$
to allow
calculations of the hard scattering coefficients
within finite-order perturbation
theory. The $\mu $ dependence of $f_{i/p}$ is given by equations of the
DGLAP and Brodsky-Lepage
kind \cite{Ji'96,Rad,OFPD,Brod'94,C.F.S'96,ours}.
The parton distributions in Eq.\ (\ref{theorem}), together with
their evolution equations, are defined using the conventions of
\cite{C.F.S'96,ours}.  They may easily be transformed into those given
in \cite{Ji'96,Rad} by a change of normalization and of kinematic
variables.

\section{Proof of Theorem}
\label{sec:proof}

The proof of our theorem Eq.\ (\ref{theorem}) can be summarized as follows
\footnote{
For a very detailed account of the basic steps and potential problems
see Ref.\ \cite{C.F.S'96}.
}:
\begin{itemize}

\item Establish the non-ultra-violet regions in the space of loop momenta
      contributing to the amplitude.

\item Establish and prove a power counting formula for these regions.

\item Determine the leading regions of the amplitude.

\item Define the necessary subtractions in the amplitude to avoid double
      counting.

\item Taylor expand the amplitude to obtain a factorized form.

\item Show that the part containing the long-distance information can be
      expressed through matrix elements of renormalized, bi-local, gauge
      invariant operators of twist-$2$.

\end{itemize}

\subsection{Regions}
\label{regx}

First let us establish the regions in the space of loop momenta contributing
to the asymptotics of the amplitude, i.e.,\ the generalized reduced graphs.
The steps leading to the generalized reduced graphs are identical to the steps
1--3 in Sec.\ IV of Ref.\ \cite{C.F.S'96}, i.e.,\ scale all momenta by a factor
$Q/m$, use the Coleman-Norton theorem to locate all pinch-singular surfaces
in the space of loop momenta (in the zero-mass limit), and finally identify
the relevant regions of integration as neighborhoods of these pinch-singular
surfaces.

In the first step, the scaling of momenta, we proceed as follows \cite{LibSt}.
We write the general momentum $k^{\mu}$ and a general mass $m$ in units of 
the large scale $Q$:
\beq
k^{\mu} = Q\tilde k^{\mu}, ~~~~~~ m = Q\tilde m.
\eeq
Due to working in the rest frame of the virtual photon both of the light like 
components are of order $Q$. Therefore, when everything is expressed in terms 
of the above scaled variables, dimensional analysis shows that the large $Q$ 
limit is equivalent to the $\tilde m \rightarrow 0$ limit. Since the amplitude 
is dimensionless we have
\beq
T(Q^2,p,q',\Delta,m;\mu) = T(1,\tilde p,\tilde q',\tilde \Delta,\tilde m,Q/\mu)
,  
\eeq
by regular dimensional analysis.

The most basic region is found where all internal lines obey $k^2\geq Q^2$, 
with the scaled momenta $\tilde k$ having virtualities of $1$ or bigger. In 
such a region one is entitled to setting the masses equal to zero, make the 
external hadrons light-like and set the renormalization scale 
$\mu$ equal to $Q$, thus avoiding large logarithms. As it turns out, however, 
this region is not only not the only one but it is even not leading. 
Nevertheless, one can now see that all other relevant regions correspond to 
singularities of massless Feynman graphs. They are neighborhoods of 
pinch-singular surfaces of massless graphs, in other words, surfaces where the 
loop momenta are trapped at singularities. The conditions for a pinch 
singularity are the Landau conditions for a singularity of a graph. Only pinch
singularities are of interest, since at a non-pinched singularity, one can 
deform the contour of integration such that at least one of the singular 
propagators is no longer near its pole. In case of a pinch singularity from 
propagator poles in the massless limit, we know then that in the real graph, 
with nonzero masses but large $Q$, the contour of integration is forced to 
pass near the propagator poles. Consequently, it is not possible to neglect 
the masses in this region. Conversely, if the contour is not trapped by the 
poles, the contour can be deformed away from the poles, and the mass may be 
neglected in evaluating the corresponding propagators.

In the second step we use the Coleman-Norton theorem \cite{colnor} which 
states that each point on a pinch-singular surface in loop momentum space, 
corresponds to a space-time diagram obtained in the following way. First one 
obtains a reduced graph by contracting to points all of the lines whose 
denominators are not pinched. Then one assigns space-time points to each 
vertex of the reduced graph in such a way that the pinched lines correspond to
classical particles.     In other words each line is assigned a particle 
propagating between space-time points corresponding to the vertices at its 
ends. The momentum of the particle is exactly the momentum carried by the 
line, with its orientation such that it has positive energy. If, for some set 
of momenta, one cannot construct such a reduced graph, one is free to deform 
the contour of integration. A reduced diagram, therefore, corresponds to a 
classically allowed space-time  scattering process. In the zero mass limit, 
the construction of reduced graphs becomes very simple, since all pinched 
lines must carry either light-like or zero momentum. Furthermore, each 
light-like momentum must be parallel to one of the, now light-like, external 
lines.

To be precise, in the zero mass limit of the process under consideration we 
have:
\begin{itemize}  
\item One light-like incoming proton of momentum 
      $p^{\mu}_A=(p^{+},0,0_{\perp})$.
\item One light-like outgoing proton line of parallel momentum\\ 
      $p'^{\mu}_A=((1-(1+\alpha)x)p^{+},0,0_{\perp})$.
\item One light-like outgoing photon line of momentum 
      $q'^{\mu}=(\alpha xp^{+},Q^2/2xp^{+},0_{\perp})$.
\item One incoming virtual photon of momentum 
      $q^{\mu}=(-xp^{+},Q^2/2xp^{+},0_{\perp})$.
\end{itemize}

\begin{figure}
\centering
\mbox{\epsfig{file=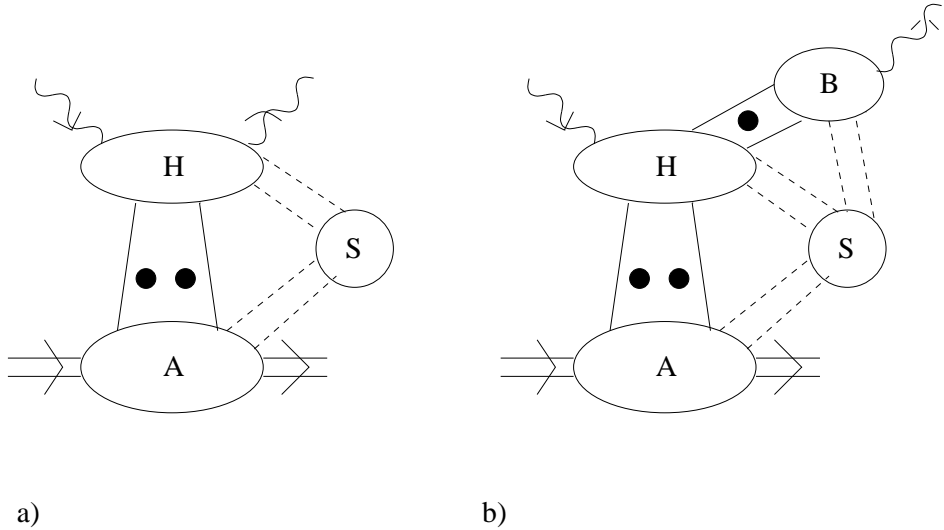,height=7cm}}
\vspace*{5mm}
\caption{a) Reduced graph for DVCS with direct coupling for
the out-going photon to hard subgraph.
b) The same without a direct coupling for the
out-going photon. }
\label{Reduced.Graphs}
\end{figure}

The results of the above construction are the two kinds of reduced graph 
shown in Fig.\
\ref{Reduced.Graphs}.  There, $A$ and $B$ denote collinear graphs
with one large momentum component in the $+$ and $-$ direction
respectively, $H$ denotes the hard scattering graph, and $S$
denotes a
graph with all of its lines soft, i.e.,\ in the center-of-mass
frame all the components of the momenta in $S$ are much smaller than
$Q$.
Note that, of the external momenta, $p$ and $p'$ belong to $A$,
$q'$ belongs to $B$ or $H$ , and $q$ belongs to $H$.

When the two external photons have comparably large virtualities,
the only reduced graphs are of the first kind, Fig.\
\ref{Reduced.Graphs}a, where the out-going photon couples
directly to the hard scattering. But when the outgoing photon
has much lower virtuality than the incoming photon, for example,
when it is real, we can also have the second kind of reduced
graph, Fig.\ \ref{Reduced.Graphs}b, where the out-going photon
couples to a $B$ subgraph. As we will see later, power counting
will show that the second kind of reduced graph, Fig.\
\ref{Reduced.Graphs}b, is power suppressed compared to the first
kind, with a direct photon coupling. This implies that we will
avoid all the complications which were encountered in \cite{C.F.S'96}
that are associated with the meson wave function.

\begin{figure}
\centering
\mbox{\epsfig{file=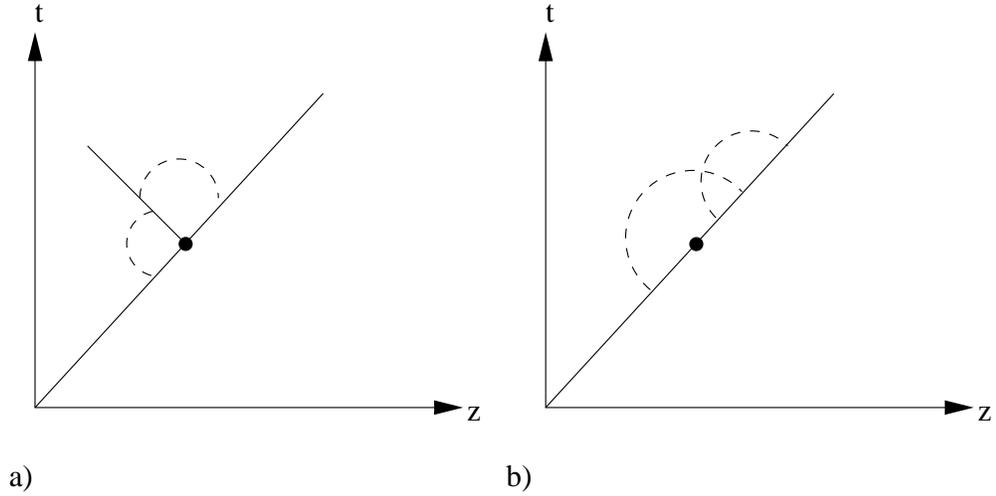,height=6.5cm}}
\caption{a) Space-time picture of the DVCS process with a collinear-to-$B$ 
part as in vector meson production. b) Space-time picture of the DVCS process 
without a collinear-to-$B$ part.}
\vspace*{5mm}  
\label{space}
\end{figure}

The corresponding space-time diagram is Fig.\ \ref{space}. In this figure, 
each solid line corresponds to a light-like line of the reduced graphs, with a
$45^{\circ}$ orientation to correspond to their light-like lines of 
propagation.
The dashed lines correspond to the soft subgraph $S$. As far as the 
Coleman-Norton theorem is concerned the lines are degenerate. In fact they are
carrying zero-momentum implying that they have no specific orientation. They 
are therefore indicated by curved lines of no particular orientation. The 
location of the endpoints of the soft lines can be anywhere along the world 
lines of the collinear lines. The hard vertex $H$ occurs at the intersection 
of the collinear lines. The world line , in the $+$ direction, of the 
collinear-to-$A$ subgraph actually consists of several lines propagating 
together and possibly interacting with each other. 

In the space-time representation of a Feynman graph, there is normally an 
exponential suppression when there are large space-time separations between 
vertices. One obtains a singularity when this suppression fails and the 
Coleman-Norton construction gives exactly the relevant configurations of the 
vertices. The singularity is generated by the possibility of integrating over 
arbitrarily large scalings\footnote{The scaling of the world lines in a 
reduced graph by a common factor does not affect the properties of that graph.} 
in coordinate space without obtaining exponential suppressions.

Note that the above discussion relies on the use of a covariant gauge. The 
use of an axial gauge, albeit convenient, leads to unphysical singularities 
in the propagators. These singularities do not give the normal rules of causal
 relativistic propagation of particles and, furthermore, make the derivation 
of the factorization theorem, beyond leading log, very difficult 
\cite{ColSter,csg}.

\subsection{Power Counting}
\label{count}

Each reduced graph codes a region of loop-momentum space, a
neighborhood of the surface $\pi $ of a pinch singularity in the
zero-mass limit.
The contribution to the amplitude from a neighborhood of $\pi $
behaves like $Q^{p(\pi )}$, modulo logarithms, in the large-$Q$
limit, with the power given by
\begin{eqnarray}
    p(\pi ) &=& 4 - n(H) - \mbox{\#(quarks from $S$ to $A$, $B$)} -
    3\mbox{\#(quarks from $S$ to $H$)}\nonumber\\
    & & - 2\mbox{\#(gluons from $S$ to $H$)}.
\label{pform}
\end{eqnarray}
where $n(H)$ is the number of collinear quarks, transversely
polarized gluons,
and external photons attaching to the hard subgraph $H$.
Such results were obtained by Libby and Sterman
\cite{LibSt,Sterman}.
The particular form of Eq.\ (\ref{pform}) was given in \cite{C.F.S'96}
together with a proof that applies without change to DVCS.

We will detail it here, nevertheless, for completeness sake. The arguments used
 in the proof will rely on general arguments about dimensional analysis and 
Lorentz boosts.

We first consider the case of only the hard and collinear subgraphs without a 
soft subgraph. Let the hard subgraph have $N_q$ external quark/antiquark lines
 and $N_g$ external gluons, as well as two photon lines. By definition, all 
components in the hard subgraph have virtuality of order $Q^2$. Since the hard
 subgraph has dimensions $d_H=2-\frac{3}{2}N_q - N_g$ and all the couplings 
are dimensionless, it contributes a power 
\beq
Q^{d_H}=Q^{2-\frac{3}{2}N_q - N_g}
\label{pocf}
\eeq
to the amplitude\footnote{The factor $3$ in Eq.\ (\ref{pocf}) is the number of 
colors and the factor $1/2$ corresponds to the spin of the quarks as the 
factor $1$ in front of $N_g$ corresponds to the spin of the gluon.}.

For the momenta collinear to the proton we have
\beq
\mbox{typical A momentum} \propto \left(Q,\frac{m^2}{Q},m \right ).
\eeq
Since $x$ is small, there are also collinear momenta with $+$ components much 
larger then $Q$. We will deal with this problem later on; let us just assume 
for the moment that $x$ is not small.

The collinear configurations can be obtained by boosts from a frame in which 
all components of all momenta are of order $m$. Since the virtualities and the
 sizes of regions of momentum integration are invariant under boosts, we start
 by assigning the collinear subgraphs an order of magnitude 
$m^{\mbox{dimension}}$, which contributes unity to the power of $Q$. Note that 
we define the collinear factors to include the integrals over the momenta of 
the loops that connect the collinear subgraphs and the hard subgraph.

In the next step, we have to take into account that the collinear subgraphs 
are coupled to the hard subgraph by Dirac and Lorentz indices. The effect of 
boosting a Dirac spinor from rest to the energy $Q$ is to make its largest 
component of order $(Q/m)^{1/2}$ bigger than the rest frame value and the 
effect on a Lorentz vector is to give a factor of $(Q/m)^1$. The exponents 
are just the spins of the corresponding fields. Multiplying by these powers 
gives
\beq
Q^{2-N_q}.
\label{pcolh}
\eeq  
This agrees with Eq.\ (\ref{pform}) in the case that all external lines of the 
hard graph are quarks but is a factor $Q^{N_g}$ larger if the external 
particles are gluons.

The well-known problem of gluons with scalar polarization (see,
for example, \cite{Muell'89,LabSt}) will be dealt with later on.  Suffice it
to say here that gluons with such a polarization can be
factorized into the parton distributions by using
gauge-invariance arguments.

For the moment we just need to define the concepts of scalar and transverse 
polarization in the sense that we will use and show how this affects the power
 counting.

Consider the attachment of one gluon, of momentum $k^{\mu}$, from the 
collinear-to-$A$ subgraph to the hard subgraph. One has a factor 
$A^{\mu}g_{\mu\nu}H^{\nu}$, where $A^{\mu}$ and $H^{\nu}$ denote the 
collinear-to-$A$ and hard subgraph respectively, and $g_{\mu\nu}$ is the 
numerator of the gluon propagator in the Feynman gauge. On can now decompose 
this factor into components:
\beq
A\times H = A^+H^- + A^-H^+ - A_{\perp}\times H_{\perp}.
\eeq
One observes now that after the boost from the proton rest frame, the largest 
component of $A^{\mu}$ is the $+$ component. The largest term is therefore 
$A^+H^-$ and this is the term which gives the power in Eq.\ (\ref{pcolh}). The 
other two terms are suppressed by one or two powers of $Q$.

One can define now the following decomposition:
\beq
A^{\mu} = k^{\mu}\frac{A^+}{k^+} + \left ( A^{\mu} - k^{\mu}\frac{A^+}{k^+} 
\right ).
\label{decomp}
\eeq
The first term will be called the scalar component of the gluon and gives a 
polarization vector proportional to the momentum of the gluon. The second term
 gives the transverse part of the gluon with no $+$ component, it, therefore, 
gives a contribution to $A\times H$ which is one power of $Q$ smaller than the
 contribution of the scalar component. The $k^{\mu}$ factor in the scalar term
 multiplies the hard subgraph and gives a quantity that can be simplified by 
the use of Ward identities.

The above decomposition will now be applied to every gluon joining the 
subgraphs $A$ and $H$. The contribution of our region to the amplitude is now 
a sum of terms in which each of these gluons is either scalar or transverse. 
Each term has a power
\beq
Q^{2 - N_q - N_g}Q^{N_s} = Q^{2 - N_q - N_T};
\label{pcols}
\eeq
where $N_s$ is the number of scalar gluons and $N_T=N_g-N_s$ is the number of 
transverse gluons that enter the hard scattering. This is exactly the power in
 Eq.\ (\ref{pform}) without a soft subgraph.

In Taylor expanding the hard subgraph, we will need to slightly modify the 
above decomposition in Eq.\ (\ref{decomp}) and also apply an exactly analogous 
argument to the couplings of soft gluons to a collinear graph.

One will also need to pick out the largest component of the Dirac structure of
 the collinear subgraphs, here however, this is not necessary to be made 
explicit since we do not have a cancelation of the highest power. Note that 
the projection of the largest Dirac component is directly reflected in the 
$\gamma^+$ factor in the definition of the quark distribution.

In the derivation of Eq.\ (\ref{pcols}) we assumed $x$ not to be small. In the 
case of large $x$ then, we have to boost parts of the collinear-to-$A$ 
subgraph to get + momenta of order $p^+$ instead of $xp^+$. This leads to 
groups of lines with very different rapidities. In Feynman graphs, the effect 
is just to give a factor $1/x$ times logarithms, however only if the exchanged
 lines between the regions of different rapidities are gluons \cite{azi}. In 
the case of exchanged quarks, there will be a suppression factor of $x$. All of
 this does not influence the power of $Q$.

If one now adds in a soft subgraph $S$, one has the problem of choosing the 
appropriate scaling of the momenta. There are several possibilities and the 
literature is not entirely clear about the best scaling. In our case we will 
choose to have all soft momentum components of $O(m^2/Q)$. This has the 
advantage of not sending any collinear-to-$A$ lines far off-shell, the trade 
off is that we introduce regions where the momenta are unphysically soft in a 
confining theory and the power counting is mass sensitive.

With this scaling the basic power in the Breit frame is $m^2/Q$ to a power 
which is the dimension of the soft subgraph, including the integration over 
the soft loop momenta that circulate between S and the rest of the graph with 
the assumption of negligible masses. The power is simply
\beq
-N_{gS}-\frac{3}{2}N_{qs}
\eeq
where $N_{gS}$ and $N_{qS}$ are the number of external gluons and quarks of 
the soft subgraph. These external lines go either into the hard subgraph in 
which case the dimension of the hard subgraph is reduced by $3/2$ for each 
extra quark and $1$ for each extra gluon or into the collinear-to-$A$ subgraph 
which does not affect the power of $Q$.  Now we have to take into account that
 we are dealing with vectors and spinors connecting $S$ to $A$, this means
 that we have to use the same factors as before meaning we gain a factor 
$Q^{1/2}$ for each quark and $Q^1$ for each gluon.

Putting all the factors together we get for the soft to hard contribution 
$-N_g -\frac{3}{2}N_q - N_g -\frac{3}{2}N_q = -2N_g - 3N_q$ and for the soft to 
collinear contribution $-N_g -\frac{3}{2}N_q + N_g + \frac{1}{2}N_q = -N_q$. This 
together with our earlier result (Eq.\ (\ref{pcols})) yields Eq.\ (\ref{pform})
.   
    
\begin{figure}
\centering
\mbox{\epsfig{file=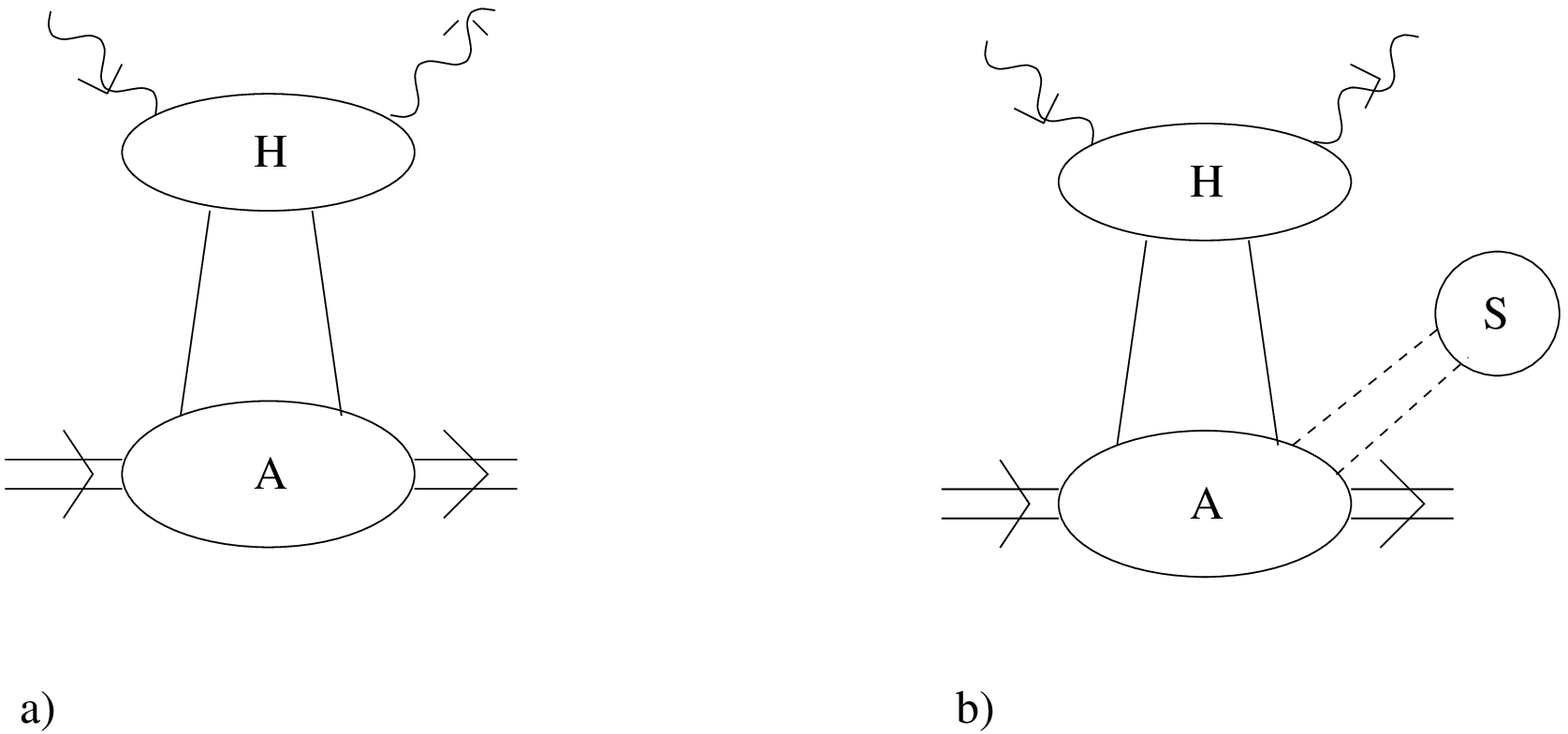,height=6cm}}
\vspace*{5mm}
\caption{Those reduced graphs that contribute to the leading
regions in DVCS. }
\label{Leading.Regions}
\end{figure}

\subsection{Leading Regions}
\label{lreg}

The leading regions for the amplitude are those with the largest
exponent $p(\pi )$ in Eq.\ (\ref{pform}).  It is easy to see that
these correspond to the reduced graphs in Fig.\
\ref{Leading.Regions}, independently of whether the out-going
photon is real or far off-shell.  The corresponding power is $Q^{0}$.
These reduced graphs have direct photon couplings to the hard
subgraph, they have exactly two parton lines connecting the
collinear subgraph $A$ to the hard subgraph $H$, and they have no
soft lines connecting to $H$. The two kinds of graph differ only
by the absence or presence of a soft subgraph that connects to
$A$ alone.

Among the other reduced graphs, which are non-leading for our
process, are those of the type
in Fig.\ \ref{Reduced.Graphs}b, which are leading in the case of
diffractive meson production, where the leading region gives
$Q^{-1}$.

In the case of a photon that is off-shell by order $Q^{2}$,
the amplitude for production
of the photon behaves like $Q^{0}$, the same as for a real photon.
However, the physically observed process includes the decay of
the time-like photon (to a $\mu ^{+}\mu ^{-}$ pair, for example), which
results in a power suppression of the observed cross section by
$1/Q^{2}$ compared with the cross section for making real photons.

\subsection{Proof of absence of a soft part in leading regions}
\label{poaos}

As mentioned in Sec.\ \ref{lreg}, there might, in principle,
be a soft part $S$ in the leading reduced graph connected solely
to the $A$ graph by gluons, as shown in Fig.\
\ref{Leading.Regions}b. Note that by Eq.\ (\ref{pform}), quarks
connecting $S$ to $A$ would lead to a power
suppression. We will now show that this soft part $S$ is indeed
absent, and so we only need to consider regions of the form of
Fig.\ \ref{Leading.Regions}a.
\begin{figure}
\centering
\mbox{\epsfig{file=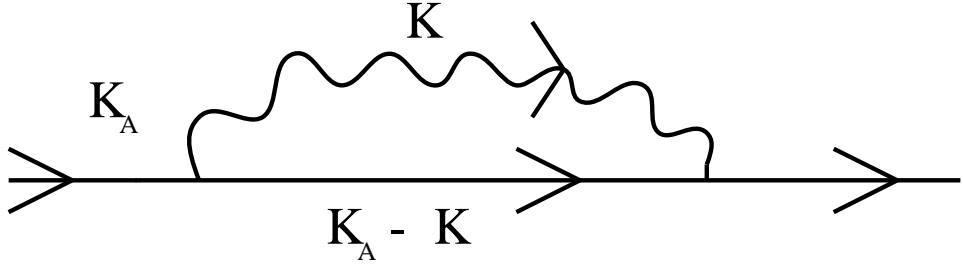,height=3.5cm}}
\vspace*{5mm}
\caption{Soft gluon loop attaching to collinear line.}
\label{Soft.to.A}
\end{figure}

We will first examine a simple one loop example, Fig.\
\ref{Soft.to.A}.  The external quark is part of the $A$ subgraph
in Fig.\ \ref{Leading.Regions}b, and the gluon is soft. So we
parameterize the momenta by:
\begin{eqnarray}
   k_{A} &=& (x_{1}p^{+}, k_{A}^{-}, {\bf k_{A,\perp }}),
\nonumber\\
   k  &=& (k^{+},k^{-},{\bf k_{\perp }}).
\label{mom}
\end{eqnarray}
where the $k^{+}_{A}$ is $O(Q)$ and all the other components are of
$O(m)$ or smaller.

If we omit irrelevant factors in the numerator,
the loop integral takes the following form:
\begin{eqnarray}
& &\int _{{\rm soft}~k}\frac {d^{4}k}{(k^{2}+i\epsilon )\left [(k_{A}-k)^{2} 
-m^{2} +i\epsilon \right ]}
\nonumber\\
& &=\int _{{\rm soft}~k}
  \frac {d^{4}k}{(2k^{+}k^{-} -k_{\perp }^{2}+i\epsilon ) \left [2(x_{1}p^{+} -
k^{+}) (k_{A}^{-}-k^{-})-(k_{A,\perp }-k_{\perp })^{2}-m^{2} +i\epsilon \right
]}
\nonumber\\
& &\simeq \int _{{\rm soft}~k}
  \frac {dk_{+}}{(2k^{+}k^{-} -k_{\perp }^{2}+i\epsilon ) \left [2x_{1}p^{+}
(k_{A}^{-}-k^{-})-(k_{A,\perp }-k_{\perp })^{2}-m^{2} +i\epsilon \right ]}.
\label{softk}
\end{eqnarray}
As before, $\simeq$ stands for ``equality up to power suppressed terms''.
As one can see, there is no $k^{+}$-pole in the second part of the denominator
and we can freely deform the contour in $k^{+}$ to avoid the pole in the soft
gluon propagator.  This takes us out of the soft region for $k$.

In the general situation, Fig.\ \ref{Leading.Regions}b, we can
use a version of the arguments in Ref.\ \cite{C.F.S'96,C97} to show
that the soft momenta ${k^{+}_{i}}$ can be rerouted in such a way
as to exhibit a lack of a pinch singularity.  The essential idea
is that one can find a path backwards or forward from one
external line of $S$ to another external line of $S$.  The loop
is completed along lines of $A$, all of which have much larger
$+$ momenta than what is typical of soft momenta, and hence there
is no pinch.

\subsection{Subtractions}
\label{sub}

The subtractions\footnote{We give examples of hands-on applications of the 
distributional ideas, as employed in this section, in the Appendix A.} 
necessary to avoid double counting in the amplitude are
exactly similar to the ones in Sec.\ VI of Ref.\ \cite{C.F.S'96}, since
the distributional arguments to construct the subtraction terms on a
pinch-singular surface $\pi $ presented there are very general in nature and
are not limited to the case of diffractive vector meson
production that was considered in \cite{C.F.S'96}.

Let us review the construction. For each graph $\Gamma$, there may be several 
different pinch singular surfaces $\pi$ in the space of loop-momentum 
contributing to the leading power. We write therefore the graph as a sum of 
those contributions
\beq
\mbox{Asy}\Gamma = \Sigma_{\pi}\Gamma_{\pi},
\eeq
where 'Asy' denotes the asymptotic behavior of the graph.

The term $\Gamma_{\pi}$ is obtained by Taylor expanding the hard and collinear
 subgraphs in powers of the small variables denoted by $T_{\pi}$. Since one can
 possibly have several regions contributing to a given graph, one must make 
subtractions to avoid double counting. This operation will be denoted by $R$. 
Therefore, we can write
\beq
\mbox{Asy}\Gamma = \Sigma_{\pi}RT_{\pi}\Gamma_{\pi}.
\eeq

This is exactly analogous to the Bogoliubov $R$-operation for renormalization.
 The easiest way to formulate the above procedure is due to Tkachov et 
al.\ \cite{tkach}. In this method the integrand of each graph $\Gamma$ is 
viewed as a distribution, in other words 
\beq
\Gamma\otimes f = \int dk \Gamma(k,p)f(k),
\label{intapprox}
\eeq
where k denotes the collection of loop momenta, p the external momenta and 
$f(k)$ is a test function. Putting the test function to $1$ will give the 
contribution of $\Gamma$ to the amplitude.

The advantages of the methods employed in Ref.\ \cite{tkach} are the control 
they give in treating different regions of momentum space separately without 
having sharp boundaries between different regions. Note that this is 
particularly convenient in our case where one has to deform contours of 
integration away from pinch singularities. If we were to use sharp boundaries 
these deformations would be impossible.

Using this language, the contribution $\Gamma_{\pi}$ to Asy$\Gamma$ from the 
neighborhood of a pinch-singular surface $\pi$ is localized on the surface, 
in other words is proportional to a $\delta$-function that restricts the 
integration to the surface. One then defines a hierarchy of regions  through 
set-theoretic inclusion: $\pi_1 > \pi_2$ means that the pinch-singular 
surface $\pi_1$ contains $\pi_2$. One than constructs any $\pi$ on the 
assumption that the terms for all bigger regions have already been 
constructed. Thus one has a recursive construction starting from the largest 
region.

Assume that one has constructed the terms $\Gamma_{\pi'}$ for all regions 
bigger than $\pi$ and then decompose Asy$\Gamma$ as
\beq
\mbox{Asy}\Gamma = \Sigma_{\pi' > \pi}\Gamma_{\pi'} + \Gamma_{\pi} 
+~\mbox{other terms}.
\eeq
The ``other terms'' correspond to three classes of surfaces. The first class
are those surfaces that are smaller than $\pi$, the second class consists of 
those that intersect $\pi$ in a subset of $\pi$, i.e.\ , they have a lower 
dimension and the third class are those that do not intersect $\pi$ at all. 

The hypothesis for our induction is that the sum of $\Gamma_{\pi'}$ over 
$\pi' > \pi$ gives a good approximation to the original $\Gamma$ except in 
the neighborhoods of the smaller surfaces for which $\Gamma_{\pi}$ has not yet
 been constructed. The integrals defining the $\Gamma_{\pi}$'s cover the whole
 space of integration variables, but are only required to give a good 
approximation if one excludes neighborhoods of smaller surfaces. This means 
that we require the test function in Eq.\ (\ref{intapprox}) to be zero on 
these smaller surfaces.
When $\Gamma_{\pi}$ is combined with the $\Gamma_{\pi'}$ for larger surfaces 
it must give a good approximation to $\Gamma$ on a neighborhood of $\pi$. 
Note that a good approximation is not necessary on the surfaces where the 
test function is zero; there is no need for $\Gamma_{\pi'}$ to be constructed 
for the smaller surfaces since they will give zero with such a test function. 
This is enough to prove the inductive hypothesis for the next recursive step.

Since $\Gamma_{\pi}$ is located on $\pi$, it is necessary to consider only a 
neighborhood of $\pi$. This combined with the statement in the previous 
paragraph ensures that there is no need for the unconstructed ``other terms'' 
in order to construct $\Gamma_{\pi}$.

One can therefore define
\beq
\Gamma_{\pi} = T_{\pi}\left [ \Gamma - \Sigma_{\pi' > pi}R\Gamma_{\pi'}\right ],
\label{eqaprox}
\eeq
where $T_{\pi}$ stands for a Taylor expansion in powers of the small variables
 on $\pi$. The first term is the Taylor expansion of the original graph and 
the remaining terms can be thought of as subtractions that prevent double 
counting of contributions to the integral over a neighborhood of $\pi$.   

This results in a sum over $\Gamma_{\pi}$ and the terms for the larger regions
\beq
\Gamma_{\pi} + \Sigma_{\pi' > \pi}R\Gamma_{\pi'},
\eeq
giving the correct contribution to the asymptotics of $\Gamma$ that originates 
from a neighborhood of $\pi$ and of all larger regions. Neighborhoods of 
smaller regions are, of course, excluded.

In general $\Gamma_{\pi}$ gives a divergence when integrated with a test 
function over a neighborhood of any of these smaller regions. Therefore, it 
is only defined when integrated with a test function which is $0$ on these
 smaller regions. We now extend this distribution to a distribution defined on
 all test functions by adding infra-red counterterms to cancel the arising 
divergences from the smaller regions. This operation is analogous to the well 
known $+$ distribution employed earlier in this thesis. The result will be 
$R\Gamma_{\pi}$ with the counterterms all being local in momentum space. 
Choosing a definition of $\Gamma_{\pi}$ on the smaller surfaces is perfectly 
satisfactory, since we have not yet considered how to approximate $\Gamma$ in 
regions smaller than $\pi$. We only require $R\Gamma_{\pi}$ to be finite and 
the counterterms to be localized on smaller surfaces than $\pi$ in order not 
to affect the good approximation we have for $\pi$ and larger surfaces. 
Continuing with the recursion, one eventually obtains appropriate 
approximations for these smaller surfaces. 

Note that the subtraction terms, as defined in Eq.\ (\ref{eqaprox}), ensure 
that changes in the choice of counterterms localized on any particular surface 
$\pi$ are cancelled by corresponding changes in the subtraction terms in the 
definition of $\Gamma_{\pi}$. Therefore, the overall result for the asymptotic
 expansion of $\Gamma$ is independent of these choices. 

The above statements lead to the following asymptotic form of the amplitude
\begin{equation}
\mbox{Asy}\, T = \Sigma_{\Gamma }\mbox{Asy}\, \Gamma = A \times H.
\end{equation}
where $\Gamma $ stands for a possible graph for the amplitude $T$.

\subsection{Taylor expansion}
\label{TE}

We now obtain the leading term in the hard subgraph, when
it is expanded in powers of the small momenta.
The arguments used are exactly analogous to the ones used
in Sec.\ VII of Ref.\ \cite{C.F.S'96} except that we do not have to deal with a
$B$ subgraph as was the case in \cite{C.F.S'96}. So we have:
\begin{eqnarray}
 A \times H &\simeq&
    \int dk_{A}^{+}
    \,
    H\left(q,q',(k_{A}^{+},0,0_{\perp }),
    (\Delta ^{+} - k_{A}^{+},0,0_{\perp }
    \right)
\nonumber\\
&&
    \int dk_{A}^{-}d^{2}k_{A,\perp }
    \,
    A(k_{A},\Delta -k_{A}),
\label{CTE}
\end{eqnarray}
where $k_{A}$ is the loop momentum joining the $A$ and $H$
subgraphs,
and again $\simeq$ means equality up to power-suppressed corrections.
Eq.\ (\ref{CTE}) has
already a factorized form.
However we still have to deal with the extra scalar gluons that
may be exchanged between the subgraphs $A$ and $H$;
this will be done in the next subsection.

Eq.\ (\ref{CTE}) can be written in the following way:
\begin{equation}
A \times H \simeq \Sigma_{i} \int dk^{+}
C_{i}(q,q',k^{+})O_{i}(p,p',k^{+}),
\label{sumeq}
\end{equation}
where the $C_{i}$ are the short distance coefficient functions and the $O_{i}$
are the matrix elements of renormalized light-cone operators.

\subsection{Gauge Invariance}
\label{gi}

\begin{figure}
\centering
\mbox{\epsfig{file=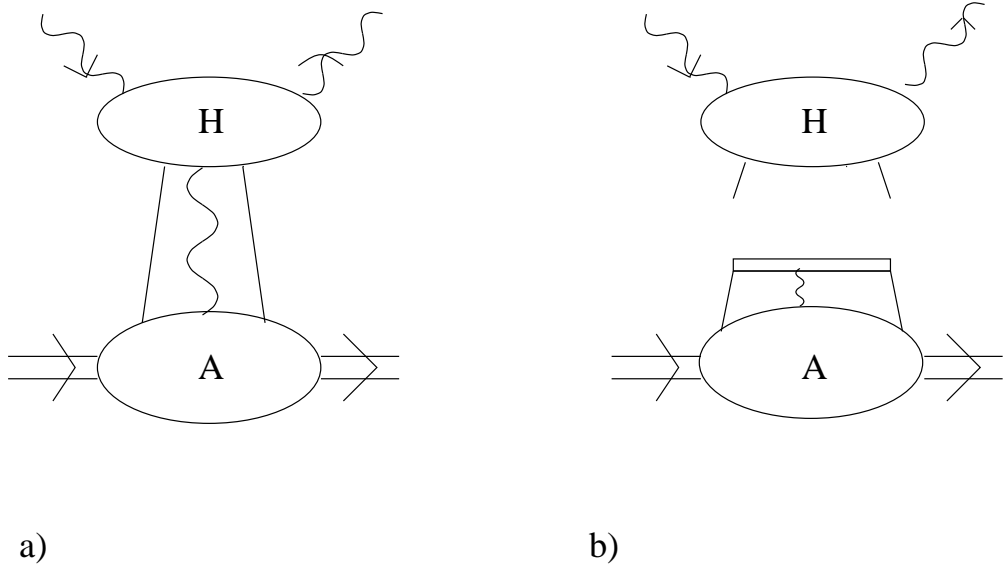,height=7.5cm}}
\vspace*{5mm}
\caption{a) A scalar gluon attaching the collinear subgraph to
the hard subgraph $H$ in the unfactorized form. b) Factorized form after
application of gauge invariance and Ward-identities. The double line
represents the eikonal line to which the scalar gluon attaches.}
\label{Scalar.Gluon}
\end{figure}

In order to identify the $O_{i}$ with the parton distributions as defined in
\cite{C.F.S'96} (for example),
it is necessary to show that all gluons with scalar polarization
attaching to the hard graph can be combined into a path-ordered
exponential.
Fig.\ \ref{Scalar.Gluon} shows the example of one scalar gluon.
We can follow the arguments of Sec.\ VII.\ D of Ref.\
\cite{C.F.S'96} step by
step since those rely on very general results obtained by Collins \cite{col}.
In this way we obtain exactly the same parton distributions as in \cite{C.F.S'96},
namely:
\begin{eqnarray}
   f_{q/p} &=& \int ^{\infty }_{-\infty }\frac {dy^{-}}{4\pi
}e^{-ix_{2}p^{+}y^{-}}
      \langle p| T\bar \psi (0,y^{-},{\bf 0_{\perp }})\gamma ^{+}{\cal P}\psi
(0)|p'\rangle ,
\nonumber\\
   f_{g/p} &=& -\int ^{\infty }_{-\infty }\frac {dy^{-}}{2\pi }\frac
{1}{x_{1}x_{2}p^{+}}
   e^{-ix_{2}p^{+}y^{-}}
   \langle p| T G_{\nu }^{+}(0,y^{-},{\bf 0_{\perp }}){\cal P}G^{\nu
+}(0)|p'\rangle .
\end{eqnarray}
Here, $\cal P$ represents a path-ordered exponential of the gluon
field that makes the operators gauge invariant.  The variable
$x_{2}$ is the same as in Eq.\ (\ref{theorem}). The evolution
equations are the same as in \cite{Rad} and \cite{C.F.S'96}.

\subsection{Partons with $k^+=0$: breakpoints and endpoints}
\label{subt}

In the factorization theorem Eq.\ (\ref{theorem}), the integral over
the fractional momenta includes the points $x_1=0$ and $x_2 = 0$.  At
these points, the hard scattering coefficient for DVCS has a pole, and
so we appear to get a logarithmic contribution to the cross section
from a region in which one of the lines joining the parton density to
the hard scattering subgraphs is soft instead of collinear.  This
apparently contradicts our power-counting result that such a region
gives a non-leading power.  This phenomenon was investigated by
Radyushkin \cite{Rad}.  In this section, we will give a general
demonstration that the region in question does not give a problem.

First, let us observe that the region of integration over $x_1$ in the
factorization formula Eq.\ (\ref{theorem}) is $-1+x\leq x_1 \leq 1$.
This is proved by the methods of light-front perturbation theory, by
requiring that the intermediate states in Fig.\ \ref{Leading.Regions}
be physically allowed.  See Ref.\ \cite{Rad,ours} for detailed
derivations and discussions.  The points $x_1=0$ and $x_2=0$ at which
the potential problem arises are what we will call ``breakpoints'',
since they occur in the middle of the range of integration where one
of the two lines changes direction.

\begin{figure}
\centering
\vspace*{0.5cm}
\mbox{\epsfig{file=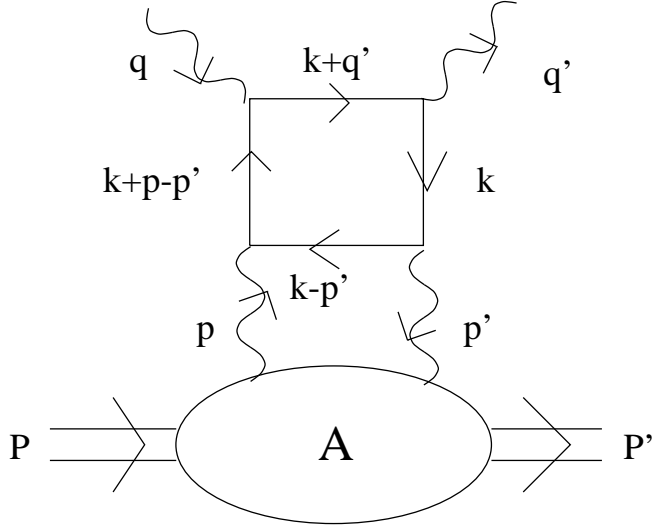,height=7cm}}
\vspace*{0.5cm}
\caption{Particular example of potentially problematic diagram.}
\label{problems}
\end{figure}

We continue by examining a particular case, illustrated in Fig.\
\ref{problems}, and showing how the argument generalizes.  To simplify
the example, let us restrict our attention to regions where the
subgraph $A$ and the lower three lines ($p$, $p'$ and $k-p'$) have
their momenta collinear to the proton. We will also require the two
quark lines, $k+p-p'$ and $k$, on the sides of the ladder to have
their momenta either collinear to the proton or soft.

We will also only need the case of the production of a real photon,
${q'}^2 = 0$, since this is where the problem arises.

The top loop of the graph has the form, omitting the $i\epsilon$ factors in the
propagators for convenience :
\begin{equation}
U = \int d^4k
   \frac{{\rm Numerator~factors}}{
       \left( k^2 - m^2 \right) 
       \left[ (k-p')^2 - m^2 \right]
       \left[ (k+p-p')^2 - m^2 \right]
       \left[ (k+q')^2 - m^2 \right]
} .
\label{probint}
\end{equation}
When both $k$ and $k+p-p'$ are collinear to $A$, the top line is
off-shell by $O(Q^2)$, and it is correct to use the collinear
approximation
\begin{equation}
    \frac{1}{(k+q')^2 - m^2 + i\epsilon}
  \to 
    \frac{1}{ x_2 Q^2 / x + i\epsilon} ,
\label{coll.approx}
\end{equation}
where $x_2 = k^+/(xp^+)$.  A corresponding replacement is also to be
made in the numerator in $U$. The right-hand side of Eq.\
(\ref{coll.approx}) exhibits the afore-mentioned pole at $x_2=0$.
The result of applying the collinear approximation is to give the
appropriate contribution to the factorization formula Eq.\
(\ref{theorem}). 

The collinear approximation becomes invalid when $k$ becomes soft,
i.e., when $x_2 \to 0$.  We must now demonstrate two facts.  The first
is that, when $k$ is in a neighborhood of the soft region, the
collinear approximation is valid after integration over $k$.  The
second fact is that the use of the collinear approximation does not
give an important contribution from some other region of $k$ that was
not permitted in the original graph.

We now examine the integral $U$ in the neighborhood of the soft region
for $k$.  It has the following form, again omitting the $i\epsilon$ factors in 
the propagators :
\begin{eqnarray}
U_{{\rm soft}~k} &\simeq& \int_{{\rm soft}~k} d^4k
\frac{1}
{ \left[ 2k^{+}k^{-}-k^2_{\perp} - m^2 \right]
  \left[ 2{p'}^{+}({p'}^- - k^{-}) - (p'-k)^2_{\perp} 
         - m^2 \right]
}
\label{U.soft}
\\
&&\frac{1}
{\left[ 2 (p^{+}-{p'}^+) (p^- -{p'}^- +k^{-})
 - (p-p'+k)^2_{\perp} - m^2 \right]
 \left( \frac{k^{+}Q^2}{xp^{+}}-k^2_{\perp} - m^2 \right)
} ,
\nonumber
\end{eqnarray}
where we have neglected $k^{+}$ in the collinear-to-$A$ lines, $k^{-}$
in the collinear-to-$B$ line and $x\neq x_1$. We have also ignored the
numerator factors, which are an irrelevant complication for our
purposes.

According to the power-counting results of Sec.\ \ref{count}, which
are obtained from \cite{C97}, the soft region for $k$ gives a
power-suppressed contribution.  This estimate assumes that all
components of $k$ are comparable (in the Breit frame), and is obtained
as follows. Let the magnitude of the components of $k$ be $m$.
Then the order of magnitude of the
soft part of $U$ is a product of factors 
$1/(m^5Q^3)$ from the denominators, $m^4$ from the phase space, and
$Q^2m^2$ from the numerator, for an overall power $m/Q$.
This result can be obtained by writing down the largest 
components in the trace and propagators of Fig.\ \ref{problems} and
Eq.\ (\ref{U.soft}). Moreover in this region it is
correct to replace the fourth propagator in Eq.\ (\ref{U.soft}) by its
collinear approximation Eq.\ (\ref{coll.approx}), so that we do not
lose the factorization theorem.

However, if the components of $k$ are asymmetric this estimate no
longer holds.  In particular if the longitudinal components of $k$ are
of order $k^\pm \sim m^2/Q$ while the transverse components remain of
order $m$, then we get contributions of order 
$1/m^8$ from the denominators, $m^6/Q^2$ from the phase space, and
$Q^2m^2$ from the numerators, for a total of $m^0Q^0$. 
This shows that the contribution from this region
is unsuppressed for large $Q$.

At this point we must appeal to the contour deformation arguments of
Ref.\ \cite{C97}.  It is only when the integration over $k$ is pinched
in the region in question that it needs to be taken into account.  In
the dangerous region we have $k^+k^- \ll k_\perp^2$, so that the only
$k^+$ dependence in Eq.\ (\ref{U.soft}) is the pole in the fourth
denominator.  We can therefore deform $k^+$ into the complex plain a
long way out of the region we are considering, indeed all the way to
the collinear-to-$A$ region.  Then the collinear approximation is
valid so that we can replace the graph by its contribution to the
factorization formula.

This contour deformation argument is completely general, as explained
in Sec.\ IIIE of Ref.\ \cite{C97}.  Whenever we have a soft momentum
with $k^+k^- \ll k_\perp^2$, the contour of $k^+$ can be deformed away
from poles in the jet subgraph associated with with the produced
photon.  Since all the relevant singularities are in the final state,
they are all on the same side of the real axis.  

Now that we have established in more detail that the only leading
regions are those symbolized in Fig.\ \ref{Leading.Regions}(a), we can
apply the collinear approximation as described earlier, and hence we
obtain the factorization theorem.  

But we still see the following problem.  In the factorization theorem,
Eq.\ (\ref{theorem}), the parton densities are non-analytic at the
breakpoints $x_1=0$ and $x_2=0$, whereas the coefficient function has
a pole at each of these points.  Again consider the collinear
approximation to Fig.\ \ref{problems} in the region we were
considering.  The parton density is non-analytic when $x_2=0$, while
the coefficient function has a pole there, as is seen from the
right-hand-side of Eq.\ (\ref{coll.approx}).  So we cannot literally
apply the contour deformation argument.  

What we will show is that the parton density is continuous at the
breakpoint, so that it can be written as the sum of a function that is
analytic at $x_2=0$ and a function that has a zero at $x_2=0$.  The
only potential leading twist contribution near the breakpoint is
associated with the non-zero analytic term to which the contour
deformation argument applies.

\begin{figure}
\centering
\vspace*{0.5cm}
\mbox{\epsfig{file=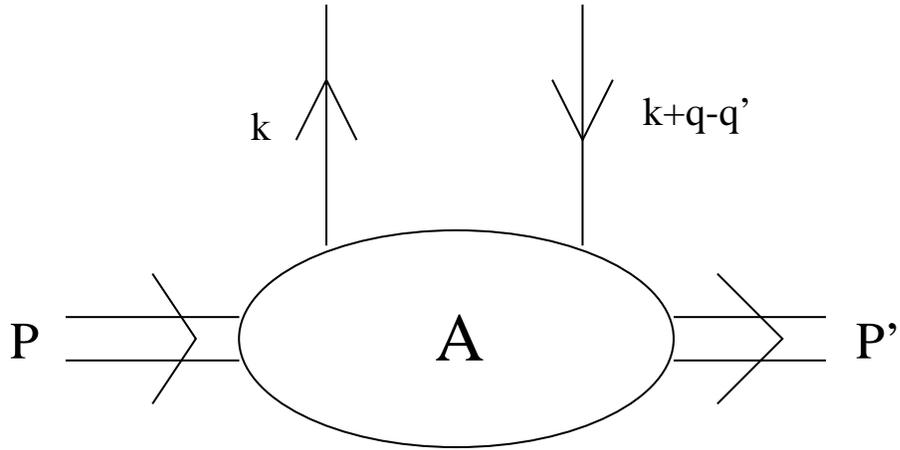,height=6cm}}
\vspace*{0.5cm}
\caption{Parton distribution amplitude.}
\label{pargra}
\end{figure}

To prove this property of the parton density at a breakpoint, consider
a general graph for the parton density, as shown in Fig.\
\ref{pargra}. We have found it convenient to change the labeling of
the momentum compared with the previous figure.
As always, the $k^{-}$ and ${\bf k_{\perp}}$ components of $k$
have been short circuited and are integrated over.
The $k$-line gives a pole at
$k^{-} = (k^{2}_{\perp} + m^2 - i\epsilon) / 2k^{+}
       = (k^{2}_{\perp} + m^2 - i\epsilon) / 2 x_1 P^{+}$, 
while the $k+q'-q$-line gives a pole at
$k^{-} = [(k+q-q')^{2}_{\perp} + m^2 - i\epsilon] / 2 (k^{+}-\xi P^{+})
       = [(k+q-q')^{2}_{\perp} + m^2 - i\epsilon] / 2 x_2 P^{+}$.
Here, $\xi$ is the fractional longitudinal momentum transfer
$1-{P'}^+/P^+$. In addition there are poles from the collinear-to-$A$
lines in the blob.  For example if the blob consists of a single line, we
have a pole at 
$k^{-} = P^- - (k^{2}+m^2_{\perp}-i\epsilon) / 2 (1-x_1) P^{+}$
or at
$k^{-} = - {P'}^- + [(k+P')^{2}+m^2_{\perp}-i\epsilon) / 2 (1-\xi+x_1)P^{+}$. 

As we vary $x_1$, the $k^-$ contour can generally be deformed to avoid
the poles, so that we have analytic dependence on $x_1$.  The possible
exceptions occur when the $k^-$ contour is pinched for finite $k^-$ or
when a singularity coincides with the endpoint of the integration at
$k^-=\infty$.  A pinch never occurs; in the general case this is a
consequence of the Landau rules.  But endpoint singularities occur,
and these are precisely at the breakpoints.

For example if $k^{+}\rightarrow 0$, then $k^{+}$ can approach $0$
from above and below. The pole giving us trouble stems from the
$k$-line, all other propagators are unproblematic in this case, since
their poles are at finite $k^-$. The pole in $k^{-}$ approaches
$+\infty - i\epsilon$ as one approaches $0^+$ and $-\infty+i\epsilon$
as one approaches $0^-$. This means that the $k^{-}$ pole crosses the
real axis at infinity. Hence the parton distribution is non-analytic
there.  Since the singularity is at $|k^-|=\infty$ the other
propagators have large denominators, and hence we get a zero for the
non-analytic part of integral at the breakpoint.  Thus the parton
density is continuous at the breakpoint, as claimed.  This result
enables the factorization formula to be valid in the neighborhoods of
the breakpoints.  Since the other poles in the $k^-$ integral are on
opposite sides of the real axis, the parton distribution is non-zero
at the breakpoints.

Effectively the crossover of the pole occurs when $k$ is in a
collinear-to-$B$ region, which we know is power suppressed.  This
indicates that the argument we have just given generalizes to all
graphs.

We also remark on the behavior at the endpoints.  Let us look at the
case $k^{+}\rightarrow p^{+}$. We find that another of the poles pole
runs off to $-\infty$ this time and crosses the real axis there. But
now all the other poles are on a single side of the real axis, so that
the sole contribution to the parton density comes from the pole at
infinity, and hence there is a zero of the parton density at the
endpoint.

\subsection{Completion of Proof}
\label{comp}

Using the definitions of the parton distributions and the hard scattering
coefficients we finally obtain Eq.\ (\ref{theorem}). Note that the theorem is
valid for the production of a real photon which directly goes into the final
state and for the production of a time-like photon that decays
into a lepton pair.

\section{Conclusion}
\label{conclx}

We have proved the factorization theorem for deeply virtual
Compton scattering up to power suppressed terms to all orders in
perturbation theory. The form of the theorem is independent of
the virtuality of the produced photon.

\chapter{Diffractive Exclusive Photon Production in DIS at HERA} 
\indent

\section{Introduction}
\label{sec:intro}

Recent data from HERA has spurred great interest in exclusive or diffractive 
 direct production of photons  in $e - p$ scattering (deeply virtual Compton
scattering or DVCS) as another source to obtain more 
information about the gluon distribution inside the proton for nonforward 
scattering. Therefore, we will start a phenomenological investigation of
this process in this chapter based on Ref.\ \cite{FFS} after having proved 
the validity of a factorization theorem for this process in the last chapter.

Exclusive, diffractive virtual Compton processes at large $Q^2$,
first investigated in \cite{Bartels}, offer a new and 
comparatively ``clean''\footnote{Clean in the sense that the wave function of 
a spatially small size configuration within a real photon is better known as 
compared to the wave functions of vector mesons thereby removing a big 
theoretical uncertainty in the determination of the gluon distribution.}
way of obtaining information about gluons inside the proton in a 
nonforward kinematic situation. We are interested in the production of a 
real photon compared to the inclusive DIS cross section. The
 exclusive process is nonforward 
in its nature, since the photon initiating the process is virtual 
($q^2<0$) and the final state photon is real, forcing a small but finite
momentum transfer to the target proton, i.e.\ , forcing a nonforward kinematic 
situation as we would like.

The chapter is organized in the following way:
In Sec.\ \ref{AJM} we estimate the amplitude in the normalization point
$Q_0^2 \sim 2 GeV^2$ using
the aligned jet model approximation and conclude 
that for such $Q^2$ the nondiagonal amplitude is larger than the diagonal one 
by a factor of $\sim 2$.
 In Sec.\ \ref{sec:hp} we 
calculate the imaginary part of the amplitude for $\gamma^* + p \rightarrow 
\gamma + p$ in the leading order
of the running coupling constant $\alpha_s$ and compare it to the 
imaginary part of the amplitude in DIS in the same order. 
In Sec.\ \ref{Bslope} we argue that at sufficiently small $x$
the  $t$-dependence of the cross section should reflect the interplay
of hard and soft physics typical of diffractive phenomena in
DIS. In other words, at fixed $x$ and increasing $Q^2$, hard physics should 
tend to occupy the dominant part of the space of rapidities. In contrast to 
this, 
at fixed $Q^2$ and decreasing $x$, hard physics should occupy a finite 
range of rapidities which increases with $Q^2$ as 
  $~\sim \ln {Q^2\over \beta M_{\rho}^2}$ and 
with $\beta\sim  0.1 - 0.2 $ in the HERA energy range
due to the QCD evolution, and that soft QCD physics occupies the rest of 
phase space. 
In Sec.\ \ref{sec:slope} and Sec.\ \ref{sec:cross} we give the total cross 
section of exclusive photon 
production and give numerical estimates of the DVCS production rate 
at HERA and find that such measurements are feasible for the current 
generation of experiments. We also show the feasibility of directly measuring 
the real part of the DVCS amplitude and hence, at least, the shape of the 
nondiagonal parton distributions through a large azimuthal angle asymmetry in 
$ep$ scattering for HERA kinematics.
Sec.\ \ref{sec:concl} finally contains concluding remarks.

\section{The amplitude for diffractive virtual Compton scattering at 
intermediate $Q^2$ }
\label{AJM}
Similar to the case of deep inelastic scattering, in real photon
production it is possible to calculate within perturbative QCD
the $Q^2$ evolution of the amplitude but not its value at the normalization
point at $Q_0^2 \sim $ {\it few GeV$^2$}
where it is given by nonperturbative effects. 
Therefore, we start by discussing expectations for this region. It was 
demonstrated 
in \cite{FS88} that the aligned jet model \cite{bj} coupled with the idea of 
color screening provides a reasonable semiquantitative description of
$F_{2N}(x \le 10^{-2}, Q_0^2)$. 
In this model the virtual photon interacts at intermediate $Q^2$ and small $x$
via transitions to a $q \bar q$ pair with small transverse momenta - 
$k_{0,t}$ ($\left<k_{0,t}^2\right> \sim 0.15 GeV^2$)
and average masses $\sim Q^2$ which thus carry asymmetric fractions of the
virtual photon's longitudinal momentum. Due to a large transverse color
separation, 
$b \sim 2\sqrt {2/3} r_{\pi}$,
  the aligned jet model components of the photon wave 
function interact strongly with the target given by the cross section
$\sigma_{tot}(``AJM''-N) \approx \sigma_{tot}(\pi N)$.
Neglecting contributions of the components of the
$\gamma^*$ wave function with smaller color
separation, one can write $\sigma_{tot}(\gamma^*N)$
 using the Gribov dispersion representation \cite{Gribov} as \cite{FS88}:
\beq
\sigma_{tot}(\gamma^*N)= {\alpha \over 3 \pi}
 \int_{M_0^2}^{\infty}{\sigma_{tot}(``AJM''-N) R^{e^+e^-}(M^2)
M^2 {3 \left<k_{0~t}^2\right>\over M^2} \over (Q^2+M^2)^2}d M^2,
\label{AJMeq}
\eeq
where the factor $M^2$ in the  numerator is due to the 
overall phase volume, 
$R^{e^+e^-}(M^2)={\sigma(e^+e^- \to hadrons) \over 
\sigma(e^+e^- \to \mu^+\mu^-)}$. The factor 
${3 \left<k_{0~t}^2\right>\over M^2}$ is the fraction of the whole phase 
volume occupied by the aligned jet model ,
and the factor $1/(Q^2+M^2)^2$ is due to 
the propagators of the photon in the hadronic intermediate state
with mass square equal to $M^2$. Based on the logic of a 
local quark-hadron  duality (see e.g. \cite{FRS} and references
therein), we take the
lower limit of integration to be $M_0^2 \sim 0.5 GeV^2 \le m_{\rho}^2$.
  In the case of real photon production, the imaginary part of the 
amplitude for $t=0$ is given by
\beq
\frac{1}{s}Im A(\gamma^*+N \to \gamma +N)_{t=0}= {\alpha \over 3 \pi}
\int_{M_0^2}^{\infty}
{\sigma_{tot}(``AJM''-N) R^{e^+e^-}(M^2)
M^2 {3 \left<k_{0~t}^2\right>\over M^2} \over (Q^2+M^2)M^2}d M^2,
\label{ImAJMeq}
\eeq 
with $s=2 q_0m_N$ being the flux factor.
The only difference from Eq.\ \ref{AJMeq} for
$\sigma_{tot}(\gamma^*+N)$ is the change of one of the propagators from 
$1/(Q^2+M^2)$ to $1/M^2$ - here $q_0$ is the energy of the virtual photon in 
the rest frame of the target.
 
Approximating $R^{e^+e^-}(M^2)$ 
as a constant for the $Q^2$ range in question \footnote{We understand this in 
the sense of a local duality of the hadron spectrum and 
the $q \bar q$ loop.} we find 
\beq
R \equiv {Im A(\gamma^*+N \to \gamma^* +N)_{t=0}
\over Im A(\gamma^*+N \to \gamma +N)_{t=0}}= {Q^2 \over Q^2+M_0^2}
\ln^{-1}(1+Q^2/m_0^2).
\label{ratiogam}
\eeq
In the following analysis we will take
$Q^2_0$ for the
perturbative QCD evolution as 2.6 GeV$^2$ to avoid ambiguities. 
It is easy to convince oneself that for $M_0^2 \sim 0.4 \div 0.6~ GeV^2$  and 
$Q^2\approx 2-3~\mbox{GeV}^2$ Eq.\ (\ref{ratiogam}) leads to $R \approx 0.5$.
A similar value of $R$ has been obtained within the generalized
vector dominance model in Ref.\ \cite{FGS}
 As we will see below, QCD evolution leads to a strong increase of
$Q^2 Im A(\gamma^*+N \to \gamma +N)_{t=0} $ for increasing $Q^2$
and fixed $x$. However, this does not change the value of $R$ appreciably.

\section{The amplitude for exclusive real photon production at large $Q^2$.}
\label{sec:hp}

\begin{figure}
\centering
\mbox{\epsfig{file=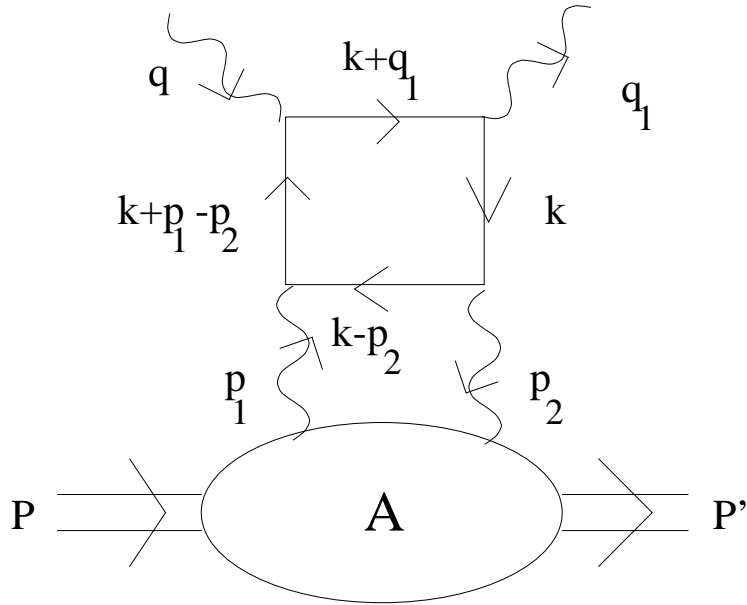,height=8cm}}
\vspace*{5mm}
\caption{Leading contribution to DVCS at small $x$.}
\label{dem}
\end{figure}

The process of exclusive, direct production of photons in 
first nontrivial order of $\alpha_s \ln {Q^2\over Q_0^2}$
at small $x$ can be calculated as the sum of a hard contribution 
calculated (see Fig.\ \ref{dem}) within the framework of QCD evolution 
equations \cite {Abram'95}
and a soft contribution which we evaluated in the previous section
within the aligned jet model. 
The hard contribution can be described through a two gluon exchange of the box 
diagram with the target proton.
In order to calculate the imaginary part of the amplitude, we need to 
find the hard amplitude from the box
as well as the gluon-nucleon scattering plus the soft contribution at $Q_0^2$
from the aligned jet model analysis. This is an example of the usual 
subtraction scheme in factorization where one subtracts out the collinear 
region in the factorization formula to have a convolution of a parton 
distribution with a hard scattering coefficient and then adds it back in  
which corresponds to the aligned jet model contribution in our case. 

Let us first give a general 
expression for the imaginary part of the amplitude and then proceed to deal 
with the gluon-nucleon scattering, followed by the calculation of the box 
diagrams. 

First, let us discuss the hard contribution which actually dominates
in the considered process. To account for the gluon-nucleon scattering, 
we work with Sudakov variables for the gluons with momenta $p_1$ and $p_2$
 attaching the box to the target and the following kinematics  for the 
gluon-nucleon scattering:
\beq
p_1 = \alpha q' + x_1 p' + p_{t}, \; \; d^{4}p_1=\frac{s}{2}d\alpha dx_1 d^{2}
p_{t},
\label{suva}
\eeq
where $q'$ and $p'$ are light-like momenta related to $p,q$ the momenta of the
target proton and the probing virtual photon respectively, by:
\bea
& &q = q' - xp', \; \; p = p' + \frac{p^2}{2p'\cdot q'}q', \nonumber\\
& &s = 2p\cdot q = 2p'\cdot q'- xp^2,
\eea
with $x$ being the Bjorken $x$ and $x_1$ the proton momentum fraction 
carried by the outgoing gluon. Note that $p'$ is not related to $P'$, the 
momentum of the outgoing proton. 
Equivalent equations to Eq.\ (\ref{suva}) apply 
for $p_2$ with the only difference being that $x_1$ is replaced by $x_2$,
the momentum fraction of the incoming gluon,
signaling that in leading log there is only a difference in the longitudinal 
momenta but not in the transverse momenta. 
Furthermore there is a simple relationship between $x_1$ and $x_2$ : 
$\Delta = x_1 - x_2 = \mbox{const.}$, following from the kinematics of the 
considered reaction\footnote{In the case of the imaginary part of the 
amplitude which we are discussing at this point, one has $x_1 > \Delta >0$ 
and we can treat the soft part as a parton distribution function (the DGLAP
regime), whereas if $0<x_1<\Delta$ one would have the situation of a 
distributional amplitude as first discussed by Radyushkin \cite{Rad'96} 
which is 
governed by the Brodsky-Lepage evolution equations.} where $\Delta$ is the 
asymmetry parameter or skewedness of the process under consideration.
Therefore, one is left with just 
the integration over $p_1$ since $p_2$ cannot vary independently of $p_1$.  
Being exclusively interested in the small $x$ region, one can safely make the 
following approximations: $s = 2p\cdot q \simeq 2p'\cdot q'$ and $p' \simeq p$. Since we 
are working in the leading $\alpha_s \ln Q^2$ approximation, neglecting 
corrections of order $\alpha _s$, the main contribution comes from the region 
$p_{t}^2<<Q^2$, hence the contribution to the imaginary part of the amplitude 
simplifies considerably. First, since 
$|p^{2}_{1}|=|\alpha x_1 s + p_{t}^2| << Q^2$, one has $\alpha << 1$ and the 
polarization tensor of the 
propagator of the exchanged gluon in the light-cone gauge $q'_{\mu}A^{\mu}=0$ 
becomes, see Ref.\ \cite{GribLip} :
\beq
d_{\mu\lambda}\simeq \frac{p'_{\mu}q'_{\lambda}}{p'\cdot q'}.
\label{prop}
\eeq
In other words it is enough to take the longitudinal polarizations of the 
exchanged gluons into account.

Using Eq.\ (\ref{prop}) one obtains the following expression for the total 
contribution of the box diagram and its permutations:
\beq
Im\, A = \int \frac{d^{4}p_1}{(2\pi)^{4}i}\frac{1}{p^2_1p^2_2}
2ImA^{ab(P)}_{\mu\nu}ImA^{ab(T)}_{\lambda\sigma}d_{\mu\lambda}(p_1)
d_{\nu\sigma}(p_2),
\label{amp}  
\eeq
where $ImA^{ab(P)}_{\mu\nu}=ImA^{ab}_{\mu \nu}(\gamma^* g 
\rightarrow q\bar q)$ is 
the sum of the box diagrams, $ImA^{ab(T)}_{\lambda \sigma}$ is the amplitude 
for the gluon-nucleon scattering, a,b  are the color indices and the 
overall tensor structure has been neglected for now. The usage of
the imaginary part of the scattering amplitude and in particular limiting
ourselves to the s-channel contribution as the dominant part in both
the forward and the nonforward case (Eq.\ (\ref{amp})) is correct 
(see Ref.\ \cite{ours} for more details) as long as we restrict ourselves to 
the DGLAP region of small $x$ and thus small $t$, where 
 $t=(p_1-p_2)^2$ is the square of the momentum transfered to the target.
The real part of the amplitude will be evaluated below by applying a
dispersion relation over the center of mass energy $s$.
Using Eq.\ (\ref{prop}) and the following Ward identity which is the same as 
in the abelian case since the box contains no gluons, i.e.\ , is color 
neutral: 
\beq
A^{ab(P)}_{\mu \nu} p_{1\mu } = 0,
A^{ab(P)}_{\mu \nu} p_{2\nu} = 0,
\eeq
yielding in the expression of the imaginary part of the amplitude 
Eq.\ \ref{amp}
\beq
\frac{ImA^{ab(P)}_{\mu \nu} p^{'}_{\mu} p^{'}_{\nu}}{4(p\cdot q)^{2}} = 
\frac{ImA^{ab(P)}_{\mu \nu} p_{t\mu} p_{t\nu}}{x_{1}x_{2}s^{2}},
\eeq
one can rewrite Eq.\ (\ref{amp}) as:
\beq
{Im A\over s}= \int^{1}_{x} \frac{dx_1}{x_1} 
E(\frac{x}{x_1},\frac{\Delta}{x_1},Q^2,p_t^2,Q_0^2)\int 
\frac{s d\alpha d^{2}p_{t}}
{(2\pi )^{4} p_{1}^{2} p_{2}^{2}} p_{t}^{2} \Sigma _{a} 
\frac{4ImA^{a(T)}_{\lambda \sigma} q_{\lambda} q_{\sigma}}{s^2},
\label{iamp}
\eeq
where we have used $<p_{t\, \mu}p_{t\, \nu}>=-\frac{1}{2}g_{\mu\nu}^{t}
p^{2}_{t}$ (the average over the transverse gluon polarization) and defined 
the imaginary part of the hard scattering to be given by:
\beq
E(\frac{x}{x_1},\frac{\Delta}{x_1},Q^2,p_t^2,Q_0^2)= - \frac{1}{2}
g_{\mu\nu}^{t} 
\frac{ImA_{\mu\nu}^{ab(P)}}{x_2 s}\delta_{ab},
\eeq
where the sum over repeated indices is implied. Up to this point, we have just
rewritten the equation for the imaginary part of the total 
amplitude but
have not identified the different perturbative and non-perturbative pieces.
In the case of a virtual photon with longitudinal polarization, this would be 
an easy task since the $q\bar q$ pair would only have a small space-time 
separation and we could follow the argument in Ref.\ \cite
{Brod'94,C.F.S'96,FRS} stating
that the box is entirely dominated by the hard scale $Q$ and thus can 
unambiguously be calculated in pQCD. However, in our case we are dealing with 
a virtual photon which is transversely polarized and thus one can have large
transverse space separations between $q$ and $\bar q$. The resolution
to this problem can be found in the following way: one accepts that 
one has a contribution from a soft, aligned-jet-model-type configuration
and  that there is no unambiguous separation of the amplitude in a 
perturbative and non-perturbative part up to a certain scale $Q_0^2$.
However, in the integration over transverse gluon momenta, one
will reach a scale at which a clear separation into perturbative and 
non-perturbative part can be made and hence we can unambiguously
calculate albeit not the imaginary part of the amplitude of the 
upper box but its $\ln Q^2$ derivative, i.e.\ , its kernel convoluted with a 
parton
distribution. At this point then, one can 
include the non-perturbative contribution of the aligned jet model
into the initial 
distribution of the imaginary part of the total amplitude and solve the 
differential equation in $Q^2$. One obtains the following solution for the 
imaginary part \cite{Abram'95}:
\beq
ImA(x,Q^2,Q_0^2) = ImA(x,Q_0^2)+
\int^{Q^2}_{Q_0^2}{dQ'^{2}\over Q'^{2}} 
\int^{1}_{x} \frac{dx_1}{x_1}P_{qg}(\frac{x}{x_1},\frac{\Delta}{x_1})
g(x_1,x_2,Q'^2),
\label{tamp}
\eeq
where $P_{qg}$ is the evolution kernel\footnote{In Ref.\ \cite{Rad'96,7} a similar 
equation was derived for the complete amplitude for larger $x\simeq 0.1$, 
where the quark distribution dominates and one only needs the $P_{qq}$ kernel.
Of course, at sufficiently small $x$ the contribution of this term is 
numerically small.} and starting from $Q_0^2$ the gluon distribution can be 
defined from 
Feynman diagrams in the leading $\alpha _s \ln Q^2$ approximation by 
realizing that in Eq.\ (\ref{iamp}) one can replace 
$p_{1}^{2}$ and $ p_{2}^{2}$ by $p_{t}^{2}$ and one finds:
\beq
\int \frac{s d\alpha d^{2} p_{t}}{(2\pi )^{4} p_{t}^2} 
\Sigma _{a} 
\frac{4ImA^{a(T)}_{\lambda \sigma} q_{\lambda} q_{\sigma}}{s^2} = 
g(x_{1},x_2,Q^2),
\label{pdis}
\eeq
where $g$ is the nondiagonal parton distribution in general.
Comparison of Eq.\ (\ref{iamp}) with the QCD-improved parton
model expression for the total cross section of charm production given in 
\cite{12} shows that $g$ in the case $\Delta=0$
is the conventional, diagonal gluon distribution.

Note that the parton distribution which
serves as an input in Eq.\ (\ref{tamp}) has to be evolved over the $Q^2$-range
covered by the $Q'^2$ integral which complicates the calculation. We will
explain below how to deal with this issue in practical situations.

At this point we would like to comment on equivalent
definitions of nondiagonal 
parton distributions in the literature which differ by kinematic factors
(see for example \cite{C.F.S'96,Ji'96,Rad'96,7}). 
Eq.\ (\ref{pdis}) corresponds to the definition used in \cite{Rad'96,7}, 
since it is given on the level of Feynman diagrams.
For the  non-perturbative input, $Im A(x,Q_0^2)$
we will be able to use the aligned jet model analysis of Sec.\ \ref{AJM}
and the standard relation between $Im A^{\gamma^*p \to \gamma^*p}(x,Q^2,t=0)$
and $F_{2p}(x,Q^2)$:
\beq
Im A^{\gamma^*p \to \gamma^*p}(x,Q^2,t=0)
=\frac{F_{2p}(x,Q^2)}{4\pi^2\alpha x}.
\eeq

Following the discussion above, we now only need to calculate $P_{qg}$ to 
leading logarithmic accuracy, in order to make predictions for the 
imaginary part of the whole amplitude. Therefore, let us now consider the 
box diagram where the two horizontal quark propagators are cut, corresponding 
to the DGLAP region, i.e.\ , neglecting the u-channel contribution.

\begin{figure}
\centering
\mbox{\epsfig{file=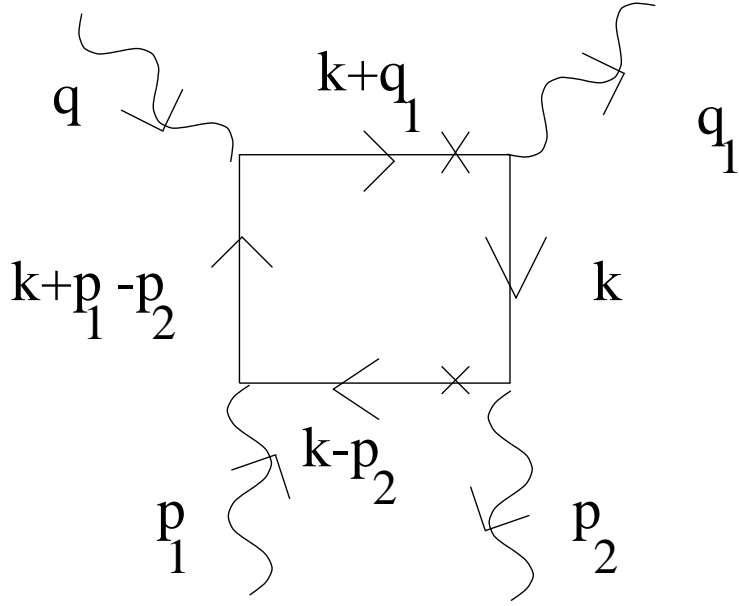,height=8cm}}
\vspace*{5mm}
\caption{Cut box diagram giving the kernel for the imaginary part of the DVCS 
amplitude.}
\label{dem1}
\end{figure}

The kinematics (see Fig.\ \ref{dem1}) for the calculation of the cut box 
diagram, again using Sudakov variables\footnote{Note the slight differences 
between the Sudakov decomposition used here and before e.g. 
$x_1 \rightarrow \beta$ etc.}, is the following. The quark-loop momentum $k$ is
given by:
\beq
k = \alpha q' + \beta p' + p_{t}, \; \; d^{4}k=\frac{\hat s}{2}d\alpha d\beta 
d^{2}k_{t},
\eeq
where $q'$ and $p'$ are light-like momenta related to $p,q$ by:
\bea
& &q = q' - xp', \; \; p_1 = p' + \frac{p^2}{2p'\cdot q'}q', \nonumber\\
& &\hat s = 2p_1q = 2p'\cdot q'- xp_{1}^2.
\eea
The momenta of the exchanged gluons, in light cone coordinates, are given by:
\beq
p_1=(x_1p_+,m^2/Q,p_t),\; \; p_2=(x_2p_+,m^2/Q,p_t).
\eeq
The probing transverse photon and the produced photon have the following 
momenta, again in light cone coordinates:
\beq
q=(-xp_+,\frac{Q^2}{2xp_+},0_t),\; \;
 q_1=(\simeq 0,\frac{Q^2}{2xp_+},0_t).
\eeq    
$P_{qg}$ is calculated in the light cone gauge yielding the following result 
for the most general case\footnote{Note that this expression is defined 
differently from the gluon $\rightarrow$ quark splitting kernel as given in 
e.g. Ref.\ \cite{Rad'96,7} by a factor of $1/x_1$ due to the fact that the additional
$x_1$ already appears in the convolution integral for the $\ln Q^2$ 
derivative.}:
\beq
P_{qg}(\frac{x}{x_1},\frac{\Delta}{x_1}) = 4 \pi^2 \alpha \frac{\alpha_s}{\pi} 
N_F\frac{
x(x-\Delta) + (x_1-x)^2} {x_1(x_1 - \Delta)^2}. 
\label{cross}
\eeq
The DIS kernel is analogous to Eq.\ (\ref{cross}) except that $\Delta = 0$
and the kernel for real photon production is obtained for $\Delta=x$.

We now can proceed to calculate the total imaginary part of the amplitude 
from Eq.\ (\ref{tamp}) where we parameterize the gluon distribution at small 
$x$ as
\beq
g(x_1,x_2,Q^2)= A_0(Q^2) x_1^{A_1(Q^2)}.
\label{gluondis}
\eeq
We neglect the $x_2$ dependence for the moment\footnote{This  effect will be 
taken into account in the actual numerical calculation - see discussion below.}
. The above parameterization is taken from CTEQ3L as well as the 
parameterization of $\alpha_s$ in terms of $Q^2$ in leading order:
\bea
A_0(Q^2)&=&\exp[-0.7631 - 0.7241\ln\left ( \frac{\ln(Q/\Lambda)}{\ln(Q_0/\Lambda)}
\right ) - 1.17\ln^2\left ( \frac{\ln(Q/\Lambda)}{\ln(Q_0/\Lambda)} \right )
\nonumber\\
& &+ 0.534\ln^3\left ( \frac{\ln(Q/\Lambda)}{\ln(Q_0/\Lambda)}\right )]
\nonumber\\
A_1(Q^2) &=& -0.3573 + 0.3469\ln\left ( \frac{\ln(Q/\Lambda)}
{\ln(Q_0/\Lambda)} \right ) - 0.3396\ln^2\left ( \frac{\ln(Q/
\Lambda)}{\ln(Q_0/\Lambda)} \right )\nonumber \\
& &+ 0.09188\ln^3\left ( \frac{\ln(Q/\Lambda)}{\ln(Q_0/\Lambda)} 
\right ),
\eea
with $\Lambda$, $Q_0$ and $\alpha_s$ given by:
\beq
\Lambda = 0.177\, \mbox{GeV},\;\;  Q_0=1.6\,\mbox{GeV},\;\; \alpha_s = 
\frac{4\pi}{9\ln(Q^2/\Lambda^2)},
\eeq
where we have taken $N_C=3$ and $N_F=3$. 

The ratio $R$ of the 
imaginary parts of the amplitudes\footnote{The tensor structure which is the 
same in both cases, namely:
\beq
-g_{\mu\nu} + \frac{p_{\mu}q_{\nu} + q_{\mu}p_{\nu}}{p\cdot q} + 2x\frac{p_{\mu}
p_{\nu}}{p\cdot q},\nonumber
\eeq
             cancels out in the ratio!} is given by:
\beq
R = \frac{ImA(\gamma^* + p \rightarrow \gamma^* + p)}
{ImA(\gamma^* + p \rightarrow \gamma + p)}.
\label{ratio}
\eeq
We give $R$ in the $x$ 
range from $10^{-4}$ to $10^{-2}$ and for a $Q^2$ of $3.5,12$ and $45\; 
\mbox{GeV}^2$ since this kinematic range 
is relevant at HERA. One  might ask, what about the contributions due to 
quarks. The answer is that the corrections are small\footnote{We found them 
to be around $10\%$ in the ratio $R$.}
but for completeness we include 
them here. Eq.\ (\ref{tamp}) is then augmented with a similar 
expression for the quark contribution where the kernel is now that of 
quark-quark splitting and the 
nondiagonal parton distribution is that of the quark:
\bea
ImA(x,Q^2,Q_0^2)&=& ImA(x,Q_0^2) + \int_{Q_0^2}^{Q^2}
\frac{dQ'^2}{Q'^2}
\int^{1}_{x}\frac{dx_1}{x_1}[P_{qg}(\frac{x}{x_1},\frac{\Delta}{x_1})
g(x_1,x_2,Q'^2)\nonumber\\
& & + P_{qq}(\frac{x}{x_1},\frac{\Delta}{x_1})q(x_1,x_2,Q'^2)],
\label{conv}
\eea 
where the general expression for the kernel, after a similar calculation as 
before, is found to be:
\beq
P_{qq}(\frac{x}{x_1},\frac{\Delta}{x_1})=4\pi^2\alpha\frac{\alpha_s}{\pi}C_F
\left[ \frac{\frac{x}{x_1} - \frac{x^3}{x_1^3} - \frac{\Delta}{x_1}
\left(\frac{x}{x_1} + \frac{x^2}{x_1^2})
\right )}{x_1(1-\frac{\Delta}{x_1})(1-\frac{x}{x_1})_+}\right ],
\eeq
and the + - prescription is the one used in Ref.\ \cite{ours}. The quark 
distribution itself is also taken from CTEQ3L\footnote{We used the u-quark 
parametrization for all light quarks for simplicity, which is surely 
unproblematic at small $x$.} and given by:
\beq
q(x_1,x_2,Q^2)=A_0(Q^2)x_1^{A_1},
\label{qdis}
\eeq
with
\bea
A_0(Q^2)&=&\exp[0.1907 + 0.04205\ln\left ( \frac{\ln(Q/\Lambda)}{\ln(Q_0/\Lambda)} 
\right ) + 0.2752\ln^2\left ( \frac{\ln(Q/\Lambda)}{\ln(Q_0/\Lambda)}\right )
\nonumber\\
& &-0.3171\ln^3\left ( \frac{\ln(Q/\Lambda)}{\ln(Q_0/\Lambda)}\right )]
\nonumber\\
A_1&=&0.465.
\eea
We chose $A_1$ to be constant since it varies only between $0.4611$ and $0.468$
in the $Q^2$ range of interest, i.e.\ , the error we make is almost negligible
since the quark distribution themselves are small in the $x$-range 
considered. Furthermore, according to our discussion in Sec.\ \ref{AJM}, we 
chose the initial distribution for the imaginary part of the DVCS amplitude 
to be twice that of the initial distribution for the imaginary part of the DIS
amplitude. In the evolved QCD part, the nonforward kinematics are taken into 
account in the  kernels of the QCD evolution equation, also the different 
$Q^2$ evolution of nondiagonal as compared to diagonal distribution has been 
taken into account as explained below.

As the calculation with MATHEMATICA showed, the amplitude of the production
of real photons is larger than the DIS amplitude over the whole 
range of small $x$ and $R$ turns out to be $0.551$, $0.573$ 
and $0.57$ for $x=10^{-4}$, $0.541$, $0.562$ and $0.557$ for $x=10^{-3}$ and 
$0.518$, $0.519$ and $0.505$ for $x=10^{-2}$ in the given $Q^2$ range. 
It has to be pointed out that for a given $Q^2$, the ratio is basically 
constant. Of course, the ratio $R$ will approach $1/2$ as $Q^2$ is 
decreased to the nonperturbative scale since this is 
our aligned jet model estimate.
The reason for the deviation from $R=1/2$ is due to the difference in the
evolution kernels.

It is worth noting that in the kinematics we discuss, 
the ratio is still rather sensitive to the nonperturbative boundary
condition. For example, assuming the same boundary
conditions for DVCS and DIS, would result in  a reduction  of $R$ of about 
$20(10)\%$ at $Q^2 \sim 12 (40)~ \mbox{GeV}^2$ and $x \sim 10^{-3}$

In Eq. (\ref{conv}) the median point of the integral, as found by a MATHEMATICA
program, corresponds to
$x_1/2 \sim x_2 \approx x$. This is due to the mass of the $q\bar q$ pair in 
the quark loop being $\propto Q^2$. For such a $x_1/x_2$ the ratio of
nondiagonal to diagonal gluon density depends weakly
on $x_2$. Therefore, with an accuracy of a few percent, we can approximate
this ratio by its value  at $x_1/x_2=2$. Therefore, in the calculation of $R$,
 we used Eqs.\ (\ref{gluondis}) and (\ref{qdis}) for both the 
diagonal and nondiagonal case but then multiplied the real photon result 
of the amplitude by a function $f(x,Q^2)$ for each $x$ and $Q^2$ to take 
into account the different evolution of the nondiagonal distribution as 
compared to the diagonal one,
\beq
ImA(x,Q^2,Q_0^2) = ImA(x,Q_0^2)+
\int^{Q^2}_{Q_0^2}\frac{dQ'^2f(x,Q'^2)}{Q'^{2}} 
\int^{1}_{x} \frac{dx_1}{x_1}P_{qg}(\frac{x}{x_1},\frac{\Delta}{x_1})
g(x_1,x_2,Q'^2).
\eeq
The function was determined by 
using our modified version of the CTEQ-package and, starting from the same 
initial distribution, evolving the diagonal and nondiagonal distribution
to a certain $Q^2$ and comparing the two distributions 
at the value $x_2=x_1/2=x$ for different $x$ and then interpolating for the
different ratios of the distribution in $Q^2$ for given $x$. 
For this median point the difference between the diagonal and nondiagonal 
gluon distribution is between $8\,- \, 25\%$ depending on the $x$ and $Q^2$ 
involved and $10\,-\,55\%$ for the quarks (see the figures in 
Ref.\ \cite{ours,AF97} for more details). 
  
As far as the complete amplitude at small $x$ is concerned, we can reconstruct
the real part via dispersion relations \cite{reim1,reim2}, which to a very good
approximation gives:
\begin{equation}
\eta \equiv {Re A \over Im A}= \frac{\pi}{2}\frac{d\,\ln(x Im A)}{d\, 
\ln{1\over  x}}.
\label{reim}
\end{equation}
Meaning that  since $Im A$ can be fitted as $x^{-1-\delta}$, 
$\eta \approx {\pi \over 2}\delta$ 
is independent of $x$ to a good precision.
Therefore, our claims for the imaginary part of the 
amplitude also hold for the whole amplitude at small $x$.
This is due to the fact that within the dispersion representation of the 
amplitude over $x$, the contribution of the subtraction constant becomes 
negligible at sufficiently small $x$.
  
One also has to note that there is a potential pitfall since the
QED bremsstrahlung - the  Bethe-Heitler process - , where the electron
interacts with a proton via a soft Coulomb photon exchange and the 
real photon is radiated off the electron, can be a considerable background. As
was shown by Ji \cite{Ji'96}, the Bethe-Heitler process will give a strong 
background at small $t$ and medium $Q^2$ and $x \ge 0.1$. We will discuss this
subject in more detail later on.

\section{The {\it \lowercase {t}}-slope of the $\gamma^*N\to \gamma N$ cross 
section}
\label{Bslope}
The slope of the differential cross section of the virtual Compton scattering
${d \sigma^{\gamma^*N\to \gamma N}\over dt} \propto \exp(Bt)$ is determined by 
three effects: (i) the average transverse size of the $q\bar q$ component of 
the $\gamma^*$ and $\gamma$ wave functions involved in the transition,
(ii) the pomeron-nucleon form factor at the nucleon vertex, and (iii)
Gribov diffusion in the soft part of the ladder.
This leads to several qualitative phenomena.
At the normalization point, $q\bar q $
configurations of an average transverse
size, comparable to that of the $\rho$-meson,
give the dominant contribution to the scattering amplitude,
 leading to a slope similar to that of the  processes 
$\gamma + p \to \rho, \omega + p$. 
The contribution of the higher mass $q \bar q$ components 
is known to result in an enhancement of the differential cross
section of Compton scattering at $t=0$ by a factor $\approx 2$
as compared to the prediction of the vector meson dominance model with 
$\rho, \omega, \phi, J/\psi$ intermediate states, see e.g. \cite{FNAL}.
 Since the diffraction of a photon to masses $M_X \ge 1.3 GeV$  has 
a smaller $t$ slope than for transitions
to $\rho$ and $\omega$, one could expect that
the high mass  contribution would lead to  
a t-slope of the Compton cross section  being
somewhat smaller than for the production of
$\rho, \omega$-mesons.  However direct experimental
comparison \cite{FNAL} of the slopes of
the  Compton scattering and the $\omega$-meson
photo-production at $\left<E_{inc}^{\gamma}\right> \approx 100 GeV$
finds these slopes to be the same within the  experimental errors.
Using these data, we can estimate the slope
of the amplitude for diffractive photon production in DIS
at HERA energies but at moderate Q-i.e. in the normalization point as
\beq 
B(s,Q_0^2) = B_{Comp.Scatt.}(s_0)+ 
 2 \alpha'\ln({s\over s_0}),
\label{Brho}
\eeq
where $\alpha' =0.25 GeV^{-2}$, $s_0=200 GeV^2$, and 
$B_{Comp.Scatt.}(s_0)=6.9 \pm 0.3 GeV^{-2}$ \footnote{Note that the data 
\cite{FNAL} can be equally well described by the fit 
$d \sigma/dt \propto \exp(Bt)$ with $B=6.9 \pm 0.3 GeV^{-2}$ and by the 
$d \sigma/dt \propto \exp(8.9t+2.2t^2)$ fit.}.   
Hence for HERA energies $B(W=200 GeV,Q_0^2) \sim 10 GeV^{-2}$.

In another limit of large $Q^2$ and large enough $x$, say 
$x \sim 10^{-2}$, the dominant $q\bar q $ configurations
have a small transverse size and the upper vertex
does not contribute to the slope. Furthermore,
the perturbative contribution occupies most of the rapidity
interval and leaves no phase space for the soft Gribov diffusion. In
this case, the slope is given by the square of the two-gluon form 
factor of the nucleon which corresponds to $B=B_{ggN} \approx 
4 \div 5 GeV^{-2}$ \cite{Brod'94}. 

An interesting situation emerges in the limit of large but fixed $Q^2$
when the energy starts to increase. In this case, the perturbative part of the 
ladder has the length  
$\sim \ln ({Q^2\over m_\rho^2 \kappa})$.
Here $\kappa=x/x_0$, where
$x_0$ is the fraction $x$ of the parent parton at a soft scale. For HERA 
kinematics $\kappa \sim 0.1-0.3$ for $Q^2 \sim 10-20 GeV^2$ and decreasing 
with increasing $s$.
Thus at high $Q^2$ one has an approximate factorization
for diffraction in the case of high masses ($M^2 \ge 100 GeV^2$, 
$M^2 \gg Q^2$) in the scattering of real and virtual photons observed at 
HERA \cite{H1}, namely 
\beq
{1 \over \sigma_{tot}(\gamma N)}{d \sigma(\gamma N\to XN)(W,M_X) 
\over dt dM^2}
\approx 
{1 \over \sigma_{tot}(\gamma^* N)}{d \sigma(\gamma^* N\to XN)(W,M_X) 
\over dt dM^2}.
\label{diffus}
\eeq
The observed slope for these processes is $B \sim 7 GeV^{-2}$ which is
consistent with the presence of a cone shrinkage at the rate
$\sim 2 \alpha'\ln(W^2/M^2)$
  as compared to the data at lower energies where
 smaller values of $W^2/M^2$ were probed.
Similarly we can expect that for virtual Compton scattering at large $Q^2$,
the slope will increase with decrease of $x$, at very small $x$,
 approximately as
\beq
B(W^2,Q^2)_{Q^2\gg \mu^2}=B_{ggN} + 2 \alpha'
(\ln(W^2\kappa/Q^2)-\ln (W_0^2/m_{\rho}^2))
\theta(W^2\kappa/Q^2-W_0^2/m_{\rho}^2),
\eeq
where $W_0^2=200 GeV^2$.  We take into account here that $B_{ggN}$
was determined experimentally from the processes at $W^2 \sim W_o^2$. 

\section{The rate of exclusive photon production at HERA}
\label{sec:slope}

To check the feasibility of measuring a DVCS signal against the DIS 
background, we will be interested in the fractional number of DIS events
to diffractive exclusive photoproduction events at HERA in DIS which will tell
us whether it will be statistically feasible to search for DVCS events among
DIS events. We define this fractional number of events as:
\beq
R_{\gamma} = \frac{\sigma(\gamma^* +p \rightarrow \gamma + p)}
{\sigma_{tot}(\gamma^*p)}\simeq
\frac{d\sigma (\gamma^* + p \rightarrow \gamma +p)}{dt}|_{t=0}\times 
\frac{1}{B}/\sigma_{tot}(\gamma^*p)
\label{events}
\eeq
Using
\beq
\frac{d\sigma}{dt} (\gamma^* + p \rightarrow \gamma + p)= 
\frac{\sigma_{tot}^2(\gamma^*p)}{16\pi R^2}(1+\eta^2)
e^{Bt},
\label{N}
\eeq
which can be derived from applying the optical theorem and using $R$, the 
ratio of the imaginary parts of the amplitudes
given by Eq.\ (\ref{ratio}), $\eta= Re\,A/Im\,A$ as given by 
Eq.\ (\ref{reim}) and where $t=-\frac{m_N^2 x^2}
{1-x}-p_t^2\simeq -p_t^2$ with $t_{min}=-\frac{m_N^2 x^2}{1-x}\simeq 0$, one 
can now rewrite Eq.\ (\ref{events}). A complete expression for DVCS will be 
given in the next section. Note that only 
$\frac{d\sigma (\gamma^* + p \rightarrow \gamma +p)}{dt}|_{t=0}$
is calculable in QCD. The $t$ dependence is taken from data fits to hard 
diffractive processes.

Using the fact that $F_2(x,Q^2)\simeq \frac{\sigma_{tot}(\gamma^*p)Q^2}
{4 \pi^2 \alpha}$ one can rewrite Eq.\ (\ref{events}) into its final form:
\beq
R_{\gamma} \simeq \frac{\pi \alpha}{4 R^2 Q^2 B}F_2(x,Q^2)(1 + \eta^2).
\label{fN}
\eeq
where $\eta^2 \simeq 0.09-0.27$ for the given 
$Q^2$ range. We computed $R_{\gamma}$, the fractional number of events given 
by Eq.\ (\ref{fN}), for $x$ between
$10^{-4}$ and $10^{-2}$ and for a $Q^2$ of $2, 3.5, 12$ and $45\,\mbox{GeV}^2$ 
 where the numbers for $F_2$ were taken from
\cite{15}. Based on our analysis of the previous section we use
Eq.\ (\ref{Brho}) for $Q^2=2 GeV^{2}$, assuming that for $Q^2=3.5 GeV^{-2}$
the slope drops by about 1 $\div $ 2 units as compared to Eq.\ (\ref{Brho})
to account for the decrease of the transverse size of the $q\bar q$-pair;
for larger $Q^2$ we use Eq.\ (\ref{diffus}).

We find $R_{\gamma}\simeq 1.1\times 10^{-3},\, 9.9\times 10^{-4}$ 
 at $x=10^{-4},\, 10^{-3}$ and $Q^2=2\mbox{GeV}^2$; $R_{\gamma}
\simeq 1.07\times 10^{-3},\, 9.3\times 10^{-4}$ 
  at $x=10^{-4},\, 10^{-3}$ and  $Q^2=3.5\mbox{GeV}^2$; 
$R_{\gamma}\simeq 4.5\times 10^{-4},\, 3.78\times 10^{-4}\, 
2.5\times 10^{-4}$ 
at $x=10^{-4},\, 10^{-3},\, 10^{-2}$ and $Q^2=12\mbox{GeV}^2$; and 
finally $R_{\gamma} \simeq 1.49\times 10^{-4},\, 1.04\times 10^{-4}$
at $x=10^{-3},\, 10^{-2}$ and $Q^2=45\mbox{GeV}^2$. 
As is to be expected, the number of events rises at small $x$ since the 
differential cross section is proportional to the square of the 
gluon distribution and the total cross section is just proportional to the 
gluon distribution, i.e.\ , the ratio in Eq.\ (\ref{events}) is expected to be 
proportional to the gluon distribution and this assumption is born out by
our calculation and falls with increasing $Q^2$ since $F_2$ does not grow as 
fast with energy.

\section{The complete cross section of exclusive photon production}
\label{sec:cross}

In order to study whether the Bethe-Heitler or DVCS process will be dominant 
in real photon production we need the expressions for the differential cross 
sections first.

We find that the differential cross section for DVCS can be simply expressed 
through the DIS differential cross section by multiplying the
DIS differential cross section by $R_{\gamma}$ (see Eq.\ (\ref{fN}))
which was calculated in the previous section.
One can see this by observing how $F_2$ is related to 
$\sigma_{tot}(\gamma^*p)$ as given in Sec.\ \ref{sec:slope} and 
$\sigma_{tot}(\gamma^*p)$ to $\sigma_{DVCS}$ via $R_{\gamma}$ in the same 
section. We then find using Eq.\ (\ref{fN}) for $R_{\gamma}$
\beq
\frac{d\sigma_{DVCS}}{dxdyd|t|d\phi_r}=\frac{\pi\alpha^3s}{4R^2Q^6}(1+(1-y)^2)
e^{-B|t|}F^2_2(x,Q^2)(1+\eta^2)
\label{dvcsc}
\eeq
with $\sigma_{DVCS} = \frac{d\sigma_{DVCS}}{dt}|_{t=0}
\times \frac{1}{B}$ using the same exponential $t$ dependence as in the 
previous section and $R$ being the ratio of the imaginary parts of the DIS to 
DVCS amplitudes as computed earlier.

\begin{figure}
\centering
\mbox{\epsfig{file=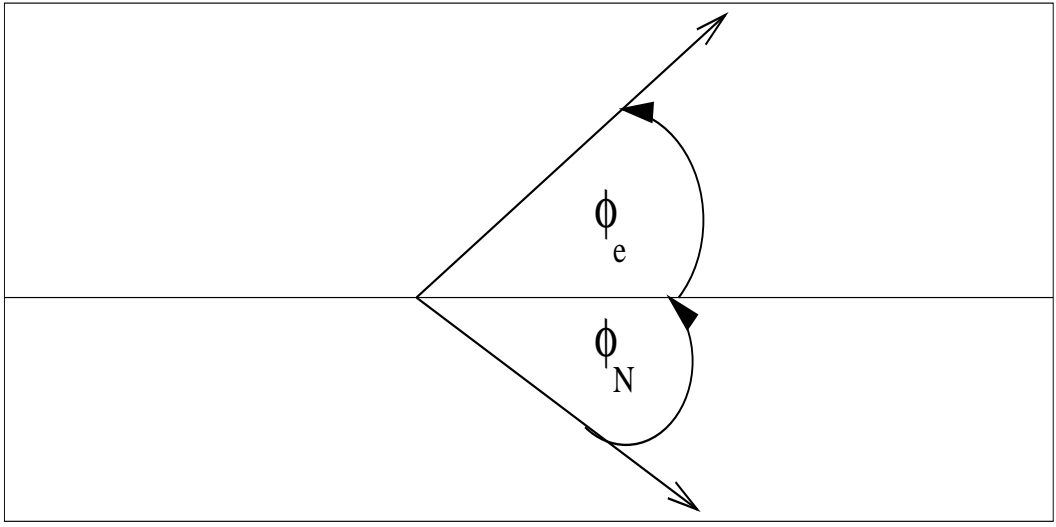,width = 14cm, height=7cm}}
\vspace*{5mm}
\caption{The azimuthal final proton and electron angle in the transverse 
scattering plane.}
\label{angle}
\end{figure}

In writing Eq.\ (\ref{dvcsc}) we neglected 
$F_L(x,Q^2)$ - the experimentally observed conservation of s channel 
helicities justifies this approximation- so that $F_2\simeq 2xF_1$.
$y=1-E'/E$ where $E'$ is the energy of the 
electron in the final state and $\phi_r=\phi_N + \phi_e$, 
where $\phi_N$ is the azimuthal angle between the plane defined by 
$\gamma^*$ and the final state proton and the $x-z$ plane and $\phi_e$
is the azimuthal angle between the plane defined by the initial and final state
electron and the $x-z$ plane (see Fig.\ \ref{angle}). Thus $\phi_r$ is nothing but
the angle between the $\gamma^*-p'$ and the electrons scattering plane.
 
In the case of the Bethe-Heitler process, we find the differential cross section at
small $t$ to be
\beq
\frac{d\sigma_{BH}}{dxdyd|t|d\phi_r} = \frac{\alpha^3 s y^2(1+(1-y)^2)}{\pi Q^4
|t| (1-y)} \left [ \frac{G_E^2(t) + \tau G_M^2(t)}{1+\tau} \right ]
\label{difcros}
\eeq
with $\tau = |t|/4m_N^2$ and $s$ being the invariant energy.
$G_E(t)$ and 
$G_M(t)$ are the electric and nucleon form factors and we describe them using 
the dipole fit
\beq
G_E(t)\simeq G_D(t)=(1+\frac{|t|}{0.71})^{-2}~~\mbox{and}~~G_M(t)=\mu_p G_D(t),
\label{nform}
\eeq
where  $\mu_p=2.7$ is the proton magnetic moment.
We make the standard assumption that the spin flip term is small in the 
strong amplitude for small $t$.

In order to write down the complete total cross section of exclusive photon 
production we need the interference term between DVCS and Bethe-Heitler. Note
that in the case of the interference term one does not have a spinflip in the 
Bethe-Heitler amplitude, i.e.\ ,  one only has $F_1(t)$, as compared to 
Eq.\ (\ref{difcros}) containing a spinflip part, i.e.\ , $F_2(t)$. The 
appropriate combination of $G_E(t)$ and $G_M(t)$ which yields $F_1(t)$ is  
\beq
\left[\frac{G_E(t) + \frac{|t|}{4m_N^2}G_M(t)}{1+\frac{|t|}{4m_N^2}}\right ].
\eeq

We then find for the interference term of the differential cross section, 
where the + sign corresponds to electron scattering of a proton and the - sign
corresponds to the positron 
\bea
\frac{d\sigma_{DVCS+BH}^{int}}{dxdyd|t|d\phi_r} &=& \pm \frac{\eta 
\alpha^3 s y(1+(1-y)^2) cos(\phi_r) e^{-B|t|/2} F_2(x,Q^2)}{2 Q^5 \sqrt{|t|}
\sqrt{1-y} R}\nonumber\\
& & \times \left [ \frac{G_E(t) + \tau G_M(t)}{1+\tau} \right ].
\label{inter}
\eea
The total cross section is then just the sum of 
Eq.\ (\ref{dvcsc}),(\ref{difcros}),(\ref{inter}).

\subsection{{\it \lowercase {t}}-dependence of Bethe-Heitler as compared
to DVCS for different $Q^2$}

\begin{figure}
\vskip-1in
\centering
\epsfig{file=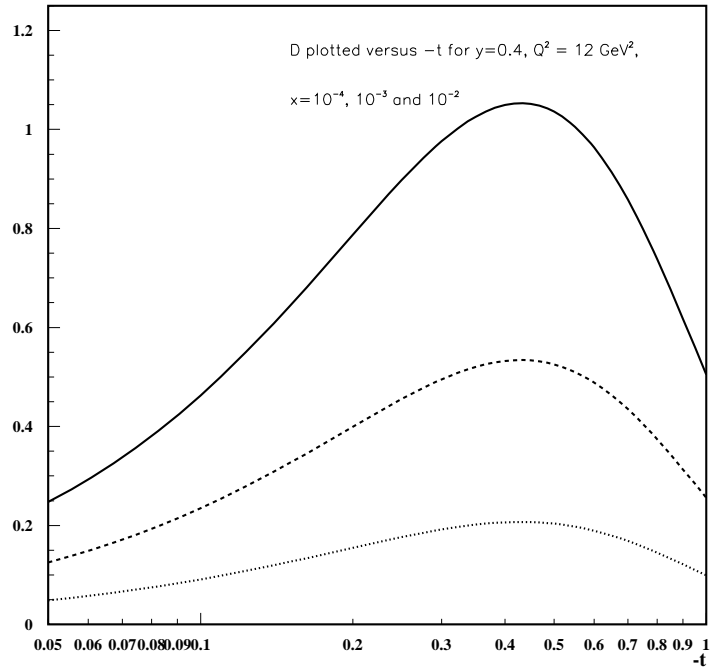,height=12cm}
\vskip-0.9in
\epsfig{file=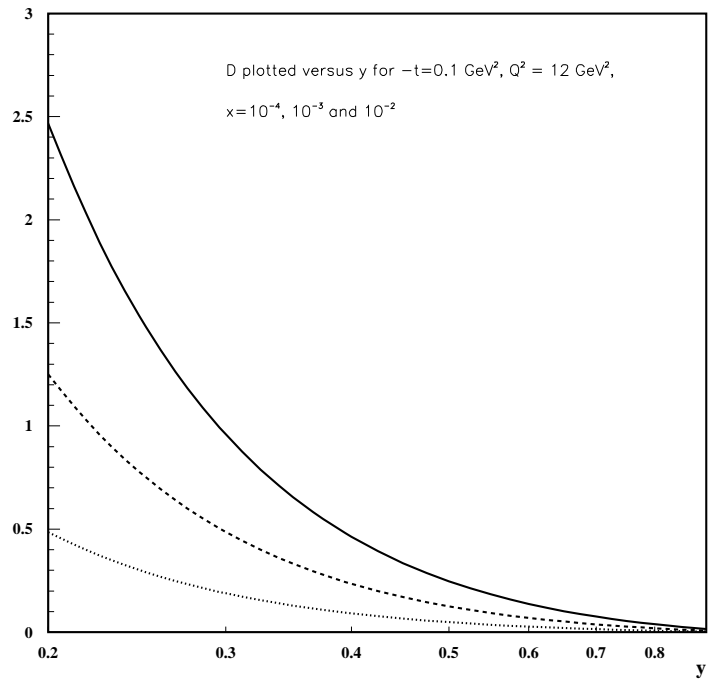,height=12cm}
\vskip-0.5in
\vspace*{5mm}
\caption{a) $D$ is plotted versus $-t$ for $x=10^{-4},10^{-3},10^{-2}$, 
$Q^2=12~\mbox{GeV}^2$, $B=5~\mbox{GeV}^{-2}$ and $y=0.4$. 
The solid curve is for $x=10^{-4}$, the dotted one for $x=10^{-2}$ and the 
dashed one for $x=10^{-3}$. b) $D$ is plotted versus $y$ for the same $x,Q^2,B$
and $-t=0.1~\mbox{GeV}^{2}$}
\label{tdep1}
\end{figure}

At this point it is important to determine how large the Bethe-Heitler 
background is as compared to DVCS for HERA kinematics, hence, in the following 
discussion, we will estimate the ratio $D$ allowing a background comparison:
\beq
D=\frac{<d\sigma_{DVCS+BH}/dxdydt>}{<d\sigma_{BH}/dxdydt>}-1.
\eeq
with $<...>=\int_{0}^{2\pi}d\phi_r$. Using the expressions from 
Sec.\ \ref{sec:cross} we compute $D$ and find that $D > 1$ 
( See Fig.\ \ref{tdep1}a, \ref{tdep2}a) for relatively small $y$ and 
$0.1\leq t \leq 0.6$ with the given values of $x$ and $Q^2$ considered.
Note, however, that this does not mean that the case for DVCS is hopeless. 
As it turns out, it is rather advantageous to have $D<1$ when looking at the 
interference term which we will do next.

\begin{figure}
\vskip-1in
\centering
\epsfig{file=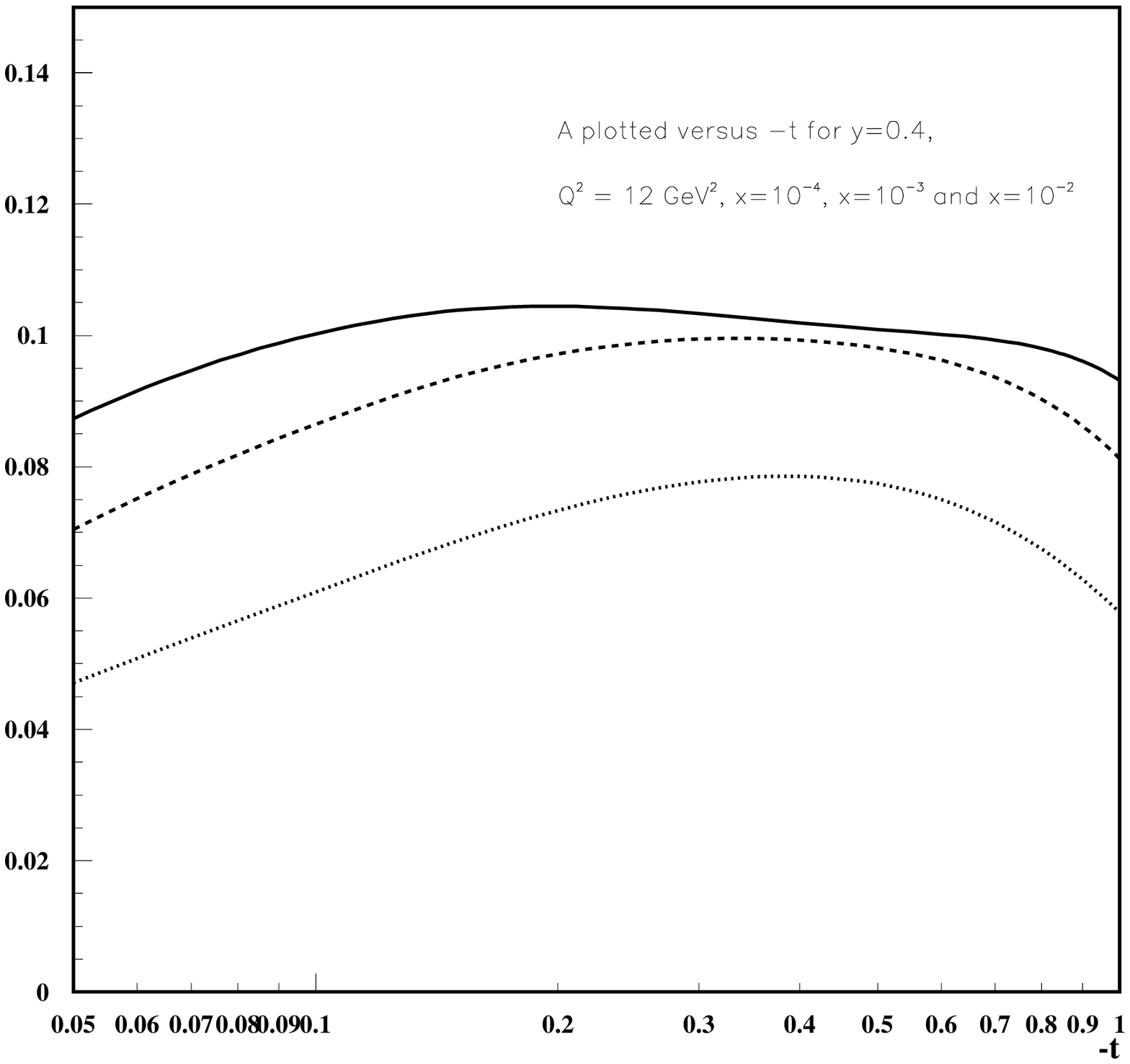,height=12cm}
\vskip-0.9in
\epsfig{file=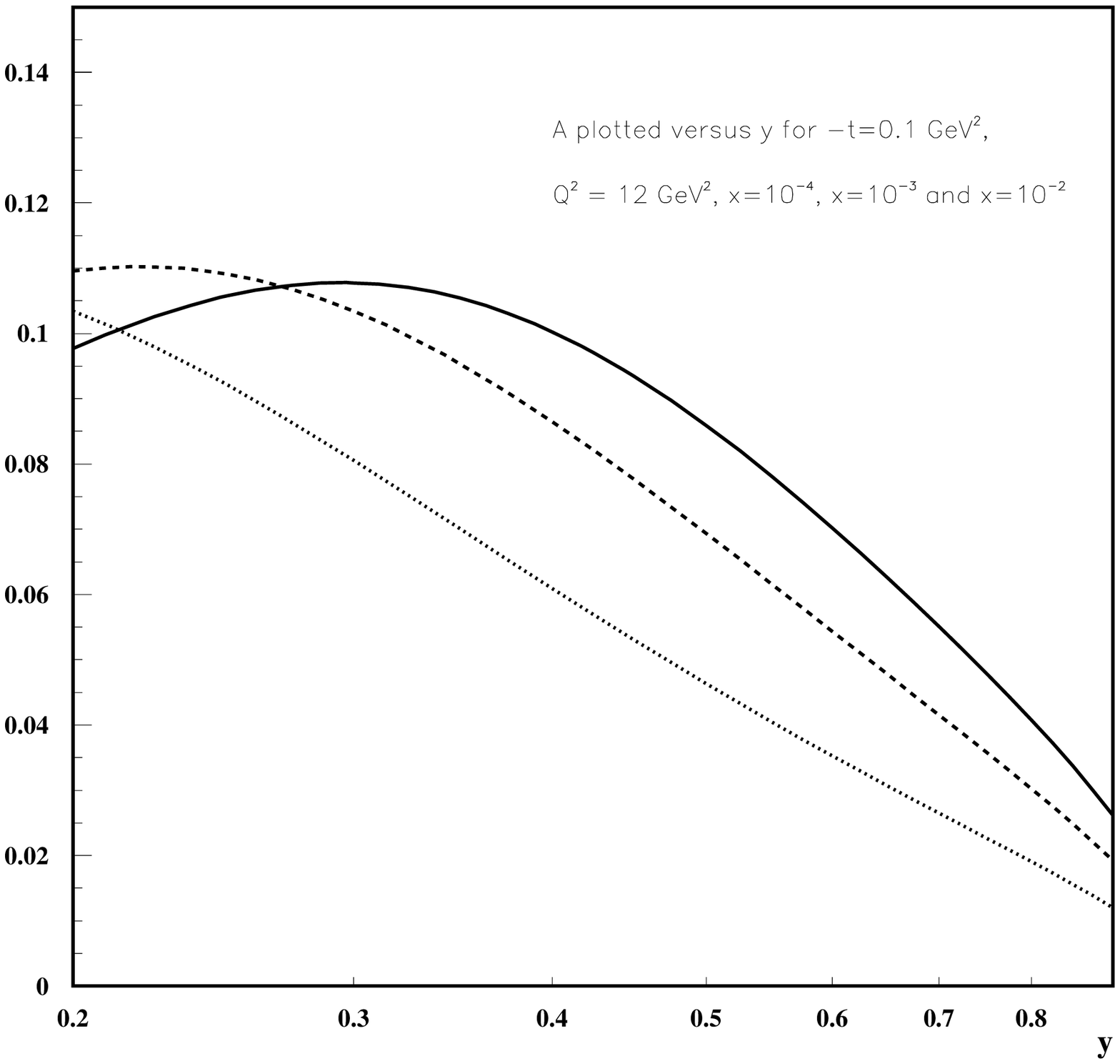,height=12cm}
\vspace*{5mm}
\vskip-0.5in
\caption{a) The asymmetry $A$ is plotted versus $-t$ for 
$x=10^{-4}$ (solid curve), 
$x=10^{-2}$ (dotted curve) and $x=10^{-3}$ (dashed curve)
again for $Q^2=12~\mbox{GeV}^2$, $B=5~\mbox{GeV}^{-2}$ and $y=0.4$.
b) $A$ is plotted versus $y$ for the same $x,Q^2,B$ and 
$-t=0.1~\mbox{GeV}^{2}$.}
\label{tdep4}
\end{figure}

\begin{figure}
\vskip-1in
\centering
\epsfig{file=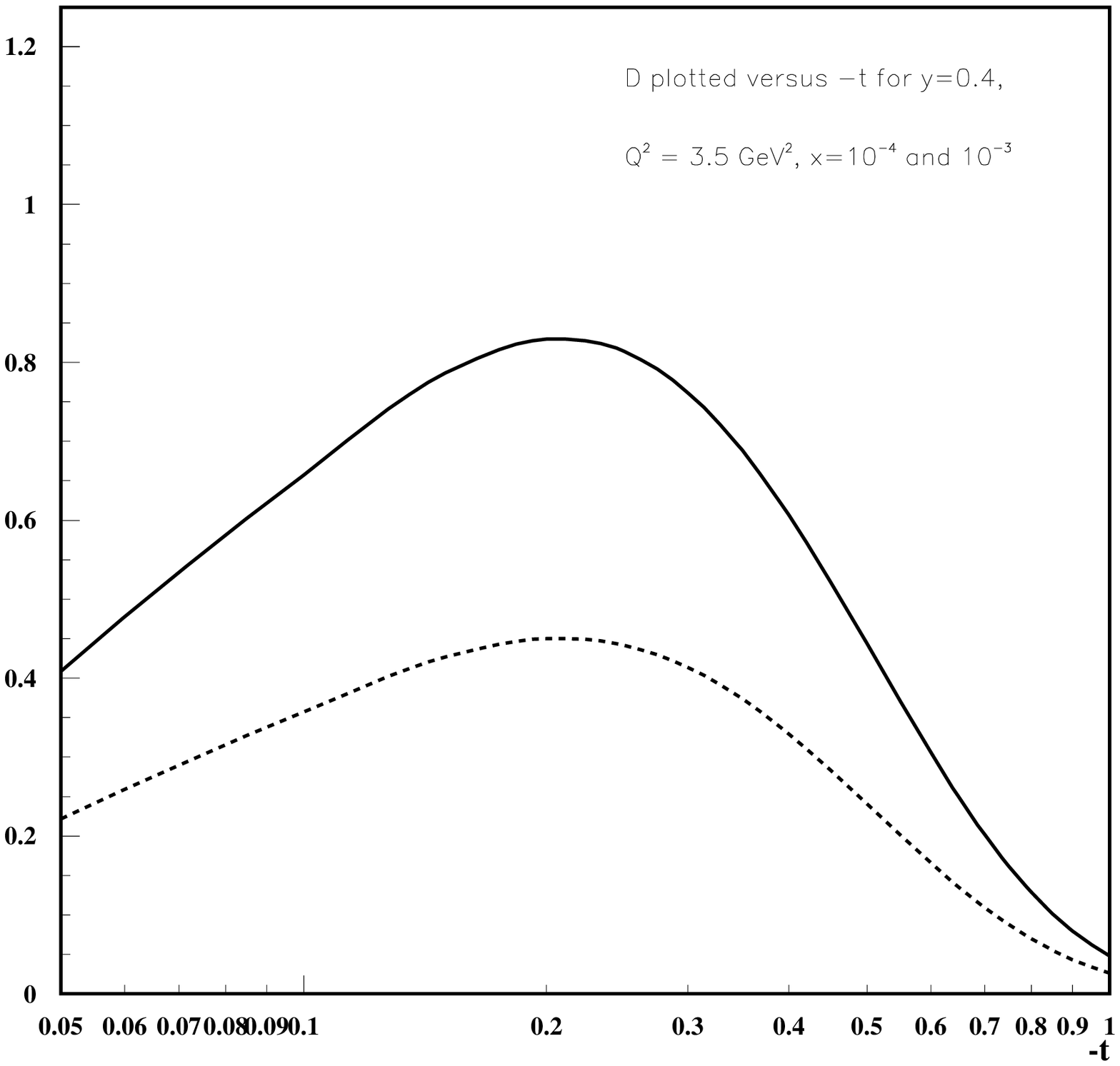,height=12cm}
\vskip-0.9in
\epsfig{file=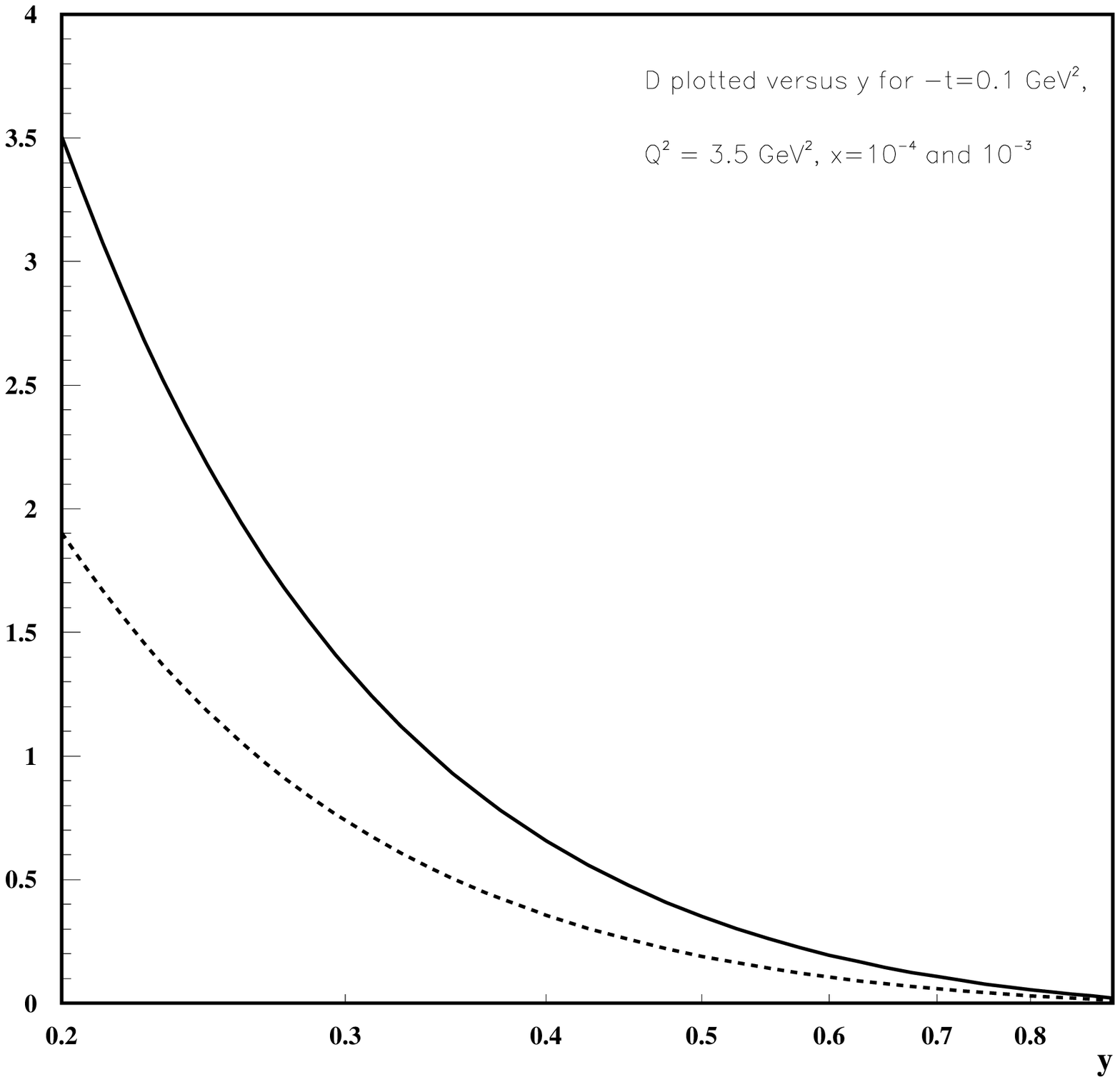,height=12cm}
\vskip-0.5in
\vspace*{5mm}
\caption{a) $D$ is plotted versus $-t$ for $x=10^{-4}$ and $10^{-3}$, 
$Q^2=3.5~\mbox{GeV}^2$, $B=8~\mbox{GeV}^{-2}$ and $y=0.4$. 
The solid curve is for $x=10^{-4}$, the dashed one for $x=10^{-3}$. 
b) $D$ is plotted versus $y$ for the same $x,Q^2,B$
and $-t=0.1~\mbox{GeV}^{2}$}
\label{tdep2}
\end{figure}

\begin{figure}
\vskip-1in
\centering
\epsfig{file=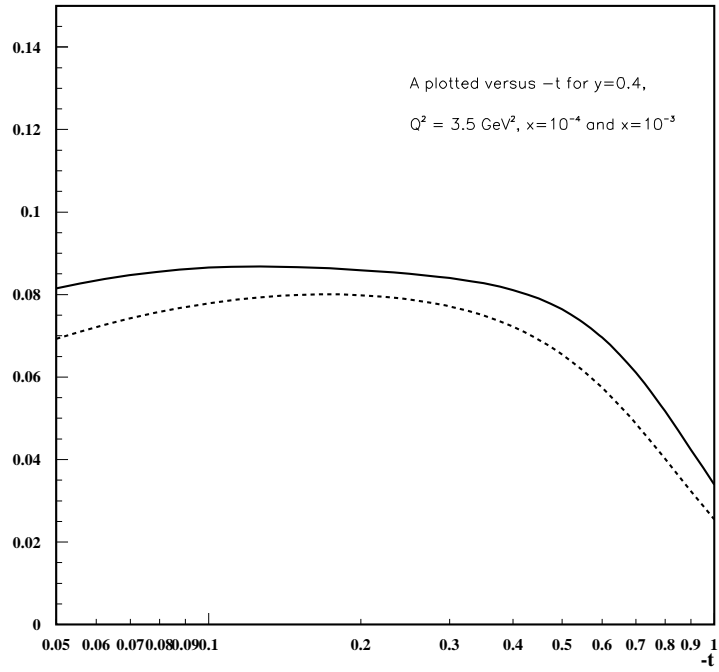,height=12cm}
\vskip-0.9in
\epsfig{file=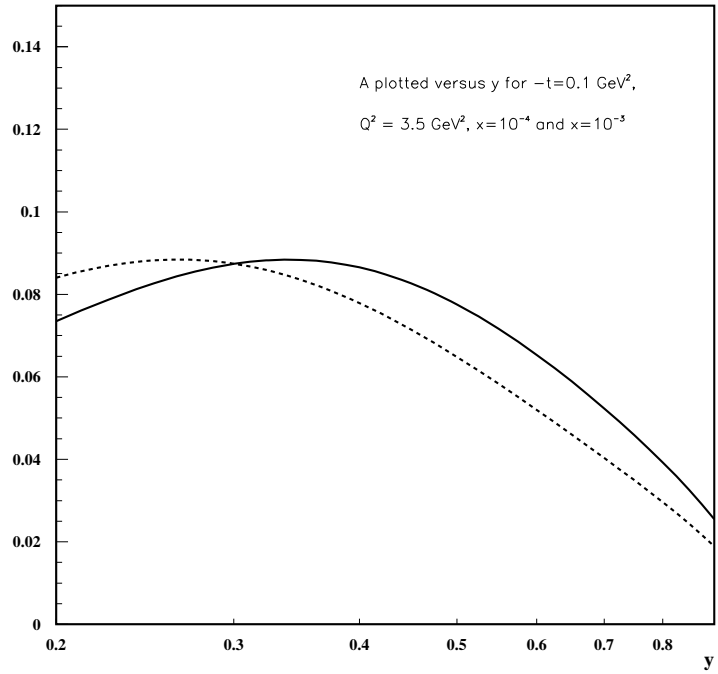,height=12cm}
\vskip-0.5in
\vspace*{5mm}
\caption{a) The asymmetry $A$ is plotted versus $-t$ for 
$x=10^{-4}$ (solid curve), 
$x=10^{-3}$ (dashed curve) again for $Q^2=3.5~\mbox{GeV}^2$, 
$B=8~\mbox{GeV}^{-2}$ and $y=0.4$.
b) $A$ is plotted versus $y$ for the same $x,Q^2,B$ and 
$-t=0.1~\mbox{GeV}^{2}$.}
\label{tdep3}
\end{figure}

It is convenient to illustrate the magnitude of the intereference term in 
the total cross section by considering the asymmetry for proton and either 
electron or positron to be in the same and opposite hemispheres ( we omit the 
rather cumbersome explicit expression but the reader can easily deduce it 
from Eq.\ (\ref{dvcsc}),(\ref{difcros}),(\ref{inter}). )
\beq
A =\frac{\int_{-\pi/2}^{\pi/2}d\phi_r d\sigma_{DVCS+BH} - \int^{3\pi/2}_{\pi/2}
d\phi_r d\sigma_{DVCS+BH}}{\int_{0}^{2\pi}d\phi_r d\sigma_{DVCS+BH}}
\label{assym1}
\eeq
in other words, one is counting the number of events in the upper hemisphere 
of the detector minus the number of events in the lower half, normalized to 
the total cross section.
Fig.\ \ref{tdep4}a,b and \ref{tdep3}a,b show $A$ for the same kinematics as 
above and we find that the 
asymmetry is fairly sizeable already for small $t$ and is strongly dependent
on the energy. Due to this fairly large asymmetry, one has a first chance to 
access nondiagonal parton distributions. 
We will discuss $A$ in more detail, in particular its energy dependence, in 
the next chapter.
 
Note, there is an increased experimental difficulty to measure DVCS if the 
recoil proton is not detected in other words if $t$ is not directly measured.
However there is a simple, practical way around this problem which we will 
discuss next.

\subsection{DVCS alternative to tagged proton in the final state}

Another interesting process, which can be studied in the context of DVCS,
 is the one where the nucleon dissociates into mass ``X''
 - $\gamma^* +p \to \gamma +X$. Perturbative QCD is applicable in this case 
as well. In particular the following factorization relation 
should be valid at sufficiently large $Q^2$:
\beq
\frac{\frac{d\sigma}{dt}(\gamma^* + p \rightarrow \gamma + X)}{\frac{d\sigma}
{dt}(\gamma^* + p \rightarrow \gamma + p)}\simeq
\frac{\frac{d\sigma}{dt}(\gamma^* + p \rightarrow J/\psi + X)}{\frac{d\sigma}
{dt}(\gamma^* + p \rightarrow J/\psi + p)}.
\label{ratio1}
\eeq
The big advantage of the dissociation process as compared to the process where
the target proton stays intact is that the Bethe-Heitler process is strongly 
suppressed for inelastic diffraction at small t due to the conservation of the 
electro-magnetic current, hence the amplitude is multiplied by an additional 
factor $\sqrt{\left|t\right|}$ which is basically $0$ for the Bethe-Heitler 
process. Thus, 
 the masking of the strong amplitude of photoproduction is small in this case.
Since there is already data available on $J/\psi$ production, this quantity 
can give us information on how different the slopes for the production of 
massless to massive vector particles are, providing us with more understanding
on how different or similar the exact production mechanisms are. Note that 
the ratio 
of the total dissociative to elastic cross section of $\rho$ meson production 
is found to be about $0.65$ at large $Q^2$ \cite{14} which is basically of 
$O(1)$. The same 
should hold true for $J/\psi$ production and in fact this ratio should be a 
universal quantity. This is due to the fact that one has complete 
factorization, hence the hard part plus vector meson is essentially a point 
and thus for the soft part, is does not matter what kind of vector
particle is produced.
The above said implies for Eq.\ (\ref{ratio1}) that it also should be of order 
unity, implying that the order of magnitude of the fractional number of events
for real photon production to DIS remains unchanged even though the actual 
number of $R_{\gamma}$ might decrease by as much as $35\%$. 
          
\section{Conclusions}
\label{sec:concl}

In the above said we have shown that pQCD is applicable to exclusive 
photoproduction by showing that the ratio of the imaginary parts of the 
amplitudes of DIS to a real photon is calculable in pQCD after specifying 
initial conditions since the derivative in energy of the hard scattering 
amplitudes can be unambiguously calculated in pQCD and all 
the non-perturbative physics can then be absorbed into a parton 
distribution. We wrote down an evolution equation for the imaginary part of 
the amplitude, which can be generalized to the complete amplitude at small $x$,
and solved for the imaginary part of the amplitude.
We also found that the imaginary part of the  amplitude
for the production of a real photon is larger than the one in 
the case of DIS in a broad range of $Q^2$ for the reasons as discussed above.
We also found the same to be true for the full amplitude at
small $x$. We also make experimentally testable predictions for
the number of real photon events and suggest that the number of events are 
small but not too small such that after improving the statistics on existing
or soon to be taken data, it would be 
feasible to access the nondiagonal gluon distribution at small $x$ from this
clean process. Finally, we demonstrated that measuring the asymmetry $A$ at HERA, which is 
fairly sizable in the kinematics in question, would allow one to determine the 
real part of the DVCS amplitude, in other words gain a first experimental 
insight into nondiagonal parton distributions, despite $D < 1$. 

\newpage

\chapter{DVCS in DIS at HERA - A Probe of Asymptotia}
\indent

\section{Introduction}
\label{introa}

In this chapter based on Ref.\ \cite{new}, we will attempt to answer one of 
the pressing questions of high-energy QCD: The energy dependence 
of the strong amplitude in the situation when one or both
colliding systems have small size. It is generally agreed that
the $x$-range currently available at HERA is not sufficient to test the current
ideas about the onset of asymptotia via measurements of the parton densities.

At the same time, the experience in studies of soft processes tells us that
the  real part of the zero angle scattering amplitude, $\eta$, provides us, 
through the dispersion representation over the invariant energy of the 
collision, with information about the energy dependence of the cross section 
well beyond the energy where $\eta$ is measured. The reason for this is that 
$\eta$, the ratio of the real to imaginary part of the amplitude essentially 
measures the $\ln s$ derivative of the cross section \cite{reim1}:
\beq
\eta = \frac{\pi}{2}\frac{d\ln(F_2(x,Q^2))}{d\ln(1/x)}.
\label{reimx}
\eeq
One can also use derivative analyticity relations to derive a more accurate
formula \cite{reim1,reim2}, leading to 
\beq
\eta = tan\left[{\pi \alpha\over 2}\right].
\label{reimbr}
\eeq
for $F_2(x,Q^2) \propto x^{-\alpha}$.

In this chapter we propose a new methodology of investigating the energy 
dependence of high-energy processes through their real parts and also the 
shapes of nondiagonal parton distributions. We use DVCS as an example which 
offers us a direct way 
to study nondiagonal parton distributions, though their actual extraction 
from the data is not possible in DVCS due to the fact that the parton 
distributions depend on $y_1$ and $y_2=y_1-x$ which are dependent variables
rather than independent as one would need and thus the inverse Mellin transform
of the factorization formula cannot be found\footnote{As was pointed out 
before this is not true for di-muon production since there, we have
two independent variables $x$ and $\xi_1$.}. 
The major new result of this chapter is that the current successful fits 
to the $F_{2N}(x,Q^2)$ HERA data lead to qualitatively different predictions  
for the asymmetry, reflecting different
underlying assumptions of the fits about the behavior of parton densities
at $x$ below the HERA range.

A recent analysis in Ref.\  \cite{FFS} and the previous chapter have shown 
that DVCS studies at HERA are feasible and we made predictions for the 
expected DVCS counting rate compared to DIS as well as the asymmetry $A$ in 
the combined DVCS and Bethe-Heitler cross section for recent H1 data \cite{15}.

The chapter is structured as follows. In Sec.\ \ref{basics} we review the 
necessary formulas of Ref.\ \cite{FFS} and the previous chapter for our 
analysis. In this context, 
the formula pertaining to the ratio of real to imaginary part of a scattering 
amplitude at small $x$ is of particular importance. We then
present the different fits to $F_2(x,Q^2)$ in Sec.\ \ref{fits} and present
the different results for the asymmetry $A$ with respect to $t$ and $y$, at
fixed $y$ and $t$ respectively. Sec.\ \ref{concla} 
contains our conclusions and outlook.

\section{Relations between DVCS and DIS}
\label{basics}

In order to compute the asymmetry $A$, we need the ratio of the 
imaginary part of the DIS amplitude to the imaginary part of the DVCS 
amplitude and the relative DVCS counting rate $R_{\gamma}$, expected at HERA 
in the interesting kinematic regime of $10^{-4}<x<10^{-2}$ and moderate $Q^2$, 
i.e.\ , $3.5~\mbox{GeV}^2<Q^2<45~\mbox{GeV}^2$. The relative counting rate 
$R_{\gamma}$ is given by \cite{FFS} 
\beq
R_{\gamma} \simeq \frac{\pi \alpha}{4 R^2 Q^2 B}F_2(x,Q^2)(1 + \eta^2).
\label{fNa}
\eeq
where $R$ is the ratio of the imaginary parts of the DIS to DVCS amplitude
as given in \cite{FFS}\footnote
{We will use the results for $R$ from \cite{FFS} in our present analysis.}
, $B$ is the slope of the $t$ dependence (for more details see 
Ref.\ \cite{FFS} and the previous chapter.)
and $\eta$ is the ratio of real to imaginary part of the DIS amplitude, i.e.\ 
, $F_2(x,Q^2)$, given by Eq.\ (\ref{reimx}).

The asymmetry $A$ is given by \cite{FFS}
\beq
A =\frac{\int_{-\pi/2}^{\pi/2}d\phi_r d\sigma_{DVCS+BH} - \int^{3\pi/2}_{\pi/2}
d\phi_r d\sigma_{DVCS+BH}}{\int_{0}^{2\pi}d\phi_r d\sigma_{DVCS+BH}},
\label{asym}
\eeq 
where $d\sigma_{DVCS+BH}$ is given by the sum of 
Eq.\ (\ref{dvcsc}),(\ref{difcros}),(\ref{inter}).
As explained in \cite{FFS} this azimuthal angle asymmetry is due to the sign 
change of the interference term in the combined DVCS and Bethe-Heitler 
cross section, when the angle $\phi_r$ is integrated over the upper hemisphere
of the detector as compared to the integration over the lower hemisphere of the
detector. Although the 
absolute value of the parton distributions cannot be extracted from DVCS, the 
shape of the distributions is nevertheless accessible 
since the real part of the DVCS amplitude is isolated through this asymmetry.
Therefore, we investigate the influence of different $F_2$ fits on the 
asymmetry through the relative counting rate which is directly sensitive to 
the ratio of real to imaginary parts of $F_2$ as shown in Eq.\ (\ref{fNa}).

\section{The different fits to $F_2(x,Q^2)$}
\label{fits}

In the calculation of the asymmetry $A$ we use the recent H1 data from Ref.\ 
\cite{15} as previously used in Ref.\ \cite{FFS}, a logarithmic fit by 
Buchm\"uller et.\ al (BH) \cite{11}, the ALLM97 fit \cite{12} and a leading order 
BFKL-fit \cite{13} for illustrative purposes.

In the H1 data, $F_2$ behaves for small $x$ as $x^{-\lambda}$ and hence $\eta$
is just $\frac{\pi}{2}\lambda$ where $\eta^2 = 0.09 - 0.27$ in the $Q^2$ range
given in the previous section. Note that $\eta$ has no $x$ dependence, 
for small enough $x$, and thus depends only on $Q^2$. This is not true for 
all of the other fits.

$F_2$ in the BH fit takes on the following form
\beq
F_2(x,Q^2) = 0.078 + 0.364\log(\frac{Q^2}{0.5~\mbox{GeV}^2})\log(\frac{0.074}
{x}),
\label{buf2}
\eeq
and hence we find for $\eta$
\beq 
\eta =\frac{\pi}{2}0.364\log(\frac{Q^2}{0.5~\mbox{GeV}^2})/F_2(x,Q^2).
\label{bueta}
\eeq
Note that this $\eta$ has not only the usual $Q^2$ dependence but depends 
rather strongly on $x$ also, which is not seen in the data for the slope of 
$F_2$.

In the ALLM97 fit $F_2$ at small $x$ takes on the following form
\beq
F_2(x,Q^2)=\frac{Q^2}{Q^2+m_0^2}(F_s^p(x,Q^2)+F_2^R(x,Q^2)),
\label{allmf2}
\eeq
with 
\bea
F_2^P(x,Q^2) &=& c_P(t)x_P^{a_P(t)}\nonumber\\
F_2^R(x,Q^2) &=& c_R(t)x_R^{a_R(t)},
\eea
where
\beq
t = \ln\left (\frac{\ln\left (\frac{Q^2+Q_0^2}{\Lambda^2}\right )}
{\ln\left (\frac{Q_0^2}{\Lambda^2}\right )}\right ),
\eeq
and
\bea
c_R(t) &=& 0.8017 + 0.97307t^{3.4942},\nonumber\\
a_R(t) &=& 0.584 + 0.37888t^{2.6063},\nonumber\\
c_P(t) &=& 0.28067 + 0.05776\left (\frac{1}{1+t^{2.1979}}-1\right ),\nonumber\\
a_P(t) &=& -0.0808 + 0.36732\left (\frac{1}{1+t^{1.1709}}-1\right ).
\eea
$x_P$ and $x_R$ are given at small $x$ by
\bea
x_P &=& x(1+\frac{m_P^2}{Q^2}),\nonumber\\
x_R &=& x.
\eea
$\Lambda^2 = 0.06527~\mbox{GeV}^2$, $m_P^2 = 49.457~\mbox{GeV}^2$, 
$Q_0^2 = 0.46017~\mbox{GeV}^2$ and $m_0^2 = 0.31985~\mbox{GeV}^2$.
$\eta$ is then given by
\beq
\eta = -\frac{\pi}{2}\frac{a_Pc_Px_P^{a_P}+a_Rc_Rx_R^{a_R}}
{c_Px_P^{a_P}+c_Rx_R^{a_R}}.
\label{etaallm}
\eeq  

In the case of BFKL where $F_2\simeq x^{-\frac{4N_c\ln(2)\alpha_s}{\pi}}$
we find $\eta$ to be
\beq
\eta = \frac{\pi}{2}\frac{4N_c\ln(2)\alpha_s}{\pi}
\label{etabfkl}
\eeq
with an $\alpha_s$ in leading order of
\beq
\alpha_s = \frac{4\pi}{3N_c\log\left ( \frac{Q^2}{\Lambda^2}\right )}.
\eeq

\section{Results for the asymmetry $A$}
\label{results}

\begin{figure}
\vskip-1in
\centering
\epsfig{file=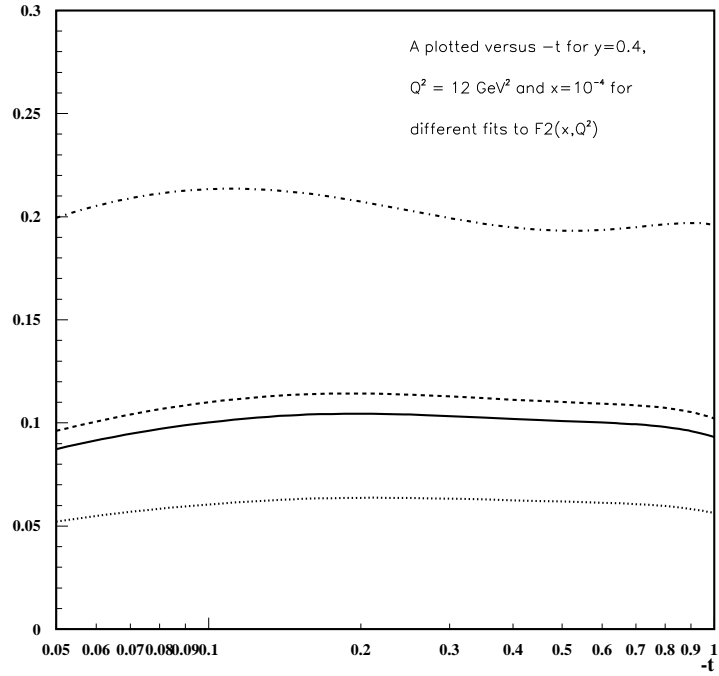,height=12cm}
\vskip-1in
\epsfig{file=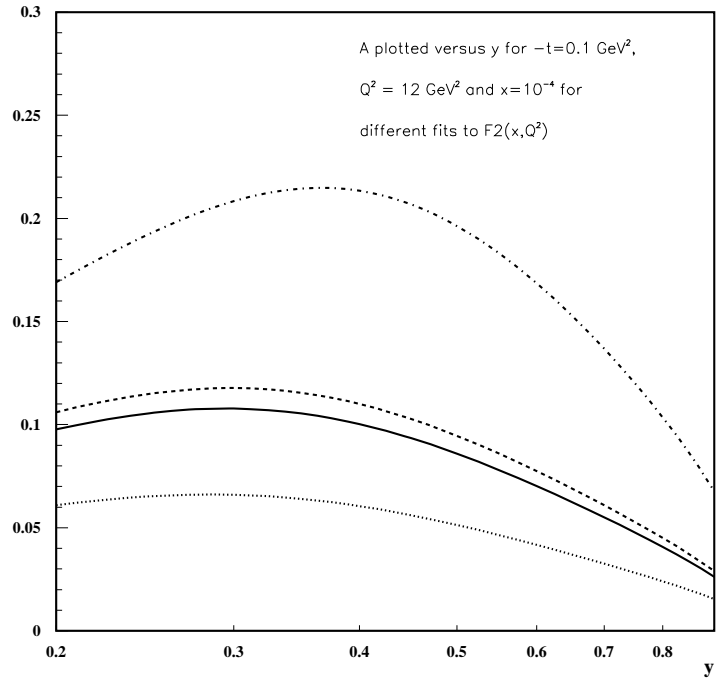,height=12cm}
\caption{H1 fit (solid curve), the BH fit (dotted curve), 
ALLM97 fit (short-dash curve) and BFKL fit (dash-dot 
curve)
for $x=10^{-4}$. a) Asymmetry $A$ versus $t$ for fixed $y=0.4$.
b) Asymmetry $A$ versus $y$ for fixed $-t=0.1~\mbox{GeV}^{2}$.}
\label{fig1}
\end{figure}

\begin{figure}
\vskip-1in
\centering
\epsfig{file=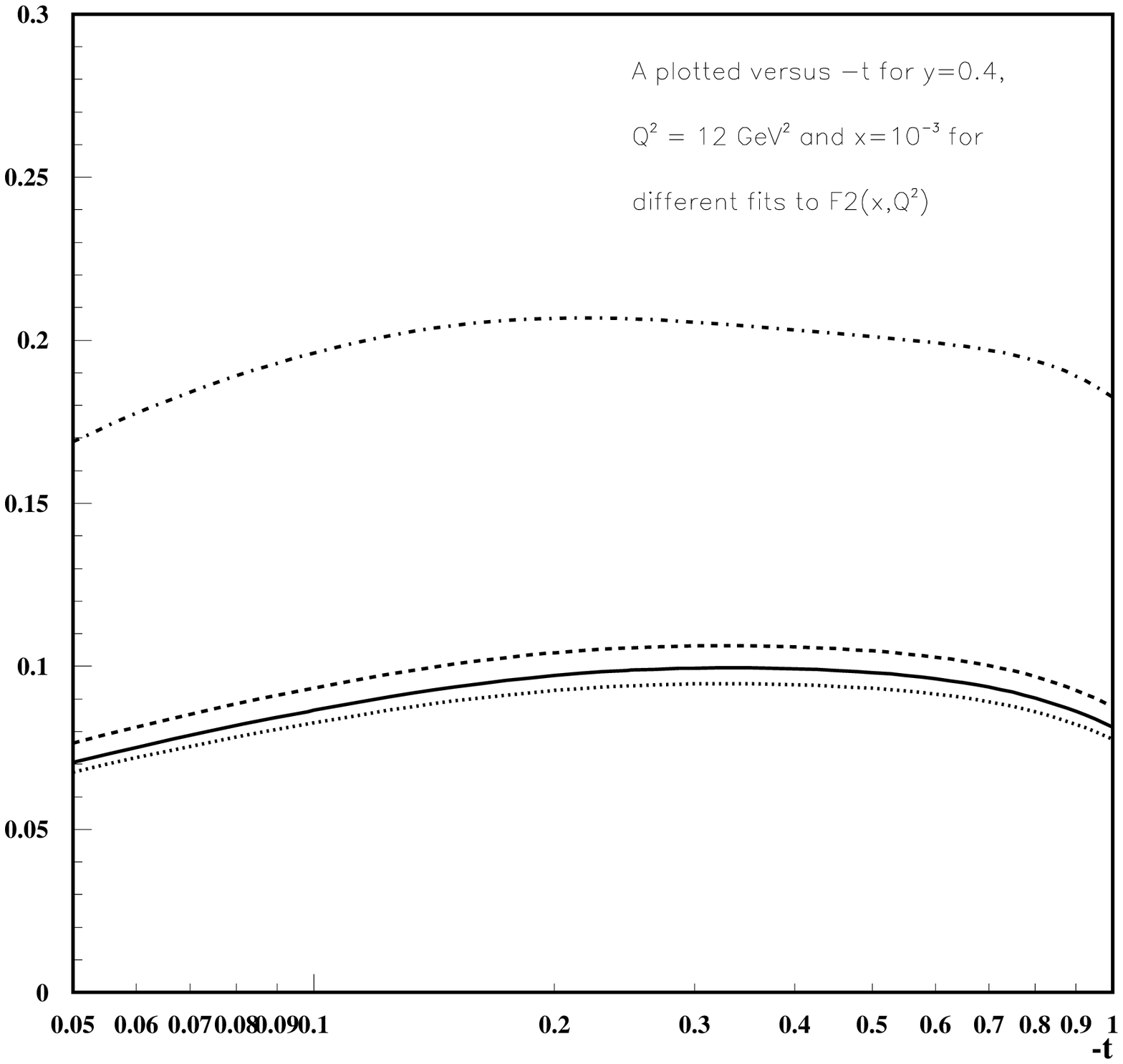,height=12cm}
\vskip-1in
\epsfig{file=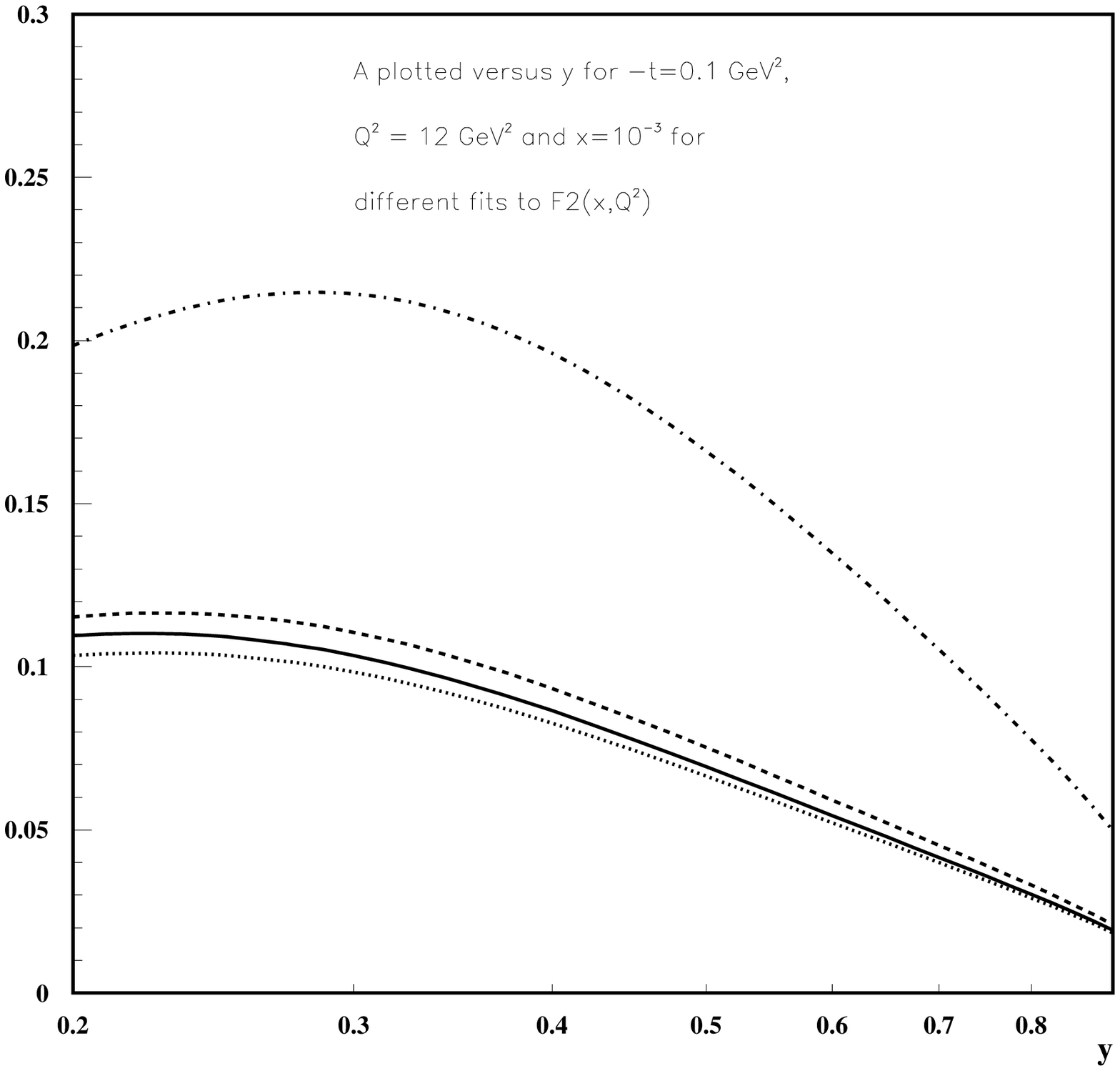,height=12cm}
\caption{H1 fit (solid curve), the BH 
fit (dotted curve), ALLM97 fit (short-dash curve) and BFKL fit (dash-dot 
curve)
for $x=10^{-3}$. a) Asymmetry $A$ versus $t$ for fixed $y=0.4$.
b) Asymmetry $A$ versus $y$ for fixed $-t=0.1~\mbox{GeV}^{2}$}
\label{fig2}
\end{figure}

\begin{figure}
\vskip-1in
\centering
\epsfig{file=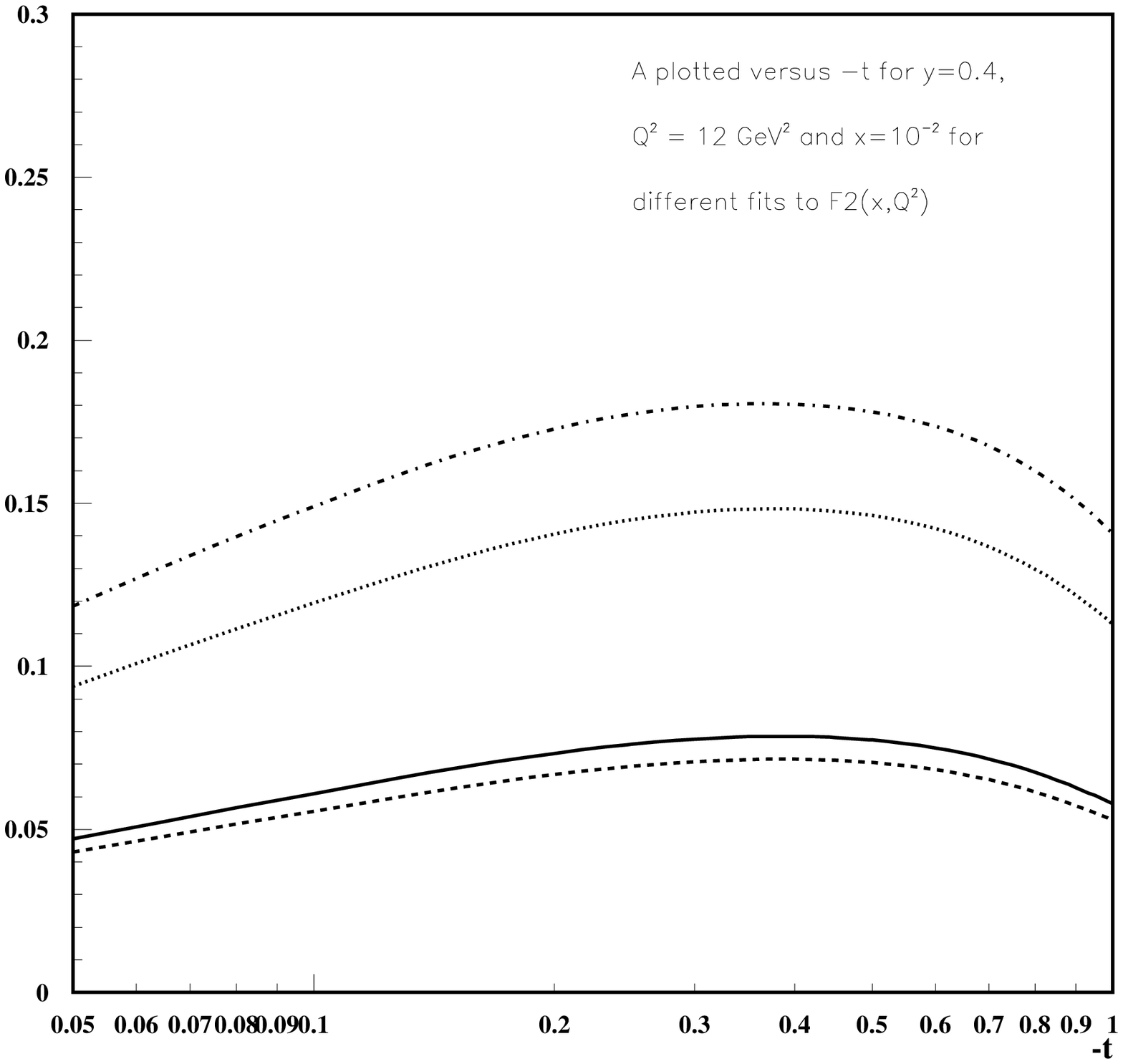,height=12cm}
\vskip-1in
\epsfig{file=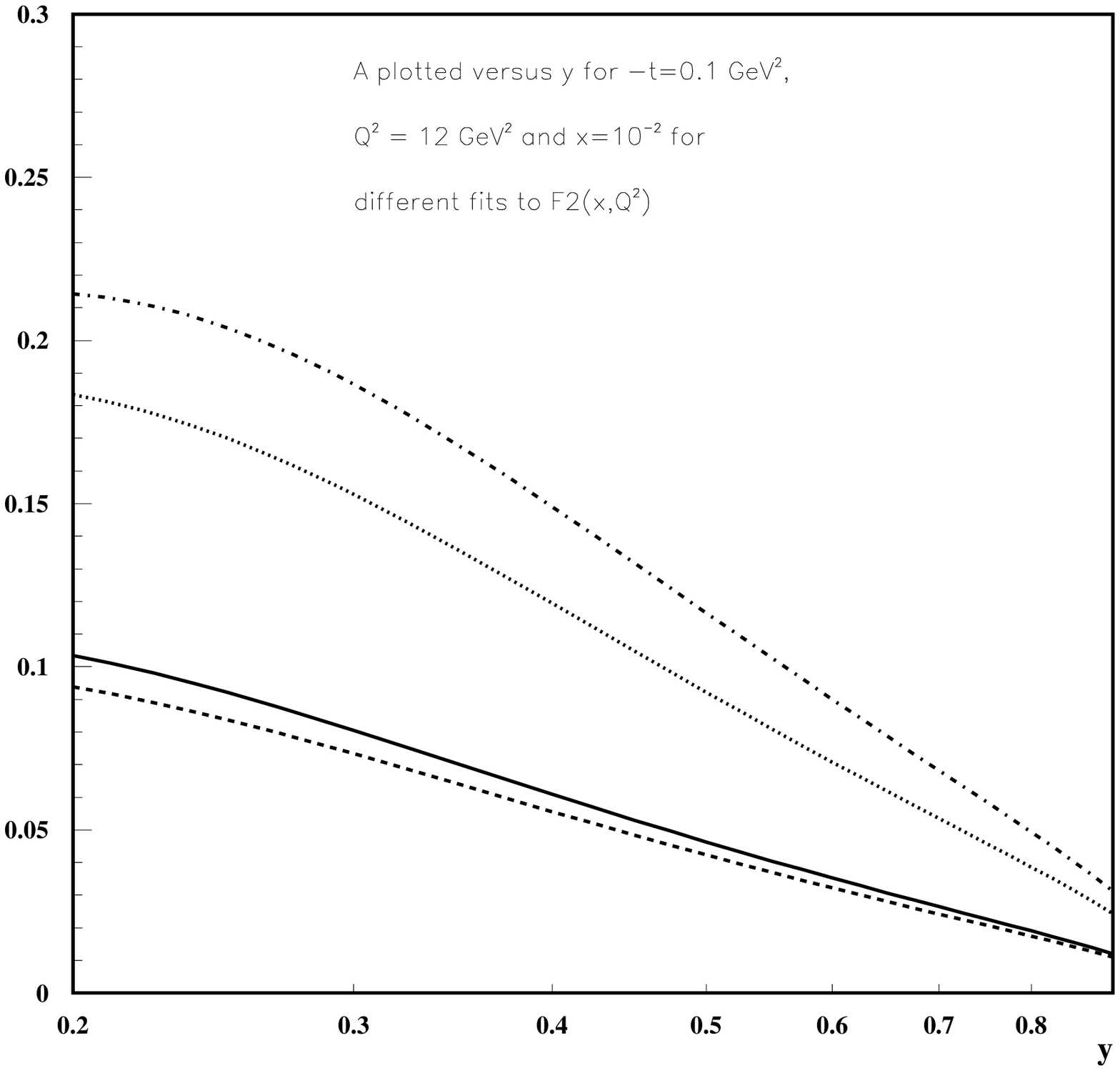,height=12cm}
\caption{H1 fit (solid curve), the BH
fit (dotted), ALLM97 fit (short-dash curve) and BFKL fit (dash-dot 
curve)
for $x=10^{-2}$. a) Asymmetry $A$ versus $t$ for fixed $y=0.4$.
b) Asymmetry $A$ versus $y$ for fixed $-t=0.1~\mbox{GeV}^{2}$ }
\label{fig3}
\end{figure}

In Fig.\ \ref{fig1} - \ref{fig3}, we plot the asymmetry $A$ as a function of 
$t$ and $y$ for fixed 
$Q^2=12~\mbox{GeV}^2$, fixed $y=0.4$ and $-t=0.1~\mbox{GeV}^{2}$ and 
$x=10^{-4},~10^{-3},~10^{-2}$. The slope $B$ of the $t$-dependence for DVCS 
was taken to be 
$B=5~\mbox{GeV}^{-2}$ whereas for the Bethe-Heitler cross section we used the 
nucleon form factor as used in chapter 5. The counting rate $R_{\gamma}$ was 
appropriately adjusted for the different fits according to Eq.\ (\ref{fNa}).
The solid curves in Fig.\ \ref{fig1} - \ref{fig3} are our benchmarks\footnote
{Though actual H1 data is used, we are still dealing with a leading order
approximation and a particular model for the nondiagonal parton distributions
at the normalization point was used in computing $R_{\gamma}$(see \cite{FFS} 
for more details on the type of model ansatz and approximations used.).}.

Comparing the BH fit (dotted curves), against our benchmarks 
we find a 
strong $x$ dependence of the asymmetry in the BH fit as well as 
different shapes and absolute values. The strong $x$ dependence of the 
BH fit in the ratio of the real to imaginary part of $F_2$ will make
it easy to distinguish this logarithmic fit from a power law fit which
yields an $x$ independent $\eta$.

As far as the ALLM97 fit is concerned (short-dash curves), there is hardly a 
difference, as compared to the H1 fit in the asymmetry as a function 
of $t$ and $y$
in absolute value, shape and $x$ dependence, except for $x=10^{-2}$ but this 
is due to the approximations we made for $x_P$ and $x_R$ which are not that 
good anymore at $x=10^{-2}$.

If one compares the BFKL fit (dash-dot curves) to the H1 fit one sees 
immediately that the BFKL fit is totally off in almost all aspects and was 
only included here as an illustrative example, not as a serious fit.     

\section{Conclusions}
\label{concla}

In the above we have shown the sensitivity of the exclusive
DVCS asymmetry $A$ to different $F_2$ fits and made comments on the 
viability of each fit. Note that even a fit which reproduces $F_2$ data, 
as well as its slope, in a satisfactory manner can be shown to lead to 
differences in the asymmetry shape. The sensitivity of the asymmetry to $y$ 
and $t$  will allow us, once 
experimentally determined, to make a shape fit and hence make a shape fit to 
nondiagonal parton distributions for the first time. 
\newpage  
\newpage

\appendix
\chapter{Direct Estimation of Sizes of Higher-Order Graphs}
\indent

\section{Introduction}
\label{sec:introduction}

In this appendix, based on Ref.\ \cite{apnd}, we want to give concrete 
examples of distributional methods which were employed in an abstract way in 
chapter 4. The starting point for this appendix is formed by the following
observations:
\begin{itemize}

\item
    The only (known) systematic method for calculating scattering
    in QCD is perturbation theory.  (Lattice Monte-Carlo methods
    work in Euclidean space-time, and are excellent for
    calculating static quantities such as masses from first
    principles.  But they are essentially useless when a
    calculation in real Minkowski space-time is needed.)

\item
    In field theory, calculations beyond low orders of
    perturbation theory are computationally complex, both because
    the calculations of individual graphs are hard and because
    there are many different graphs.

\item
    Hence it is important to make the most efficient use of
    low-order calculations.

\end{itemize}
Since the coupling in practical calculations is not very weak,
the accuracy of predictions can be ruined by uncalculated
higher-order terms.  It follows that there is a need to estimate the
sizes of the errors.
For this one wants quick estimates of terms in perturbation
theory.  The computational complexity of the estimates should
increase as little as possible with the size of the graphs.
Indeed, our aim is that one only calculates integrals of the form
\begin{equation}
      \int _{l}^{u} dx \, x^{n} \, \ln^{p}x .
\end{equation}
With suitable methods:
\begin{itemize}

\item
    One can determine good values for renormalization and
    factorization scales, by asking how to minimize the error
    estimates.

\item
    When the estimates get substantially larger than some
    appropriate ``natural'' size, one would get a diagnosis of a
    need for resummation of classes of higher-order corrections.
    The diagnosis would include an explanation of the large terms
    and thus indicate the physics associated with the
    resummation.

\end{itemize}

In this appendix, we explain how to start such a program.  It
builds on work first reported in Ref.\ \cite{Durham}.  Our
methods treat properties of the integrands of Feynman graphs, and
are therefore directly sensitive to the physics of the process
being discussed.  Some other treatments of these issues discuss
the problems in terms of the mathematics of series expansions in
general, without asking what is causing the graphs to be the
sizes they are.  A particular exception is the work of Brodsky,
Lepage, and Mackenzie \cite{BLM}\footnote{
   See also the more recent work of Brodsky and Lu \cite{BL} and
   of Neubert \cite{Neubert}.
}.
They use heavy quark loops to
probe the actual momentum scales that dominate in a particular
calculation; this is then used to provide a suitable value for
the renormalization/factorization scale.
But we believe that our methods provide a more direct route to
answering the question of why the scales are what they are and
why a calculation gives a particular order of magnitude.
The issues addressed by methods involving the Borel transform and
Pad\'e summation address complementary issues \cite{Pade}.

Now, most cross sections in QCD cannot be directly computed by
perturbation theory; this can only be used to
compute the short-distance coefficients that appear in the
factorization theorem.  So we will also treat the specific problems
that arise in estimating the sizes of the coefficient functions.
These functions have the form of a sum over Feynman graphs
(typically massless), with subtractions to cancel some infra-red
(IR) divergences. Remaining IR divergences are cancelled between
different graphs or between different final-state cuts. Two
characteristic features appear.  First, we can obtain estimates
for sums of particular sets of graphs, but not for the individual
graphs, which are divergent. Secondly, the coefficient functions
are not in fact genuine functions. They are normally singular
generalized functions (or distributions) and an estimate can only
be made for the integral of a coefficient function with a smooth
test function.

We show how to estimate the sizes of graphs by a direct
examination of the integrands.
An important part of our technique is an implementation of
subtractions directly in the integrands,
both for the subtractions that implement counterterms for
ultra-violet (UV)
renormalization and for the IR subtractions that are used in
short-distance coefficient functions.
In order to explain our ideas, we will examine two examples:
(1) the one-loop self-energy graph in $(\phi ^{3})_{4}$ theory,
and (2) a particular set of graphs for the Wilson coefficient for
deep-inelastic scattering.
Our estimates are in the form of approximations to
ordinary integrals that are absolutely convergent.
This is in contrast to the original integrals, which
are typically divergent in the absence of a regulator.
Thus a by-product of our work will be algorithms
for computing graphs numerically in Minkowski space-time, which
may have relevance to work such as Ref.\ \cite{Catani.Seymour}.
As an illustration of how estimations can be carried out, even
analytically, for a measurable quantity, we will estimate the size of
the Wilson coefficient for the structure functions $F_{T}$ and $F_{L}$
in the last section.

\section{Euclidean self-energy in $\phi ^{3}$ theory}
\label{sec:self.energy}

\begin{figure}
\centering
\mbox{\epsfig{file=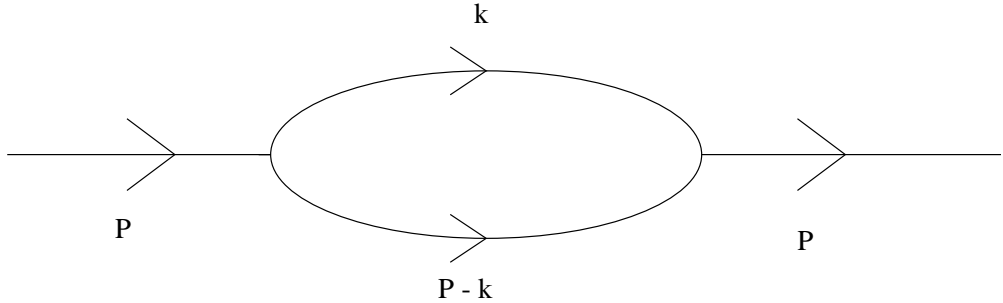,height=4cm}}
\vspace*{5mm}
\caption{One-loop self-energy graph.}
\label{fig:self.energy}
\end{figure}

In this section, we give a representation of the one-loop
self-energy graph of
Fig.\ \ref{fig:self.energy} in $\phi ^{3}$ theory in
four dimensions.
A particular renormalization scheme is used, which we relate to
ordinary $\MSbar$ renormalization.
Then we show how to estimate the size of the graph from
elementary integrals and hence how to choose the renormalization
scale suitably.

\subsection{Renormalization}

The only complication in computing the graph of Fig.\
\ref{fig:self.energy} is its UV
divergence, given that we
choose to work in Euclidean space.
We give a
representation of the graph that is an absolutely convergent
integral over the loop momentum itself in four dimensions and in
which the renormalization is explicitly ``minimal''.  This last
term means that the counterterm for the (logarithmic) divergence is
independent of the mass and the external momentum.  It is a
very useful property when one wants to take zero-mass limits,
etc.  For a general divergence, the counterterm would be
polynomial in masses and external momenta.

Our representation of the graph is:
\begin{equation}
   I(p) = \frac {1}{2} \frac {g^{2}}{(2\pi )^{4}} \int  d^{4}k
          \left\{
            \frac {1}{(k^{2}+m^{2}) \, \left[ (p-k)^{2}+m^{2}\right]}
            - \frac {\theta (k>\mu _{c})}{k^{4}}
          \right\} .
\label{Ren.self.energy}
\end{equation}
We recognize in the integrand a term that is given by the usual
Feynman rules (in Euclidean space-time), and a subtraction term.
The subtraction term is the negative of the asymptote of the
first term as $k\to \infty $, so we term our procedure\footnote{
    See \cite{Durham} for a previous account.  A formalization of
    such ideas (to all orders of perturbation theory) was given
    earlier by Ilyin, Imashev and Slavnov \cite{Slavnov}, and
    later by Kuznetsov and Tkachov \cite{KT}.
}
``renormalization
by subtraction of the asymptote''.  A cut-off is applied to
prevent the subtraction term giving an IR divergence at
$k=0$; the cut-off does not affect the $k\to \infty $ behavior and
therefore does not affect the fact that the UV divergence is
cancelled.
To see that Eq.\ (\ref{Ren.self.energy}) is
equivalent to standard renormalization, one simply applies a
UV regulator, after which each term can be integrated
separately.  The first term is the unrenormalized graph and the
second term is a $p$-independent counterterm.

Evidently, the integral is absolutely convergent, and
can therefore be computed by any appropriate numerical method.
(It can also be evaluated analytically.  But this is not
interesting to us, since we wish to obtain methods that work for
integrals that are too complicated for purely analytic methods to
be convenient or useful.)

The counterterm is in fact the most general one that is
independent of $m$ and $p$, since any other renormalization
counterterm can differ only by a finite term added to the
integral that is independent of $m$ and $p$, and a change of the
cut-off $\mu _{c}$ is equivalent to adding such a term.  We can relate
the counterterm to the commonly used $\MSbar$ one simply
by computing the counterterm alone, with dimensional
regularization:
\begin{equation}
    \mbox{standard prefactor} \times
    \int _{|k|>\mu _{c}} d^{n}k \, \frac {1}{k^{4}} .
\end{equation}
The result is that setting $\mu _{c}$ equal to the scale $\mu $ of the 
$\MSbar$ scheme gives exactly $\MSbar$ renormalization.  In
general, we would find that $\mu _{c}$ would be a factor times $\mu $, or
equivalently that we should set $\mu _{c}=\mu $ and then add a specific
finite counterterm to the graph.

Notice that the integral to relate our renormalization scheme to
the $\MSbar$ scheme is algorithmically simpler to compute
analytically than the original integral.
There is always the possibility of adding finite counterterms.
Moreover, the precise form of the cut-off is irrelevant to the
general principles.  One can, for example, change the sharp
cut-off function $\theta (k>\mu _{c})$ to a smooth function $f(k/\mu _{c})$
that
obeys $f(\infty )=1$ and $f(0)=0$.  Such a function would probably be
better in numerical integration.

Of course, our method as stated is specific to one-loop graphs.
But it is an idea that has been generalized \cite{Slavnov,KT} to
higher orders.

\subsection{Estimate}

We now show how to estimate the size of the integral Eq.\
(\ref{Ren.self.energy}).  To give ourselves a definite case, let
us choose $p^{2} \leq m^{2}$.  We obtain the estimate as the sum of
contributions from $k<m$ and from $k>m$.  Since the
renormalization counterterm is designed to subtract the $k\to \infty $
behavior of the unrenormalized integrand, we regard it as a
$\delta $-function at infinity and therefore to be associated
completely with the $k>m$ term in our estimate.

In the region $k<m$, our estimate is obtained by replacing each
propagator by $1/m^{2}$ so that
\begin{eqnarray}
  \mbox{Contribution from $k<m$}
  &\simeq& \frac {g^{2}}{32\pi ^{4}} \int _{k<m} d^{4}k \frac {1}{m^{4}}
\nonumber\\
  &=& \frac {g^{2}}{64\pi ^{2}} .
\end{eqnarray}
This factor is the product of $g^{2}/32\pi ^{4}$ for the prefactor and
$\pi ^{2}/2$ for the volume of a unit 4-sphere.  As advertised, we have
had to calculate no integral that is more complicated than a simple
power of $k$.  The approximation of replacing the propagators by
$1/m^{2}$ leads us to an over-estimate of the integral, but not by a
great factor, since we are in a region of small momentum.

The estimate for $k>m$ is obtained by replacing the propagators
by their large $k$ asymptote:
\begin{eqnarray}
  \mbox{Contribution from $k>m$}
  &\simeq& \frac {g^{2}}{32\pi ^{4}} \int _{k>m} d^{4}k
            \left[ \frac {1}{k^{4}} - \frac {\theta (k>\mu _{c})}{k^{4}}
\right]
\nonumber\\
  &=& \frac {g^{2}}{32\pi ^{4}} 2\pi ^{2}
  \left[
     \int _{m}^{\infty }\frac {dk}{k} - \int _{\mu }^{\infty }\frac {dk}{k}
  \right] .
\end{eqnarray}
Since we are taking the difference of two terms, we must be
careful about the errors, which are of order
\begin{equation}
   \int _{k>m} d^{4}k \frac {m^{2}}{k^{6}} = \frac {g^{2}}{32\pi ^{2}}.
\end{equation}

To understand the structure of the result, let us examine how the
original integral (\ref{Ren.self.energy}) appears after
integrating over the angle of $k$:
\begin{equation}
    \frac {g^{2}}{16\pi ^{2}}
     \int _{0}^{\infty }\frac {dk}{k} \left[ A(k,p,m) - \theta (k>\mu _{c})
\right] .
\label{log.integral}
\end{equation}
The function $A$ is the angular average of $k^{4}$ times the two
propagators.  It approaches 0 as $k\to 0$, so that the integral
is convergent there, and it approaches unity as $k\to \infty $, which
would give a UV divergence were it not for the
subtraction.

\begin{figure}
\centering
\mbox{\epsfig{file=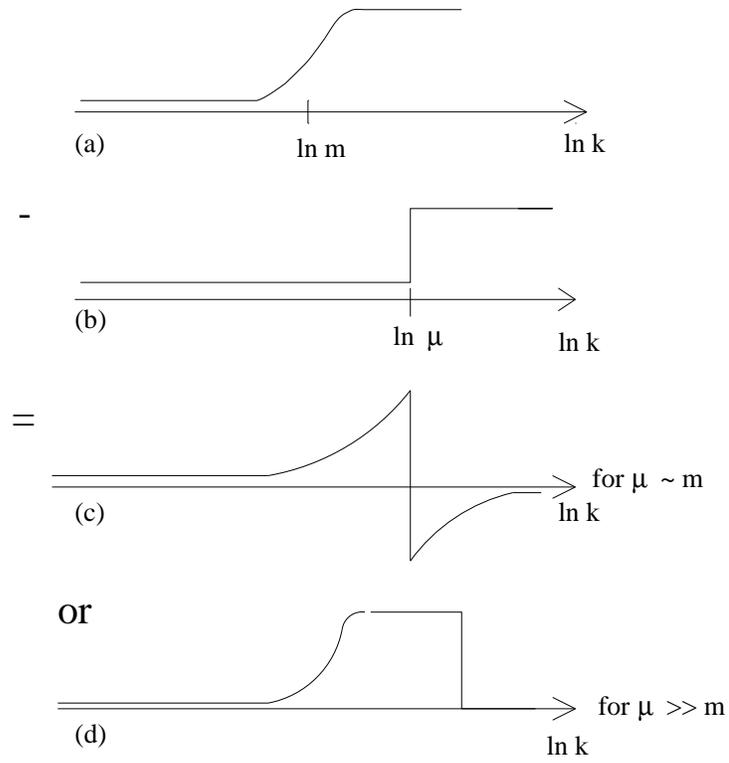,height=10cm}}
\vspace*{5mm}
\caption{(a) Scaled integrand for unrenormalized self-energy graph.
    (b) Subtraction term.
    (c) Total if $\mu \sim m$.
    (d) Total if $\mu  \gg m$.}
\label{fig:difference}
\end{figure}

We can represent this situation by the graphs of Fig.\
\ref{fig:difference}.  The transition region for the
unsubtracted integrand, where it changes from being 0 to 1,
is around $k=m$, to within a factor of 2 or so; this is evident
by examining the denominators.  By setting $\mu _{c}$ to be $m$ to
within a factor of 2, we achieve the following:
\begin{itemize}

\item
    The integrand is less than unity everywhere.

\item
    It is concentrated in a shell of thickness of order $m$
    around $k=m$.

\end{itemize}
Thus we can say that the natural size of the graph is the product
of
\begin{itemize}

\item The prefactor $g^{2}/32\pi ^{4}$.

\item The surface area of a unit 4-sphere: $2\pi ^{2}$.

\item The width of the important region, about unity, in units of
    $m$.

\item A factor of $m$ to the dimension of the integral, i.e., 1.

\item A factor less than unity, say $1/2$, to allow for the fact
    that, after subtraction, the integrand in Eq.\
    (\ref{log.integral}) is smaller than unity and that there is
    a cancelation between negative and positive pieces.

\end{itemize}
That is, the natural size is $g^{2}/32\pi ^{2}$, if $\mu $ is reasonably
close to $m$.

If $\mu $ is not close to $m$, then we get a long plateau in the
integrand---see Fig.\ \ref{fig:difference}.  The height of the
plateau is unity, and its length is $\ln (\mu /m)$ to about $\pm 1$ in
units of $\ln k$.  This clearly gives a larger-than-necessary
size, and an optimal choice of the $\MSbar$ scale is around $m$.

One should not expect to get an exact value for the scale $\mu $.
A physical quantity in the exact theory is independent of $\mu $,
and any finite-order calculation differs from the correct value
by an amount whose precise value is necessarily unknown until one has done a
more accurate calculation.
If one is able to estimate the size of the error, as we are
proposing, then an appropriate value of $\mu $ is one that minimizes
the error.
Given the intrinsic imprecision of an error estimate, there is a
corresponding imprecision in the determination of $\mu $.
One can expect to identify, without much work,  an appropriate
scale $\mu $ to within a factor 2, and, with a bit more work, to
within perhaps $50\%$.  These estimates just come from asking
where the transition region in the integrand is, and by then
obtaining an answer by simple examination of the integrand.  But
one cannot enter into a religious argument of the wrong kind as
to whether the correct scale is $1.23m$ as opposed to $1.24m$,
for example.\footnote{
    Brodsky and Lu \cite{BL} obtain very precise estimates of a
    suitable scale.  Their rationale is the elimination of
    IR renormalons in the relations between IR-safe
    observables.  This is a concern with very high-order
    perturbation theory, an issue that we do not address.
}
By definition an error estimate is approximate.

Since we have not yet investigated how to estimate even
higher-order graphs, we are making the reasonable conjecture that the
properties of graphs do not change rapidly with order.  Then our
estimate that $\mu $ should be close to $m$ will ensure that
higher-order graphs are of the order of their natural size.

\subsection{Implication for QCD}

The same arguments applied to similar graphs in QCD show that the
natural expansion parameter in QCD is
\begin{equation}
   \frac {\alpha _{s}}{4\pi } \times  (\mbox{group theory})
      \times  (\mbox{factor for multiplicity of graphs}).
\label{natural.size.QCD}
\end{equation}
These arguments rely on being able to show that in the dominant part
of the range of integration all lines have approximately a
particular virtuality and that the relevant range of integration
is a corresponding volume of momentum space.

In the general case we cannot expect to get a much smaller
result, but we can expect that in some situations the properties
of the dominant integration region(s) will not be so good.
So what we need to do next is to analyze more interesting graphs
in QCD.  This we will do in the next section.

In general, when we get corrections in QCD that are substantially
larger than the natural size given above, it must be either
because the integrand is excessively large, or because there
is no single natural scale, or because we have not chosen a
good scale.\footnote{
    Our use of the word ``natural'' may suggest that we are
    proposing to estimate higher-order corrections simply by
    multiplying the appropriate power of the natural expansion
    parameter by the number of graphs.  This is not what we mean.
    We are arguing first that the sizes of graphs can actually be
    estimated fairly simply, and that the natural expansion
    parameter is a useful {\em unit} for these estimates.
    Secondly, we show that in the most favorable cases, graphs
    are less than or about unity in these natural units. Finally,
    we argue, in the next section, that general kinematic
    arguments about the physics of a graph are useful in
    diagnosing cases where graphs are large in natural units.
}
If the integrand is especially large relative to the natural
unit, or if there is no single natural scale, then we should
investigate in more detail the reasons, and derive something like
a resummation of higher-order corrections \cite{resum}.
In precisely such
situations, one does indeed have to compute high-order graphs. At
the same time, there is no need to compute the complete graphs
in all their gory detail, but only their simple parts.

\section{Wilson coefficient for deep-inelastic scattering}
\label{sec:Wilson}

The factorization formula for the leading-twist part of
a deep-inelastic structure function $F(x)$ is
\begin{equation}
   F(x) = \int  \frac {d\xi }{\xi } f(\xi ) \hat F(x/\xi ).
\end{equation}
Here, $f(\xi )$ is a parton density, and $\hat F$ is the
short-distance coefficient (``Wilson coefficient'').  We have
suppressed the indices for the different structure functions
($F_{1}$, $F_{2}$, etc.) and for the parton flavor.  The coefficient
function $\hat F(x/\xi )$ is obtained from Feynman graphs for
scattering on a parton target with momentum $\xi p$, where $p$ is
the momentum of the hadron target.  Subtractions
for initial-state collinear singularities
are applied to
the Wilson coefficient and the massless limit is taken.

\begin{figure}
\centering
\mbox{\epsfig{file=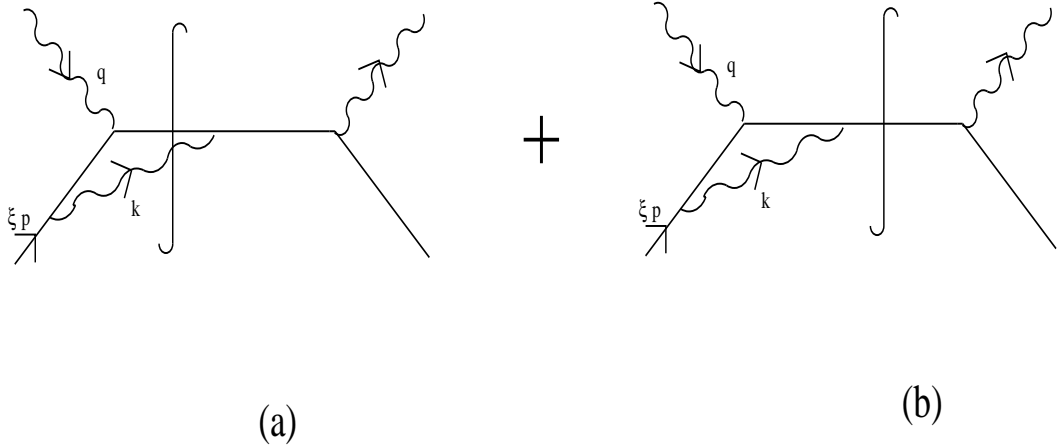,height=6cm,width=14cm}}
\vspace*{5mm}
\caption{The cuts of a one-loop graph for the Wilson coefficient
         for deep-inelastic scattering.}
\label{fig:1-loop.DIS}
\end{figure}

As an example, we will examine the contribution to the Wilson
coefficient from the  diagrams in Fig.\ \ref{fig:1-loop.DIS}.
These diagrams are the two possible cuts of a particular uncut
one-loop graph, and since we will need to use a cancelation of
final-state interactions, we must consider the sum of the two cut
graphs as a single unit.  Note:
\begin{itemize}

\item We will apply a subtraction to cancel the effect of
    initial-state collinear interactions where the incoming quark
    splits into a quark--gluon pair which are moving almost
    parallel to the incoming particle.\footnote{
     According to the factorization theorem, the subtractions cancel
     all the sensitivity to small
     momenta, i.e., to the initial-state collinear interactions.
     The subtractions are of the form of terms in the perturbative expansion
     of the distribution of a parton in a parton
     convoluted with lower-order terms in the coefficient
     function.  (See, for example, \cite{Colsop,Muell'89} for details.)
     Of course, both the partonic cross section and the subtraction term
     have to be properly renormalized. We will call the subtraction terms
     eikonal because of the particular rules involved in their
     calculation, for graphs such as Fig.\ \ref{fig:1-loop.DIS}.
}

\item There will be soft-gluon interactions and collinear
    final-state interactions.  These will cancel after the sum
    over cuts.  (In a more general situation, a sum over a
    gauge-invariant set of graphs is necessary to
    get rid of all soft-gluon interactions.)

\item The Wilson coefficient is a distribution (or generalized
    function) rather than an ordinary function of $x/\xi $.  Thus it
    is useful to discuss only the size of the coefficient
    after it is integrated with a test function, but not the size
    of the unintegrated coefficient function.  The parton density
    $f(\xi )$ provides a ready-made test function that has a physical
    interpretation.

\item The first graph of Fig.\ \ref{fig:1-loop.DIS},
    which has a virtual gluon, has a 4-dimensional
    integral, but the second graph, with a real
    gluon, has only a 3-dimensional integral.  Thus the
    cancelations associated with the sum over cuts can only be
    seen after doing at least a 1-dimensional integral.

\end{itemize}

\begin{figure}
\centering
\mbox{\epsfig{file=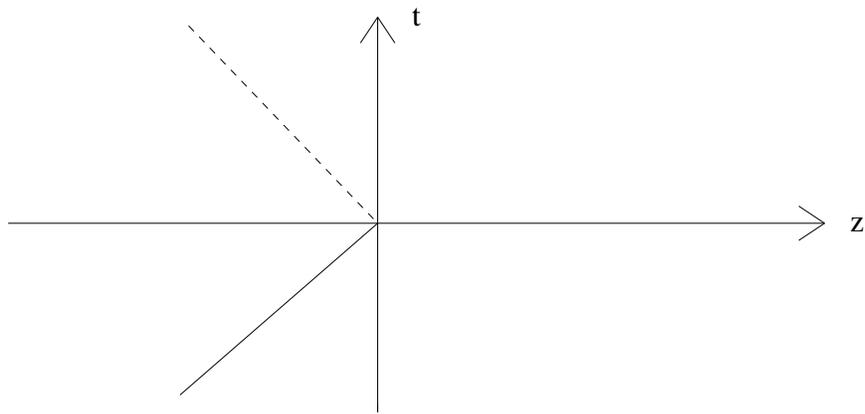,height=8cm}}
\vspace*{5mm}
\caption{Space-time structure of deep-inelastic scattering, in
    the center-of-mass frame of the virtual photon and the struck
    quark.  The solid line is the almost (light-like) world line
    of the incoming quark.  The dashed line is the world line of
    the single struck quark in the lowest-order (Born) graph for
    the hard scattering.
    }
\label{fig:sp-time}
\end{figure}

It is useful to visualize the process in space-time, Fig.\
\ref{fig:sp-time}, and to use light-front coordinates $(+,-,T)$
(defined by $V^{\pm }=(V^{0}\pm V^{3})/\sqrt 2$).  Our axes are such that the
incoming momenta for the hard scattering are:
\begin{equation}
  \xi  p^{\mu } = (\xi p^{+},0,{\bf 0}_{T}),
  ~~~~
    q^{\mu } = \left( -x p^{+},\frac {Q^{2}}{2xp^{+}},{\bf 0}_{T} \right) .
\end{equation}
Also, we find it convenient to parameterize the gluon momentum in Fig.\
\ref{fig:1-loop.DIS} in terms of two longitudinal momentum
fractions, $u$ and $z$, and a transverse momentum ${\bf k}_{T}$, as
follows:
\begin{equation}
    k^{\mu } = (u \xi (1-z) p^{+}, \, zq^{-}, \, {\bf k}_{T}).
\label{u.z.def}
\end{equation}
Thus $z$ is exactly the fraction of the total incoming minus
component of momentum that is carried off by the gluon, while $u$
is a scaled fraction of the plus component.
The scaling is somewhat unobvious, but it has the effect that positive
energy constraints on the final state restrict each of $u$ and
$z$ to range from 0 to 1.

Now we summarize how the
calculation of the real and virtual graphs and of the
subtraction graphs \cite{Colsop}, contributing to the
Wilson coefficients, is to be carried out:
\begin{itemize}

\item Using the Feynman rules for cut diagrams, we write down the
     momentum integral with the appropriate
     $\delta $-functions.   Then we contract the trace over Dirac matrices
    with the appropriate transverse, longitudinal
    or asymmetric tensor on the photon indices to obtain the
    structure function $F_{1}$, $F_{2}$, etc.
    For the sake of a definite simple example
    we choose to contract with $-g_{\mu \nu }$,
    which in fact gives the combination $3F_{1} - F_{2}/2x$.

\item We express the results of the calculations in terms of
    the light-cone components of $k$ (the internal gluon
    momentum), of $p$ (the incoming
    quark momentum), and of $q$ (the incoming photon momentum).

\item In the graphs with real gluon emission, we use the two
    $\delta $-functions to perform the integrals over ${\bf k}_{T}$ and over
    the
    fractional momentum $\xi $ entering from the parton density.
    This leaves a 2-dimensional integral.

\item In the virtual graphs, we use the one $\delta $-function to
    perform the integral over the fractional momentum $\xi $ entering from
    the parton density.  We then perform the ${\bf k}_{T}$ integral
    analytically.  Again we have a 2-dimensional integral.

\item We change the integration variables to the scaled dimensionless
    variables $u$ and $z$ defined in Eq.\ (\ref{u.z.def}).  This
    gives us an overall factor, just like the $g^{2}/16\pi ^{2}$ in the
    self energy, times an integral over roughly the unit square
    in $u$ and $z$.

\end{itemize}

\begin{figure}
\centering
\mbox{\epsfig{file=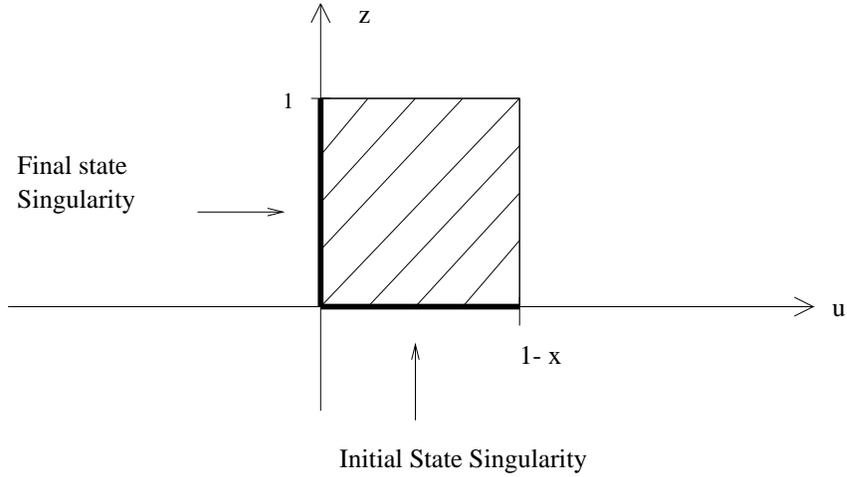,height=6.3cm}}
\vspace*{5mm}
\caption{Integration region and position of singularities for
    Fig.\ \protect\ref{fig:1-loop.DIS}(a).}
\label{fig:1-loop.Sing}
\end{figure}

\subsection{Real gluon}

For the real gluon graph Fig.\ \ref{fig:1-loop.DIS}(a) it is well
known that the integrand has the following singularities in the massless
limit:
\begin{itemize}

\item Initial-state collinear singularity on $z=0$.

\item Final-state collinear singularity on $u=0$.

\item Soft singularity $k=0$ at the intersection of the previous
    two singularities, i.e., at $u=z=0$.

\end{itemize}
(See Fig.\ \ref{fig:1-loop.Sing}.)  In accordance with the
standard recipe for constructing the Wilson coefficient, we
subtract the initial-state collinear singularity.
The subtraction term itself has a UV divergence, which
we choose to cancel by using the same method as
we used for the UV divergence of
the self-energy graph.
This gives\footnote{
  We have chosen the overall normalization of the graphs to be
  such that the lowest-order Born graph gives just $f(x)$.
}
\begin{equation}
   \frac {g^{2}}{8\pi ^{2}} C_{F} \int _{0}^{1} dz \int _{0}^{1-x} du
\,
f\left( \frac {x}{1-u} \right) \;
   \left[
        \frac {1-z}{zu}
        - \frac {\theta (z<z_{\rm cut})}{zu}
   \right] .
\label{Graph.a}
\end{equation}
The cut-off $z_{\rm cut}$ on the subtraction term is analogous to
the cut-off $\mu _{c}$ we used for the UV counterterm for the
self-energy graph.
The value of $z_{\rm cut}$ needed to reproduce the $\MSbar$
prescription can be found by a simple calculation from the
Feynman rules for parton densities.  But we will not use this
result here.  Rather, we will aim at calculating an appropriate
value for $z_{\rm cut}$ to keep the one-loop correction down to a
``normal size'' (and, most importantly, whether it is possible to
find such an appropriate value at all).
This effectively amounts to a choice of factorization scheme.
Once a suitable value for $z_{\rm cut}$
has been obtained, it is a mechanical matter to translate
it to a value for $\mu _{\MSbar}$ (or to a value of the scale $\mu $ in
any other chosen scheme).  The calculation may also result in a
need for an extra {\em finite} counterterm.

The subtraction in Eq.\ (\ref{Graph.a}) has evidently
accomplished its purpose of cancelling the initial-state
singularity.  But we are still left with the singularity on the
line $u=0$.
This singularity will cancel against a singularity in the virtual
graph, as we will now see.

\subsection{Sum of virtual and real graphs}

Next we compute the virtual graph of Fig.\ \ref{fig:1-loop.DIS}(b),
following the same line as in the previous subsection.
We will construct an integral in the same variables as the real
graph.  The reason why we do this is that the cancelation of
the divergence at $u=0$ will be point-by-point in the integrand.
This can be seen from the proof by Libby and Sterman \cite{LibSt}.
They treat a general case of final-state interactions, of which
our example is a particular case. They first treat one
integration analytically, with the aid of the mass-shell
conditions for the final state, in such a way that the
integrations for graphs related by different positions of the
final-state cut then have the same dimensions. After that, the
cancelation between the different graphs is point-by-point in
the integrand.

This implies that we need to perform the $\xi $ and ${\bf k}_{T}$ integrals.
We do the convolution with $\xi $ by the mass-shell $\delta $-function,
which now gives $\xi =x$.
Then we do the
${\bf k}_{T}$-integral analytically. (This is not the most trivial
integral, but it works conveniently with our choice of variables.)
An example of the type of integral encountered is:
\begin{equation}
 \int _{0}^{\infty }\frac {dk_{T}^{2}}{[Q^{2}(z-1)u-k_{T}^{2}+i\epsilon ] \,
[Q^{2}(u-1)z-k_{T}^{2}+i\epsilon ] \, [Q^{2}uz-k_{T}^{2}+i\epsilon ]} .
\nonumber
\end{equation}
In the virtual case, no longer does a positive energy condition restrict
the range of $u$ and $z$.  Nevertheless, it is convenient to split up the
result into a piece inside the square $0 \leq  u,z \leq  1$ and a piece
from outside the square.  It is sufficient to examine the
contribution within the square.  After subtracting the collinear
singularity and adding the result from the real graph we
get\footnote{
    In (\ref{Graph.a}) the upper limit on $u$ is $1-x$, but
    we replace the limit by $1$ when we copy the formula into
    (\ref{Wilson.1}).  This
    change is innocuous since the limit $1-x$ arises from the fact
    that the parton density $f(\xi )$ is 0 when $\xi  > 1$, and
    hence that $f(x/(1-u))$ is 0 when $u > 1-x$. The limits on
    $u$ that result from positivity of energy of the two
    final-state lines in Fig.\ \ref{fig:1-loop.DIS}(a) are
    just $0 < u < 1$.
}
\begin{eqnarray}
   && \frac {g^{2}}{8\pi ^{2}} C_{F} \int _{0}^{1} \int _{0}^{1} du \, dz \;
   \left[ f\left( \frac {x}{1-u} \right) - (1-u)f(x)
   \right]
   \; \left[
        \frac {1-z}{zu} - \frac {\theta (z<z_{\rm cut})}{zu}
   \right]
\nonumber\\
    &&
    + \ \mbox{contribution from {\em outside} $0 \leq  u,z \leq  1$} .
\label{Wilson.1}
\end{eqnarray}
The UV divergence ($k\to \infty $) is from outside the square and is
cancelled by subtractions just as for the self-energy graph.

\begin{figure}
\centering
\mbox{\epsfig{file=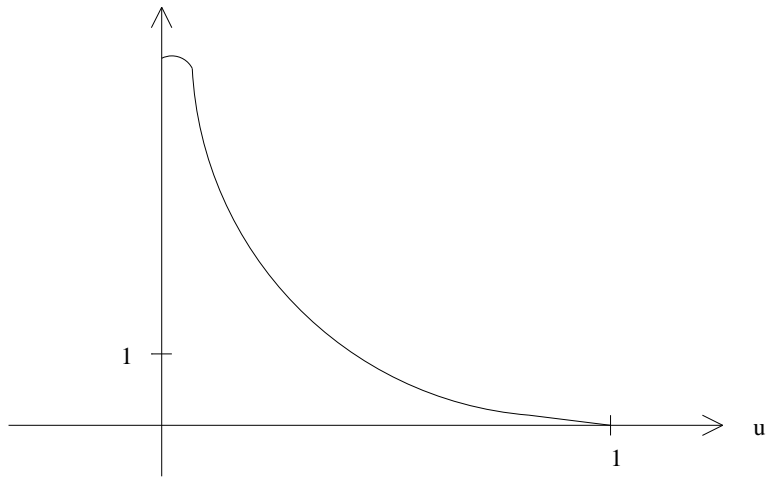,height=6.3cm}}
\vspace*{5mm}
\caption{The integrand of Eq.\ (\protect\ref{Wilson.1}) as a
         function of $u$ at fixed $z$ when $f(x)$ is a steeply
         falling function of $x$.}
\label{fig:integrand}
\end{figure}

We see that the singularity at $u=0$ for fixed $z$ has cancelled,
even at the intersection of the two singular lines.
However, how good the cancelation is depends on the test
function.  If $f(x)$ is slowly varying, the cancelation is good
over a wide range of $u$.  But if $f(x)$ is steeply falling as
$x$ increases, the cancelation is good only over a narrow range
of $u$, and we are left with an integrand that behaves like $1/u$
for larger $u$.
For fixed $z$ we then have an integrand like
that in Fig.\ \ref{fig:integrand}.

This kind of behavior is quite typical when singularities are
cancelled between different graphs with different final states.
The integrand is a non-trivial distribution, and how good the
cancelation of the singularity is depends on properties of the
test function. The cancelation only occurs after integration
over a range of final states.  This is in distinct contrast with
the case of the initial-state singularity for which we have
constructed an explicit subtraction.  The cancelation of the
$z \to  0$ singularity occurs {\em before} the integration with the
test function.

Immediately we also get complications in computing the typical
size of the graph.
In particular, when $f(x)$ is steeply falling, we should
therefore expect a size for the integral that is much larger than
the natural size we defined earlier.

\subsection{Estimate of size}
\label{subsec:estimate}
First, as a benchmark case, let us assume that $f(x/(1-u))$ is slowly
and smoothly varying when $u$ increases from 0 to $1/2$, and
that\footnote{
   \label{z.equals.half}
   The reason for using `$u=1/2$' in these criteria rather than,
   say, $u=1$ is the same as for using $\mu =m$ in the calculation
   of the self-energy.  It is a rough attempt to optimize the
   errors without using the details of the integrand, since
   $u=1/2$ is midway between the singularity at $u=0$ and the
   approximate edge of the region of integration.  The
   integration region for the real graph extends to $u=1-x$,
   while that for the virtual graph extends beyond $u=1$.
   Analogous reasoning applies to $z$ and the relation between
   the real graph and its eikonal approximation even though, in this
   case, one has to look at the whole integrand since both test functions
   are of the form $f \left (\frac {x}{1-u} \right )$.
}
$z_{\rm cut}$ is around $1/2$.  Then the size of
(\ref{Wilson.1}) can be estimated as
\begin{equation}
   \frac {g^{2}}{8\pi ^{2}}
   \times  \left( \mbox{group theory} \right)
   \times  \left( \mbox{area of unit square in $(u,z)$} \right)
   \times  f(x)
   \simeq \frac {g^{2}}{6\pi ^{2}} f(x) .
\label{Wilson.nat}
\end{equation}
A way of obtaining this result with the same method as we used
for the self-energy is to write
\begin{equation}
   \frac {g^{2}}{(2\pi )^{4}} C_{F}
   \times  \Bigg( \mbox{range of $k$, i.e., $2\pi ^{2}Q^{4}$}
     \Bigg)
   \times  \Bigg( \mbox{size of integrand, $\displaystyle \frac {f(x)}{Q^{4}}$}
     \Bigg) .
\end{equation}
Given that the lowest-order graph is $f(x)$, and that the
integrand varies in sign, all this implies that the contribution
of this graph (with all the cuts and subtractions) is probably
somewhat smaller than $g^{2}/6\pi ^{2}$ times the lowest-order graph.
In other words, the simplest estimate for the graphs, Eq.\
(\ref{natural.size.QCD}), is a valid estimate in this situation.

There are a modest number of graphs, so this result would have
very nice implications for the good behavior of perturbation
theory: the real expansion parameter of QCD would be $\alpha _{s}/\pi $,
which is a few per cent in many practical situations.

Unfortunately, the conclusion is vitiated when $f(x)$ is steep,
as is often the case.  Consider the parton density factor times
the $1/u$ factor, relative to the lowest-order factor $f(x)$.  In
the limit $u \to  0$, this is
\begin{equation}
   \frac {f \left( \frac {x}{1-u} \right) - f(x)}{u f(x)}
   \sim \frac {xf'(x)}{f(x)} .
\label{u.to.0}
\end{equation}
This factor should be of order unity, if the previous estimate of
the size, Eq.\ (\ref{Wilson.nat}), is to be valid.

But if the logarithmic derivative $xf'(x) / f(x)$ is much bigger
than 1, then we have to change our estimates.  Consider a
typical ansatz for a parton density:
\begin{equation}
   f(x) \propto  (1-x)^{6} ,
\end{equation}
for which
\begin{equation}
   \left|
        \frac {xf'}{f}
   \right|
   = \frac {6x}{1-x} .
\end{equation}
This is 6 when $x=1/2$, and goes to infinity as $x \to 1$.
Clearly our estimate in Eq.\ (\ref{Wilson.nat}) is bad.  Moreover
the $u \to 0$ estimate, Eq.\ (\ref{u.to.0}), is only approximately
valid when $f(x)$ does not change by more than a factor 2
(roughly).  Once $u$ gets larger than the inverse of the
logarithmic derivative, the $f(x/(1-u))$ term
in Eq.\ (\ref{Wilson.1})
is no longer
important, and we get the result pictured in Fig.\
\ref{fig:integrand}: we have basically a
$1/u$ form with a cut-off at small $u$.  This is a recipe for a
large logarithm, with the argument of the logarithm being the
large logarithmic derivative.

To make the estimate, it is convenient to define
\begin{equation}
    \delta u = -\frac {f(x)}{2f'(x)x} .
\label{div.u}
\end{equation}
{}From Eq.\ (\ref{u.to.0}), we see that $\delta u$ is approximately the
change in $u$ to make $f(x/(1-u))$ fall by a factor of 2.
We interpret $\delta u$ as the value of $u$ at which the final-state
cancelations become ``bad''.

Next we note that for normal parton densities $f(x)$ is a
decreasing function of $x$.  So a simple useful estimate can be
made by making the following approximation:
\begin{equation}
    f\left ( \frac {x}{1-u} \right ) \simeq
    \left\{
    \begin{array}{l l}
        \displaystyle
        f(x)\left [1-\frac {u}{2\delta u}\right ] & {\rm if~} u < 2\delta u ,
    \\
        0                        & {\rm if~} u > 2\delta u
    \end{array}
    \right . .
\label{exp.u}
\end{equation}
Therefore, in Eq.\ (\ref{Wilson.1}), we
can replace $\int _{0}^{1} \frac {du}{u}f(x/(1-u))$ by
$\int _{0}^{2\delta u} \frac {du}{u} f(x)$.

Our estimate for Eq.\ (\ref{Wilson.1}) is the sum of
contributions from the following regions:
\begin{itemize}

\item
    $0<u<2\delta u$:
    The value of the integrand is about $-f(x)/2\delta u$ times a
    function of $z$.  For good choices of $z_{\rm cut}$, the
    function of $z$ is less than about unity, but with an
    indefinite sign.  Thus we obtain a contribution of about
    $f(x)$ in size.

\item
    $2\delta u<u<1$:
    The integrand is now approximately $-f(x)/u$,
    again times a mild function of
    $z$.  We therefore obtain a contribution of about
    $f(x)\ln(1/2\delta u)$.

\item
    Exterior of unit square:
    Here, the only contribution is from the virtual graph, and we
    have no final-state singularity. Hence the naive estimate of
    unity is valid.
    (The precise value, when $\mu =Q$, is in fact somewhat larger.)

\end{itemize}
All of these are to be multiplied by the prefactor $g^{2}C_{F}/8\pi ^{2}$.
So we obtain a total contribution of
\begin{equation}
   \frac {g^{2}}{8\pi ^{2}} C_{F} f(x) \left[ \pm 2 \pm  \ln \left( \frac
{1}{2\delta u} \right)
               \right] ,
\label{estimate.vertex}
\end{equation}
where each term represents an estimate, valid up to a factor of 2
or so.  The contributions from within the unit square may have either
sign, depending on the cut in the collinear subtraction, while
the contributions from outside the unit square, from virtual
graphs only, have a negative sign.
As an explicit indication that our estimates are valid for the
sizes but not the signs of the graphs, we have inserted a $\pm $
sign in front of each term.
This estimate assumes that renormalization of the UV divergence
of the virtual graph is done at the natural scale $\mu  \simeq Q$,
and that renormalization of the parton densities is done so that
it corresponds to $z_{\rm cut}$ of about 1/2.
It also assumes that $f(x)$ falls steeply enough for $\delta u$ to be
less than about 1, as is typically true.

\subsection{Interpretation}

It is obvious that there is a logarithmic enhancement in Eq.\
(\ref{estimate.vertex}) whenever $\delta u$ is small. Our calculation
is, of course, no more than a rederivation of the standard
observation that there are large logarithms in the $x \to 1$ region.
What our derivation adds is to show that it is not so much the
limit $x \to 1$ that is causing the problem as the steepness of the
parton densities. Moreover we have given a numerical criterion
for when the correction begins to be larger than what we called
the natural size for higher-order corrections.  In other words we
have shown how to estimate the constant term that accompanies the
logarithm.  Moreover this is all presented in the context of a
general method for obtaining estimates of the sizes of graphs.

It is perhaps clear that, with sufficient foresight, one could
have predicted the large corrections merely from the observation
that the derivation of the factorization theorem requires the
cancelation of final-state divergences between different final
states.

There is in fact another source of large corrections that will
make its effect felt in even higher order.  This is a mismatch
in the scales needed for renormalizing the parton densities.  We
have renormalized these by using a value of
$z_{\rm cut}$ that must be about 1/2 to avoid making the
contribution of the graph unnecessarily large.  In the case of
the real graph, we can translate this to a scale of transverse
momentum by using the mass-shell condition for the gluon:
\begin{equation}
   k_{T}^{2} = Q^{2} u z (1-z) \frac {\xi }{x} = Q^{2} \frac {u z (1-z)}{1-u} .
\end{equation}
Evidently, whenever small values of $u$ are important, small
values of $k_{T}$ (relative to $Q$) will be important. This does not
affect our one-loop calculation. But in higher-order correction,
the virtuality of some internal lines will be controlled by the
value of $k_{T}$, and hence there will be mismatches between the
scales needed at different steps in the calculation.

Once one has diagnosed the problem, we see that a proper solution
lies in more accurately calculating the form of the Wilson
coefficient near its singularity.  This subject goes under the
heading of resummation of large corrections \cite{resum}.

\section{More detailed estimation of $F_{T,L}$ to one-loop order}
\label{sec:full.estimate}

In the following we will estimate the sizes of the
one-loop corrections to
the structure functions $F_{T}$ and $F_{L}$ {\em without} doing actual
calculations of Feynman diagrams by giving a recipe of how to construct
the estimates from general principles and kinematic considerations.
However, we will present the recipe in the context of an actual set of
Feynman graphs.
By using the calculations of the graphs in Sec.\ \ref{sec:appendix}, we will
verify that these ``simple-minded'' estimates are actually valid.

One obtains $F_{T,L}$ from the well-known hadronic tensor
$W^{\mu \nu }$ by
projecting out the ``transverse'' and ``longitudinal'' pieces via:
\begin{eqnarray}
F_{T} = -g_{\mu \nu }W^{\mu \nu }&=& 3F_{1} - \frac {F_{2}}{2x} \nonumber\\
F_{L} = \frac {Q^{2} p_{\mu }p_{\nu }}{p\cdot q^{2}}W^{\mu \nu }&=& -F_{1} +
\frac {F_{2}}{2x} ,
\end{eqnarray}
which give, for example,
\begin{equation}
F_{1} = \frac {1}{2} \left [-g_{\mu \nu } - \frac {Q^{2} p_{\mu }p_{\nu
}}{p\cdot q^{2}} \right ]W^{\mu \nu } .
\end{equation}

\subsection{Estimation of $F_{L}$}

The recipe for estimating $F_{L}$ is the following:
\begin{itemize}

\item The singularities we encounter (UV, collinear and soft) are
      all in the
      form of a factor times the Born graph, and the Born graph has no
      longitudinal part. Therefore the one-loop graphs for $F_{L}$ have no
      UV, soft or collinear singularities.

\item Since the parton densities are falling with increasing $x$, the size
      of a graph is:
      \begin{equation}
        \frac {g^{2}}{8\pi ^{2}}\times C_{F}\times f(x)\times \mbox{{\rm range
of $u$}}
        \times \mbox{{\rm range of $z$}}.
      \end{equation}

\item The range of $z$ is 1.

\item The range of $u$ is $\delta u$.

\end{itemize}
It is elementary to show that the self-energy and vertex graphs
(whether real or virtual) give a zero contribution to $F_{L}$.
Therefore, our result for $F_{L}$ is:
\begin{equation}
    F_{L} = \frac {g^{2}}{6\pi ^{2}}f(x)\delta u .
\label{FL.estimate}
\end{equation}

Let us now check whether our intuition has guided us in the right way.
We use Eq.\ (\ref{FL.ladder}) for the contribution to the
coefficient function for $F_{L}$.  We approximate the $u$ integral
by
\begin{equation}
   \int  du f\left( \frac {x}{1-u} \right)
   \simeq \delta u f(x) ,
\end{equation}
which is appropriate for a typical parton density, which falls
with increasing $x$.  The $z$ integral gives a factor $1/2$, and
we recover Eq.\ (\ref{FL.estimate}), which we obtained by more
general arguments.

Notice that $F_{L}$ is generally rather smaller than what we have
termed the natural size, because of the $\delta u$ factor: i.e.,
because of the restricted phase space available.  There are no
enhancements due to final-state singularities.

Our estimation methods can be applied in two ways.  One is to
estimate the graphs without having explicit expressions for the
graphs; one just searches for the possible singularities that
would prevent the natural size of a graph from being its actual
size.  The second way of using the methods is to examine the
expressions for graphs with the knowledge of the singularity and
subtraction structure and to perform a more direct estimate of
the sizes.  This is useful since the integrals must often be
performed numerically, e.g., whenever one convolutes with parton
densities that are only known numerically. Additional information
that is now obtained concerns the typical virtualities, etc., of the
internal lines of the graphs (compare Neubert's work
\cite{Neubert}).  This enables a diagnosis to be made of the
extent to which a problem is a multi-scale problem and therefore
in need of resummation.

A point that we have not addressed is the estimation of the size
of the trace of a string of gamma matrices, for example.  In the
standard formula for such a trace involves a large number of
terms.  Nevertheless, it is evident from the above calculations
that there are cancelations.  The final result is that we
obtain, relative to corresponding numbers for a scalar field
theory, a small factor (1 to 4) times a standard
Lorentz-invariant quantity for the process.

Evidently, more work on this subject is needed.  But the issue of
the size of the numerator factors from traces, etc., affects
completely finite quantities, such as the one-loop coefficient
function for $F_{L}$, just as much as quantities with divergences
that are cancelled.  However, it is the latter quantities that
have the potential for especially large corrections.  Numerator
factors result in magnitudes common to all graphs.

\subsection{Estimation of $F_{T}$}

We have already calculated one graph, Fig.\ \ref{fig:1-loop.DIS},
for $F_{T}$, so we only need to summarize the method and apply it to
the remaining graphs.  The general procedure is:
\begin{itemize}

\item Since we have singularities in our graphs, they have to be
    cancelled.  In Wilson coefficients such as we are
    calculating, there are explicit subtractions for the collinear
    initial-state singularities.  Then there are explicit
    subtractions for UV divergences (both those associated with
    the interactions and those needed to define the parton
    densities and that therefore enter into the initial-state
    subtractions).  Finally there are final-state singularities
    that cancel between real and virtual graphs.

\item We consider separately the integrations inside and outside
    the square $0 < u, z <1$.

\item A term of the order of the natural size arises from outside the
    square $0 < u, z < 1$.  This comes from purely virtual
    graphs, with their collinear subtraction.

\item Inside the square, a real graph without a final-state
    singularity contributes an amount of the order of the natural size
    times $\delta u$.  This reflects the restriction on the range of
    integration imposed by the parton density.

\item Similarly a virtual graph without a final-state singularity
    contributes a term of the natural size.

\item Finally, the sum of real and virtual graphs with a
    final-state singularity contributes a term of the order of the
    natural size enhanced by a factor $2 + \ln (1/2\delta u)$, just as
    in our estimate of Fig.\ \ref{fig:1-loop.DIS}.

\end{itemize}

We have the graphs of Fig.\ \ref{fig:1-loop.DIS}, their Hermitian
conjugates, the cut and uncut self-energy graph of Fig.\
\ref{fig:qself.energy}, below, and the ladder graph, Fig.\
\ref{fig:remaining.graph}.  Since $\delta u \leq 1 $ typically, the
ladder graph gives a small contribution, and it is sufficient to
multiply the estimate of the Fig.\ \ref{fig:1-loop.DIS} by 3:
\begin{equation}
F_{T} = \frac {g^{2}}{2\pi ^{2}} f(x) \left[ 2 + \ln \left( \frac {1}{2\delta
u} \right)
               \right] .
\label{eq.est}
\end{equation}

To see how this compares with the results from the actual graphs, we
use those in Sec.\ \ref{sec:appendix}.

The cut and uncut self-energy graphs, Fig.\
\ref{fig:qself.energy} below, have final-state singularities, as
can be seen in Eq.\ (\ref{FT.se}).
We found it convenient to use $z$ and $k_{T}$ as integration
variables.  There is a singularity at $k_{T}=0$ in each individual
graph.  In analogy to the definition of $\delta u$, Eq.\ (\ref{div.u}),
we define
\begin{equation}
   \delta k_{T}^{2} = \frac {-Q^{2} f(x)}{2 f'(x)},
\end{equation}
and by following the same steps as we applied to the ladder
graph, we find the following estimate for the contribution of
Fig.\ \ref{fig:qself.energy}:
\begin{equation}
   - \frac {g^{2}}{16\pi ^{2}} C_{F} f(x) \left[ 1 + \ln \left( \frac {\mu
^{2}}{2\delta k_{T}^{2}} \right)
               \right] .
\label{se.estimate}
\end{equation}
We have inserted a $-$ sign in this estimate; it is fairly easy
to see that the coefficient of the logarithm is negative.

For the ladder graph, whose contribution to $F_{T}$ is in Eq.\
(\ref{FT.ladder}), there is only an initial-state singularity and
that is cancelled by an explicit subtraction.  If $z_{\rm cut}$ is
around $1/2$, then we can apply the same reasoning as for the
longitudinal part of the ladder graph, and we find an estimate
\begin{equation}
    \pm \frac {g^{2}}{6\pi ^{2}}f(x)\delta u ,
\label{ladder.T.estimate}
\end{equation}
where, as in Eq.\ (\ref{estimate.vertex}), we use the $\pm $ to
indicate that our estimates do not determine the sign of the
contribution.

We have already examined the cut and uncut vertex graphs, in
Eq.\ (\ref{estimate.vertex}).  When we multiply this by 2 (to
allow for the Hermitian conjugate graphs) and add the self-energy
and ladder contributions, Eqs.\ (\ref{se.estimate}) and
(\ref{ladder.T.estimate}), we get somewhat less than our original
estimate, Eq.\ (\ref{eq.est}), provided the renormalization
mass $\mu $ is in a reasonable range.  This lower value is
because the self-energy graph is simpler than the vertex graph.


\section{The remaining longitudinal and transverse contributions
         to the Wilson coefficient}
\label{sec:appendix}

\subsection{Transverse part of the self-energy}

\begin{figure}
\centering
\mbox{\epsfig{file=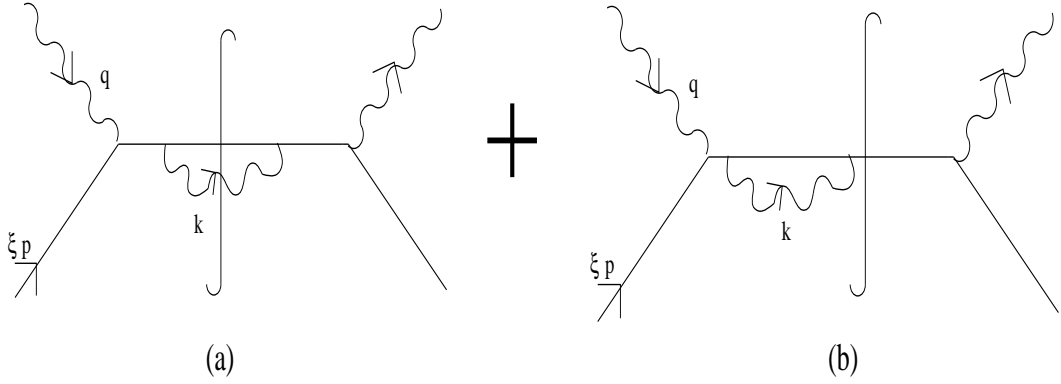,height=5cm,width=14cm}}
\vspace*{5mm}
\caption{(a) Real self-energy graph, (b) virtual self-energy graph.
    There is a second virtual graph that is the conjugate of
    graph (b).
}
\label{fig:qself.energy}
\end{figure}

The real part of the self-energy is computed along the same lines
as mentioned in the main body of the text, except that we chose to
integrate over $u$ instead of $k_{T}$ as a matter of convenience.
The graph and its possible cuts are shown in Fig.\
\ref{fig:qself.energy}.
As far as the virtual part is concerned, one notes first that the general
structure of the quark self-energy for zero mass is \cite{Pokorski}:
\begin{equation}
  \Sigma (\hat {p}) = B(0)\hat {p} ,
\end{equation}
where $B(0)$ is computed via:
\begin{equation}
               B(0) = \frac {1}{4}\left. {\rm Tr}\left(
                  \gamma ^{+}\frac {\partial  \Sigma(\hat p)}{\partial p^{+}})
                \right )   \right \vert _{\hat p=0} .
\end{equation}
The result for the virtual graph is then given by simply multiplying the
Born result by $B(0)$ and convoluting with the test function.
The complete result according to our prescription is:
\begin{equation}
\frac {g^{2}}{8\pi ^{2}} C_{F}\left[ \int ^{1}_{0} dz \int ^{k^{2}_{T,max}}_{0}
d{\bf k}_{T}^{2} \frac {z}{k_{T}^{2}}f\left (\frac {x}{y}\right )
- \int ^{1}_{0} dz \int ^{\mu ^{2}}_{0} d{\bf k}_{T}^{2} \frac {z}{k_{T}^{2}}
f(x) \right ] ,
\label{FT.se}
\end{equation}
where $\frac {1}{y} = 1 + \frac {k_{T}^{2}}{Q^{2}}$, after having made
the change of variable $k_{T}^{2} \rightarrow k_{T}^{2}z(1-z)$,
which also gives $k^{2}_{T,max} = \frac {Q^{2}(1-x)}{x}$.
Renormalization by subtraction of the asymptote is used in our
formula with a cut-off on the momentum integral for the virtual
graph of $\mu ^{2}$.
As one can see there are no initial-state singularities simply because
propagator corrections do not induce initial-state singularities as vertex
corrections do. The final-state singularities are still there;
thus one needs
to look at both the real and the virtual graphs together.

\subsection{The transverse part of the ladder diagram}

\begin{figure}
\centering
\mbox{\epsfig{file=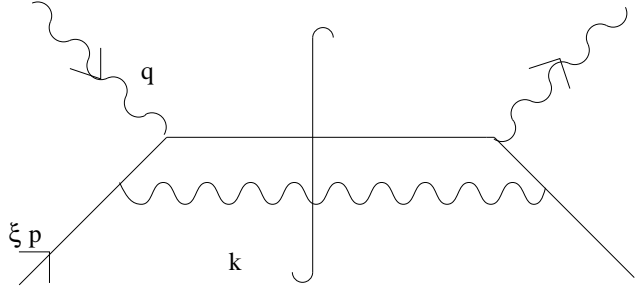,height=3.8cm}}
\vspace*{5mm}
\caption{The ladder graph for the Wilson coefficient.}
\label{fig:remaining.graph}
\end{figure}

The calculation of the ladder graph has been carried out as
outlined in the main text, yielding:
\begin{eqnarray}
\frac {g^{2}}{8\pi ^{2}} C_{F}\int ^{1-x}_{0} du \int ^{1}_{0} dz
\frac {u}{ z(1-u)}
f\left (\frac {x}{1-u}\right ) \left [1 - \theta (z - z_{\rm cut})
\right ] .
\label{FT.ladder}
\end{eqnarray}
The second term, with its $\theta $-function, is the collinear
subtraction, whose UV divergence is cancelled by subtraction of
the asymptote.  The above formula assumes that $z_{\rm cut} < 1$.
If $z_{\rm cut} > 1$, then we must extend the $z$ integration in the
second term beyond the limit $z=1$, of course.

\subsection{Longitudinal part of the vertex correction, self-energy and
ladder diagram}

The calculation has been carried out as outlined in the main text and
yields the following for the ladder graph:
\begin{equation}
\frac {g^{2}}{4\pi ^{2}} C_{F}\int ^{1-x}_{0} du \int ^{1}_{0} dz
(1-z) f\left (\frac {x}{1-u}\right ) .
\label{FL.ladder}
\end{equation}
The real and virtual graphs give $0$ for both the self-energy and the vertex
correction. There is no virtual graph for the ladder diagram.

\section{Conclusions}
\label{sec.conclusions}

\begin{itemize}

\item We have a systematic method for estimating the sizes of
    higher-order graphs.

\item The natural expansion parameter in QCD is of the order of
    \begin{equation}
       \frac {g^{2}}{8\pi ^{2}} \times  \mbox{group theory}
       \times  \mbox{number of graphs}.
    \end{equation}
    In answer to a question asked when this work was presented at
    a conference, let us observe: {\em The above number may not
    be the actual expansion parameter, but we argue that this
    natural size sets a measure of whether actual higher-order
    corrections are of a normal size or are especially large. }

\item The method allows an identification of appropriate sizes
    for renormalization and factorization scales.

\item By an examination of the kinematics of graphs,
    we can identify contributions that are large compared with
    the estimates based merely on the size of the natural
    expansion parameter.

\item Our method should give a systematic technique of diagnosing
    the reasons for the large corrections, and hence of
    indicating where one should work out resummation methods.

\item The method gives an algorithm for the numerical integration
    of graphs, both real and virtual.

\end{itemize}

Of course, there is also
a lot of overlap with work such as that of Catani and Seymour
\cite{Catani.Seymour}, of Brodsky and Lu \cite{BL}, and of
Neubert \cite{Neubert}.

\end{document}